\documentclass[preprint, 10pt]{elsarticle}
\usepackage{amsmath}

\usepackage{graphicx}
\usepackage[version=4]{mhchem}
\usepackage{amssymb}

\usepackage{color}

\let\mpar=\marginpar
\renewcommand\marginpar[1]{\mpar{\raggedright \scriptsize #1}}
\usepackage{booktabs}
\usepackage{ulem}
\usepackage{xr}
\usepackage{color}

\usepackage[labelfont=bf,labelsep=period,justification=raggedright]{caption}

\makeatletter
\renewcommand{\@biblabel}[1]{\quad#1.}
\makeatother
\date{}

\usepackage{graphicx}
\usepackage{dcolumn}
\usepackage{bm}
\usepackage{hyperref}

\def\be{\begin{equation}}   \def\ee{\end{equation}}

\usepackage{xr}
\begin{document}

\begin{flushleft}
{\Large
\textbf{Figure 1 Theory Meets Figure 2 Experiments in the Study of Gene Expression}
}\\
\bf{Rob Phillips}$^{1,2,\ast}$,\bf{Nathan M. Belliveau}$^2$, \bf{ Griffin Chure}$^2$,  \bf{ Hernan G. Garcia}$^3$, \bf{Manuel Razo-Mejia}$^2$, \bf{Clarissa Scholes}$^4$
\\
\bf{1} Dept. of  Physics, California Institute of Technology, Pasadena, California, U.S.A
\\
\bf{2} Division of Biology and Biological Engineering, California Institute of Technology, Pasadena, California, U.S.A
\\
\bf{3} Department of Molecular \& Cell Biology, Department of Physics, Biophysics Graduate Group and Institute for Quantitative Biosciences-QB3, University of California, Berkeley, California, U.S.A
\\
\bf{4} Department of Systems Biology, Harvard Medical School, Boston, Massachusetts, U.S.A
\\
$\ast$ E-mail: phillips@pboc.caltech.edu
\end{flushleft}

\noindent {\it ``hic rhodus, hic salta''   - Aesop's fables}\\


\section*{Abstract} It is tempting to believe that we now own the genome. The
ability to read and re-write it at will has ushered in a stunning period in
the history of science. Nonetheless, there is an Achilles heel exposed by
all of the genomic data that has accrued:  we still don't know how to interpret it. Many
genes are subject to sophisticated programs of transcriptional regulation, mediated by  DNA sequences that harbor binding sites for transcription factors which can up- or
down-regulate gene expression depending upon environmental conditions. This gives rise to
an input-output function describing how the level of expression depends upon
the parameters of the regulated gene -- for instance, on the number and type of binding sites in its regulatory sequence.  In recent years, the ability to make
precision measurements of expression, coupled with the ability to make increasingly sophisticated theoretical predictions, have enabled an
explicit dialogue between theory and experiment that holds the promise of
covering this genomic Achilles heel. The goal is to reach a predictive
understanding of transcriptional regulation that makes it possible to calculate
gene expression levels from DNA regulatory sequence. This review focuses on the canonical simple repression motif to ask how well the models that have been used to characterize
it actually work. We consider a hierarchy of
increasingly sophisticated experiments in which the minimal parameter set
learned at one level is applied to make quantitative predictions at the next.
We show that these careful
quantitative dissections provide a template for a predictive understanding of
the many more complex regulatory arrangements found across all domains of life.\\

\section{Introduction}

The study of transcriptional regulation is one of the centerpieces of modern
biology.  It was set in motion by the revolutionary work of Jacob and Monod in the postwar
era, which culminated in their elucidating the concept of transcriptional
regulation in the early 1960s \cite{Jacob1961, Monod1961, Monod1963}, and it has continued
apace ever since. Based on their study of the {\it lac} operon and regulation of the life cycle of bacterial viruses, Jacob and Monod hypothesised that transciption was controlled using a mechanism sometimes known as the repressor-operator model, in which repressive factors bind to promoters at sites called operators to prevent activation of genes. Here, we will refer to this regulatory architecture as the simple-repression motif.

\begin{figure}
\centering{\includegraphics[width=5.5truein]{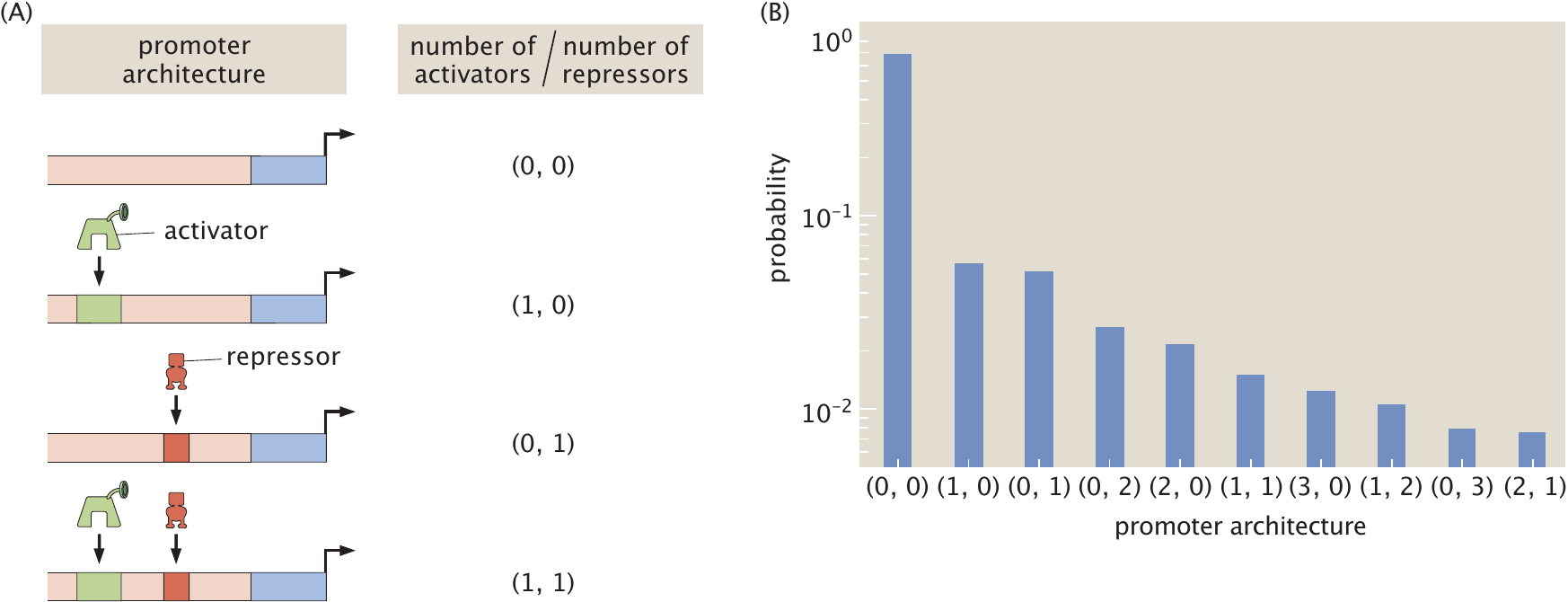}}
\caption{The distribution of regulatory architectures in {\it E. coli}.  (A) Several of the simplest
regulatory architectures are shown, featuring activator and repressor binding sites.  We adopt the notation $(m,n)$ to characterize these architectures where the first number $m$ tells us how many activator binding sites there are for our gene of interest and the second number
$n$ tells us how many repressor binding sites are controlling that same gene.   Within this
notation, a $(0,0)$ architecture is unregulated, a $(1,0)$ architecture is a simple activation motif and a $(0,1)$ architecture is a simple repression
motif and is the central focus of the present article.  (B)  Relative probability of different classes of regulatory architecture
for those genes that have been annotated in {\it E. coli}~\cite{Rydenfelt2014, GamaCastro2016}. For transcription factors that can act both as activators and
repressors, we consider their specific mode of action in the context of each regulatory architecture. For example, if a transcription factor binds to a single site near a promoter and acts as an activator, we consider it to fall within the (1,0) nomenclature even if this same protein can act as a repressor on other regulatory units.
\label{fig:ArchitecturesSummarized}}
\end{figure}

\begin{figure}
\centering{\includegraphics[width=4.0truein]{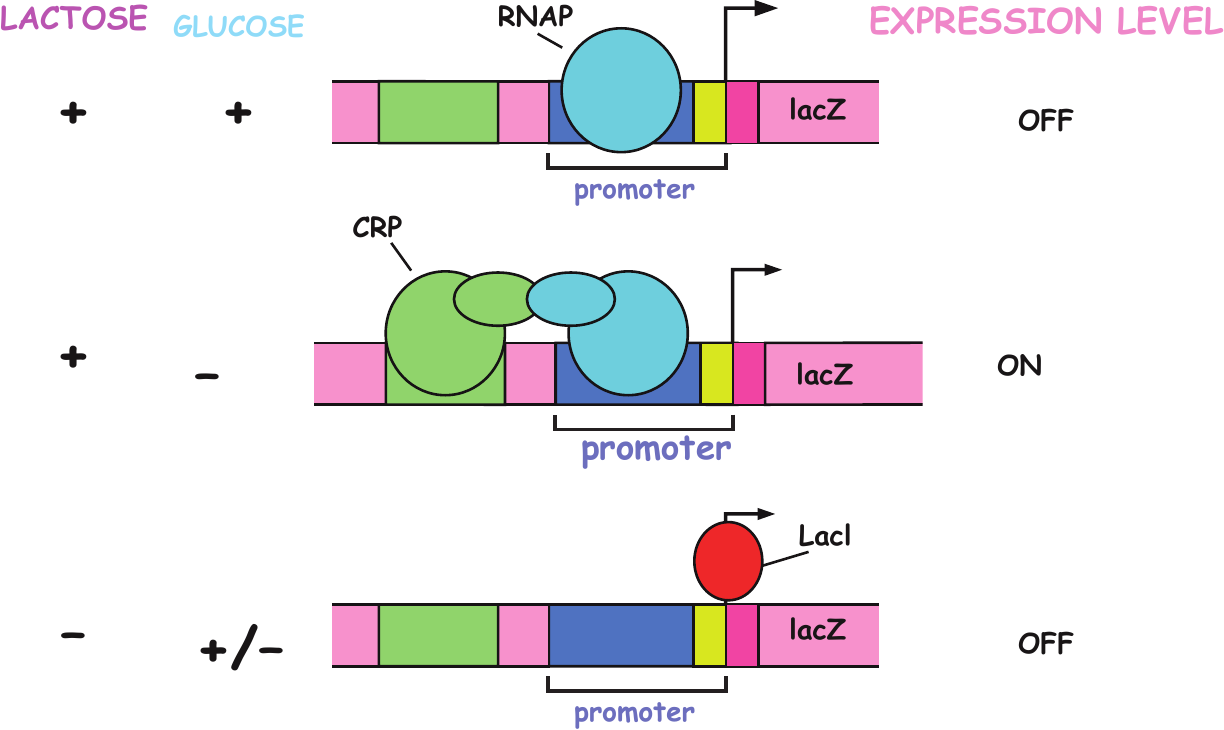}}
\caption{The high school {\it lac} operon.  The classic story of how bacteria utilize lactose rather
than glucose as a carbon source is the canonical example used to teach the concept of genetic regulation. The figure shows that only when lactose is present and glucose is absent will
the gene for the enzyme used to digest lactose be turned on.   The activator is shown in
green, the repressor is shown in red, and RNA polymerase is depicted in blue.
\label{fig:HighSchoolLac}}
\end{figure}

Jacob and Monod suspected that there would be a universal mechanism for transcriptional regulation that followed the strictures of the repressor-operator model; indeed, simple-repression, defined diagrammatically in Figure~\ref{fig:ArchitecturesSummarized}(A), has since been shown to have widespread applicability as seen in Figure~\ref{fig:ArchitecturesSummarized}(B). However, transcriptional reality is -- as is usually the case in biology -- far more
complicated~\cite{Britten1969}, and, as Figure~\ref{fig:ArchitecturesSummarized} reveals, many genes are in fact subject to both negative and
positive regulation. Ironically, the genetic circuit used by Monod to formulate
the repressor-operator model -- the {\it
lac} operon shown in Figure~\ref{fig:HighSchoolLac} --  is itself subject to  positive regulation,  which shows the repressor-operator model to be incomplete \cite{Zubay1970,
Emmer1970}.

The {\it lac} operon is one of the canonical case studies learned by high-school and college students alike when they are first introduced to the
logic of gene regulation in modern molecular and cellular
biology~\cite{Muller-Hill1996,Alberts2017}. Figure~\ref{fig:HighSchoolLac} shows in cartoon form how the gene that encodes the enzyme for digesting lactose is activated only when lactose is present and glucose is absent. This textbook case of transcriptional regulation has been studied to death, but how well do we really understand it?  The sketch in Figure~\ref{fig:HighSchoolLac} is a broad-brush view of transcriptional control at the {\it lac} operon, but it gives us no sense of how the level of gene expression is affected by, for example, changing the
copy numbers of the LacI and CRP transcription factors, changing the positioning of the operator, or
titrating the  relative concentrations of glucose and lactose. We argue that
achieving real understanding of this system requires that we are capable of making precise and
quantitative predictions about its regulatory response as a function of all
these parameters, and then that we are able to confirm these predictions experimentally.

How could we achieve this mastery?  First, we would need
theoretical models able to provide quantitative predictions that can be tested
with careful experiments. Importantly, both the predictions and the experiments themselves would need
to access the same underlying knobs to control the level of gene expression.
Second, we would need to start with the simplest of regulatory architectures. If we are unable to
understand the most basic regulatory kernel, we have no hope of doing so for more
complex regulatory circuits. Third, to dissect more subtle features of a regulatory circuit -- for instance,
to understand how expression noise depends on changing parameters -- we must be able to use  quantitative information
gleaned from one type of experiment to formulate further predictions that are tested in subsequent experiments of a different type.
Therefore, we would need to conduct all these experiments in the same system and under standardized conditions.

\begin{figure}
\centering{\includegraphics[width=4.0truein]{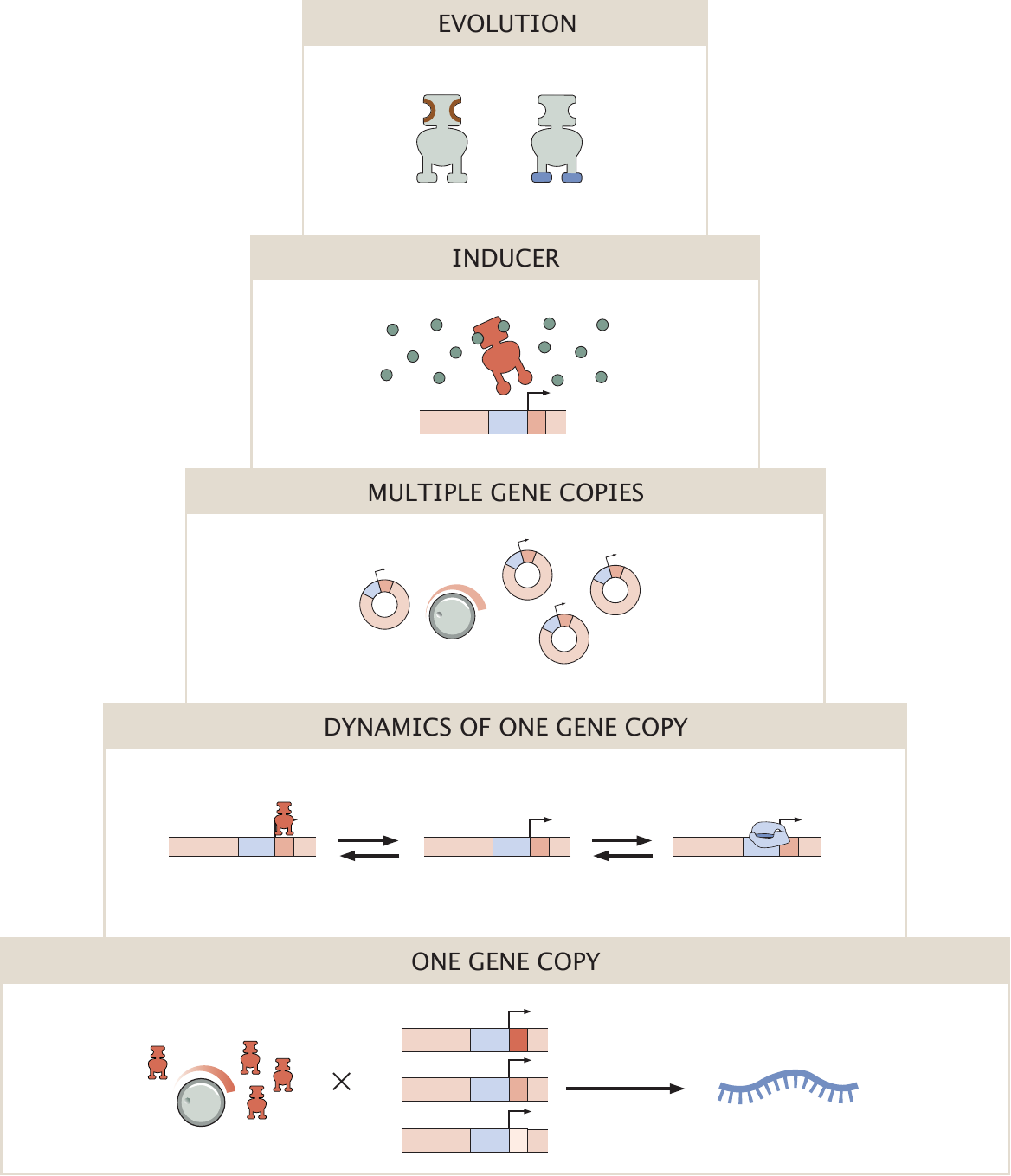}}
\caption{The simple repression pyramid.  A progression of different experiments
makes it possible to assess increasingly subtle regulatory effects for the simple repression motif.
Parameters inferred from lower levels in the pyramid are used in the analysis of
the experiments at the next level. The repressor (and its binding site) is shown in red,
RNAP polymerase (and its binding site) is shown in blue, and inducer in green.
\label{fig:AztecPyramid}}
\end{figure}

This review summarizes such an approach, which we have taken in our own laboratory over the past decade.
We discuss how, working with a set of specifically-designed synthetic constructs and challenging theoretical
models with experiments, we have been able to tackle increasingly subtle behaviors of the simple repression
architecture in {\it E. coli}. The strategy we have taken results in a pyramidal structure, as shown in
Figure~\ref{fig:AztecPyramid}, in which parameters inferred at one level are used to make quantitative predictions
about gene expression behavior in successive, more sophisticated experiments.

At the foundation of the simple repression pyramid are experiments to determine how gene expression responds to changes in
operator strength and repressor copy number. With this information in hand, we can then consider
the entire distribution of expression levels among a population of cells, as opposed to simply the average expression.
At the next level in the hierarchy, we address a number of subtle and beautiful
effects that arise when there is more than one copy of our gene of interest
or competing binding sites for the repressor elsewhere on the genome (or on plasmids).
This repressor titration effect provides a very stringent test of our understanding
of the simple repression motif.  Of course, much of gene expression is
dictated by the presence of environmental signals and the next level in
the simple repression pyramid is to ensure that these same kind of
predictive models can describe induction of transcription.
Further, changes in the environment such as media quality or growth temperature
certainly have an effect on the bacterial doubling rate. The next challenge is then to retain
predictive power by describing how
these different conditions affect the magnitude of parameters that are the basis of these
models, such as repressor copy number and binding energy. Finally, evolution of transcription
acts both at the level of transcription factor binding sites and the transcription factors
that bind to them.  Ultimately, the simple repression pyramid will be topped off
by learning the rules that relate transcriptional regulation to fitness~\cite{Gerland2002,
Berg2004a, Lassig2007, Lynch2015}.  At every level in
the pyramid, we demand that the parameters be self-consistent. That is,
regardless of how our experiments are done or which new question we ask,
the same minimal parameter set is used without recourse to new fits for
each new experiment.

Note that this paper is not a review of a field; rather it is a review of a concept,
in which one minimal parameter set is asked to describe all measurements
on a particular realization of the simple-repression motif.
This objective is not served by an approach in which
different measurements are taken from disparate sources on different strains
under different conditions.  We focus instead on measurements made
using the same strains under the same growth conditions throughout, and this renders the discussion
highly self-referential. But everything we have done was enabled by beautiful work that has come before and inspired
by wonderful experiments since; we point the reader towards as much of this literature as possible.

The goal of this review is to address
whether, for simple repression, we have reached a self-consistent theoretical picture
that stands up to careful experimental scrutiny.  After an overview of regulatory architectures in {\it E. coli},
and the simple repression motif in particular, we describe our systematic effort to make the strains, tune the
relevant knobs, and make the high precision measurements that enable us to test theoretical predictions about
how the simple repression architecture behaves. In the following sections we then address the key critiques of the theoretical framework,
before stepping back to discuss what our results entail for future efforts in understanding gene regulation.
We argue that we have achieved significant success using this hierarchical approach and that it provides hope for
understanding other, more complex, gene regulatory circuits. Indeed, the great work done by others
in {\it lac}~\cite{Saiz2005,Kuhlman2007, Saiz2013}, MarA \cite{Alekshun1997,Martin2008},
GalR \cite{Semsey2002, Semsey2004,SwintKruse2009}, Lambda \cite{Dodd2004, Dodd2005, Ptashne2004,Zong2010},
and AraC \cite{Schleif2010} lends itself to providing the fundamental stepping stones for building other transcriptional pyramids.

\begin{figure}
\centering{\includegraphics[width=6.0truein]{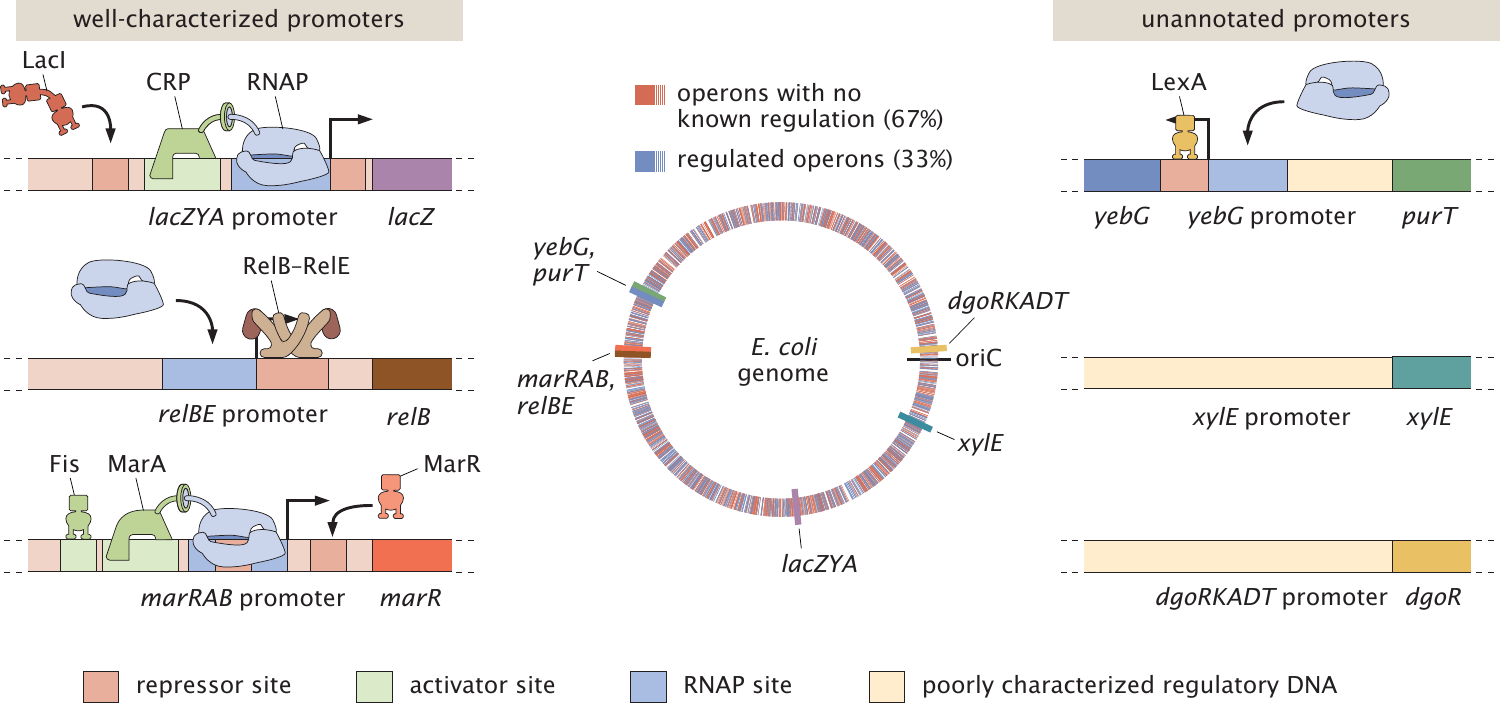}}
\caption{Regulatory ignorance in {\it E. coli}.  The central figure which schematizes the {\it E. coli} genome shows the fraction of the operons for which we know nothing about how they are regulated.
The left panels show examples of the knowledge of regulatory architectures required
to unleash the kind of theory-experiment dialogue described here.  The right side panel
shows the more common situation which is complete regulatory ignorance.
\label{fig:RegulatoryIgnorance}}
\end{figure}

\section{The Regulatory Landscape in \emph{E. coli} and the Ubiquitous Simple Repression Architecture}

Despite the dominance of {\it E. coli} as a model system for studying gene regulation, we remarkably have little or no idea how most
of its genes are controlled. As Figure~\ref{fig:RegulatoryIgnorance} demonstrates, for the majority of genes we don't know the identity of the transcription
factors that turn them on/off, where the binding motifs for those
transcription factors are, or what the regulatory logic is (at the most basic level, whether they are controlled by repressors, activators or a combination of both).
Figure~\ref{fig:ArchitecturesSummarized} provides an incomplete, but state-of-the-art
picture of our current knowledge of the regulatory landscape  by showing the distribution of different architecture types in
{\it E. coli}.
Shortly after the elucidation of the repressor-operator model (the $(0,1)$ motif) that introduced
the simple repression architecture we focus on here,
the idea of activation as a regulatory mechanism also took root.
But as we see in Figure~\ref{fig:ArchitecturesSummarized}(B), at the time of this writing, most genes in {\it E. coli} are annotated as unregulated.
This sounds counterintuitive, but for many genes it likely reflects ignorance of the binding motifs and regulators as opposed to actual
lack of any regulation. Simple repression (along with simple activation) comes in as the next most prevalent
architecture, and we now turn our attention there.

%

%

Simple repression is a common regulatory motif in \textit{E. coli} \cite{Rydenfelt2014}, but
we know little of the general principles by which it is used. To tackle this, we used
annotated regulatory information from RegulonDB~\cite{GamaCastro2016} to survey 156
promoters with a simple repression architecture, controlled by 50 different transcription factors.
We first wanted to know how the concentrations of these regulators change under different growth conditions,
and how this relates to their probability of binding to the promoters in question.

\begin{figure}[bt!]
\centering
\makebox[\textwidth][c]{\includegraphics[scale=0.9]{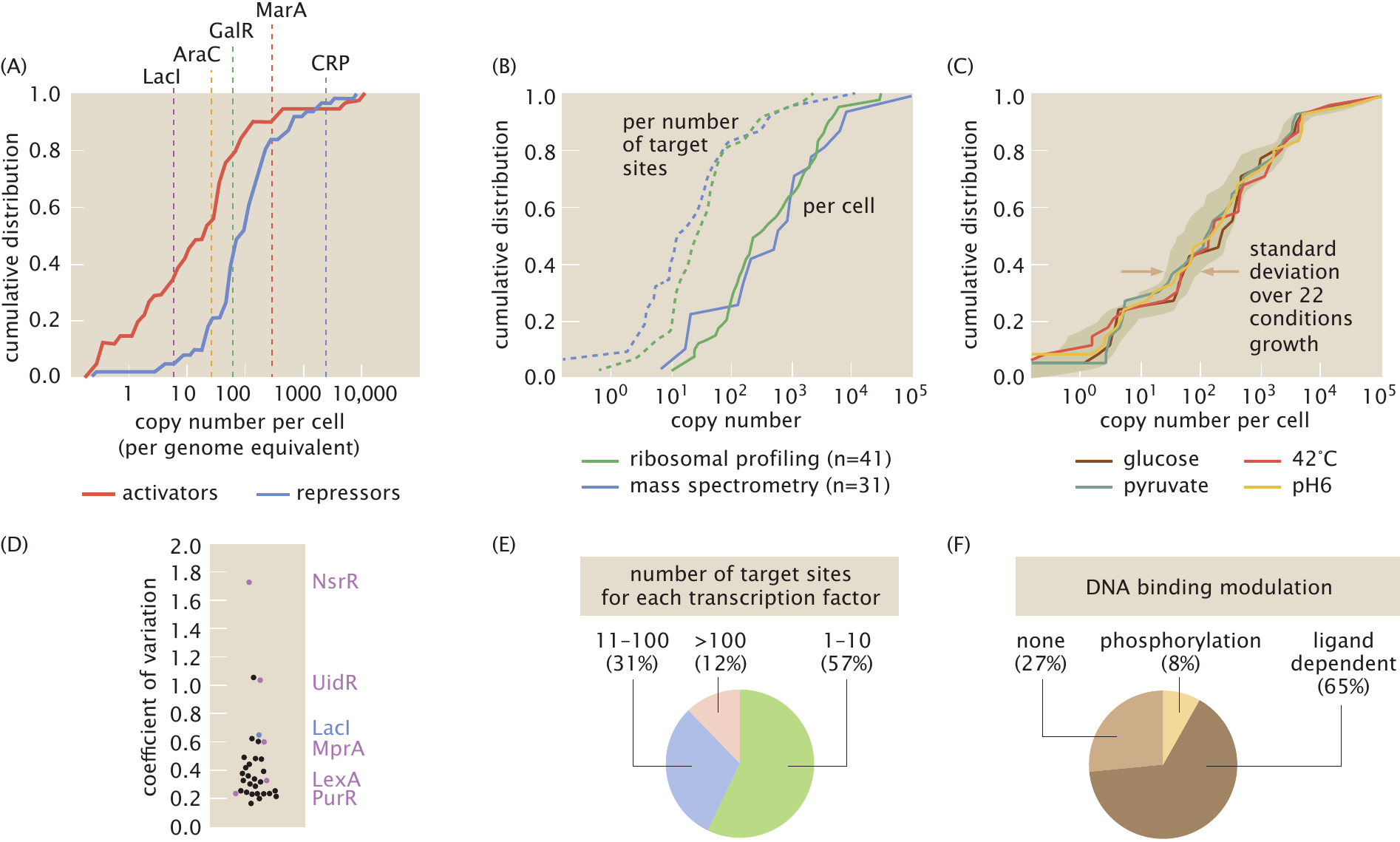}}
\caption{Summary of transcription factors that target the (0,1) simple repression architecture.  (A)  Transcription factor copy numbers in {\it E. coli}~\cite{Li2014}.  The cumulative distribution of
transcription factor copy numbers indicates
that activator copy numbers are generally lower than repressor copy numbers.  Roughly half of
the activators have copy number less than 10 while roughly half of all repressors have copy number
less than 100.  Several representative examples of well known transcription factors are shown
for reference.
    (B) Cumulative distributions are shown for transcription factors that target the
    (0,1) simple repression architecture. Data is shown from measurements using
    ribosomal profiling (41 of  the 50 identified repressors were measured in MOPS minimal media with 0.2\% glucose;
    \cite{Li2014}) and mass spectrometry (31 of  the 50 identified repressors were measured in M9 minimal media with
    0.5\% glucose; \cite{Schmidt2016}).
    (C) The variability in cumulative
    distribution is shown for the 31 transcription factors
    regulating the (0,1) architecture measured across 22
    different growth conditions, using mass spectrometry. The shaded region
    represents the 95th percentile region in cumulative distributions across growth
    conditions, with the distributions
    for four growth conditions shown explicitly.
    (D) Coefficient of variation for copy numbers of transcription factors regulating the (0,1)
    architecture across the 22 different growth conditions, measured by mass spectrometry.
    Several examples are identified along with LacI and the complete
    list is summarized in Table~\ref{tb_summary_O1}.
    (E) Number of target
    binding sites for each of the transcription factors that target a (0,1)
    architecture (using annotated information from RegulonDB \cite{GamaCastro2016}).
    (F) Mechanisms of target binding modulation for transcription factors that target a
    (0,1) architecture. Ligand-dependent transcription factors contain a known or
    predicted protein domain for binding by a ligand (using information
    from EcoCyc \cite{Keseler2010}).}
\label{fig_01_architectures}
\index{figures}
\end{figure}

To characterize each promoter, we used published data that quantified protein
abundance across the bacterial proteome under various growth conditions using either ribosomal profiling or mass spectrometry ~\cite{Li2014, Schmidt2016}. Figure~\ref{fig_01_architectures}(A) shows the distribution of repressor
and activator copy numbers genome wide, while Figure~\ref{fig_01_architectures}(B) shows
the copy numbers for just those
repressors that target the (0,1) architectures in which we are interested.  The
transcription factors vary in copy number from 0 to about 10,000
per cell. Of the repressors, just over half of them bind 10 or fewer
binding sites, while some targeted over 100 binding sites across the genome
(Figure~\ref{fig_01_architectures}(E)). Given the wide range in repressor copy number, we wondered whether it related to the number of target binding sites that exist for each of these repressors in the genome. Indeed, when we calculated the ratio between protein copy number and number of target binding sites for each transcription factor (as indicated by the dashed lines in
Figure~\ref{fig_01_architectures}(B)) we found a median ratio of about 15
transcription factor copies per binding site. The majority of the transcription factors (about 80\%) have no more than 100 copies per binding site. Given that the number of transcription factors per binding site is on the order of 10 - 100, we can infer that their typical effective binding constants (defined in detail below) are in the 10 - 100~nM range, since 1 copy of a protein per bacterial cell corresponds to a concentration of roughly 1~nM.

%

\begin{figure}
\centering{\includegraphics[width=6.0truein]{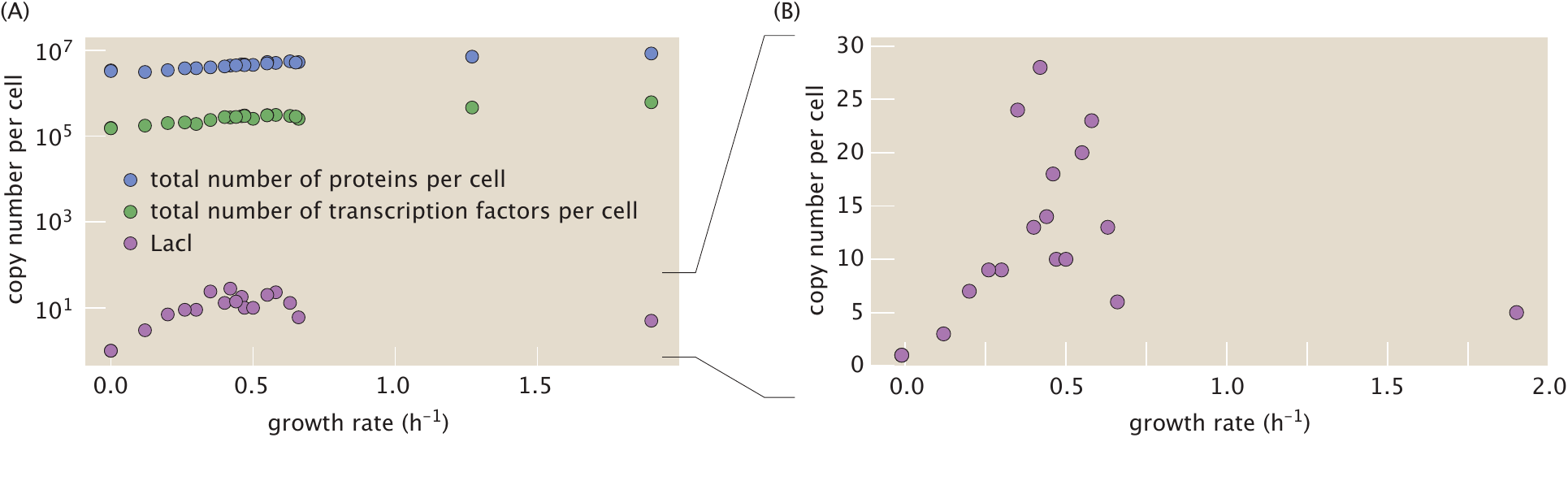}}
\caption{Protein census in {\it E. coli} as a function of growth rate. The figure shows the cellular copy
number for all proteins, only transcription factors, and for LacI. The low
copy number observed for LacI exemplifies the low protein counts that are
commonly observed for such regulatory proteins.  Each growth rate represents a
different growth condition that was considered in the work of Schmidt
{\it et al.} \cite{Schmidt2016}.
\label{fig:schmidt_lac}}
\end{figure}

We next asked how these simple repression promoters are regulated by the transcriptional repressors that control them. It might be the case that the promoters respond to changes in repressor copy number; alternatively, the copy number may remain constant but a repressor be induced by an external signal to switch to an active state. Using mass spectrometry measurements of protein copy number across 22 growth conditions
(varying carbon source, minimal versus rich media, temperature, pH, growth
phase, osmotic shock, and growth in chemostats), Schmidt
{\it et al.}, 2016~\cite{Schmidt2016} had found that most repressor copy numbers do not vary dramatically
as a function of growth condition (Figure~\ref{fig_01_architectures}(C)).
Figure~\ref{fig_01_architectures}(D) gives a quantitative picture of the variability in transcription
factor copy number for the repressors that target simple repression architectures.  Most repressors exhibit
a low coefficient of variation (standard deviation/mean copy number) in their
abundance across these growth conditions (median coefficient of variation of
0.33, compared with an 0.51 across the entire proteome).
In Figure~\ref{fig:schmidt_lac} we replot these data to show how the total proteome changes as a
function of growth rate, as compared to how the total number of transcription factors or copies of LacI do.  This plot provides a more nuanced picture of the challenges theoretical models must face in treating
expression levels over all growth conditions as will be discussed in the final section of the paper.


While it is possible that the
growth conditions considered were not appropriate
to elicit major changes in the copy number of each repressor, an alternative explanation
of the low variability in repressor copy number is that these transcription
factors, instead of relying on the modulation of their copy number, depend on ligand binding
and allosteric transitions to alter their
potency in regulating transcription.
Ligand binding followed by conformational changes between inactive and
active conformations provide allosteric control of the repressors by altering their DNA
binding strength, allowing for immediate changes in gene expression without relying on the much
 slower process of changing the transcription factor copy number through protein synthesis or degradation. As shown in Figure~\ref{fig_01_architectures}(F), we indeed find
that the majority of these repressors (65\%) are either known to bind DNA in
response to binding to a ligand, or for those less well-characterized, predicted
to have a ligand binding domain. In addition, several of the other repressors
that were identified are part of two-component systems that bind DNA in a
phosphorylation-dependent manner.


\section{Simple Repression as the Hydrogen Atom of Gene Regulation: \emph{Hic Rhodus, Hic Salta}}

%


In physics, when we establish
some model system that shows our complete command of an
area, it is often christened  ``the hydrogen atom'' of that subject.  This badge of honor
refers to the far-reaching power of the hydrogen atom in the context
of the modern quantum theory of matter. The theory not only informs the classic analysis of spectral lines in hydrogen, but also many more nuanced behaviors ranging from  the Stark and Zeeman effects
to some of the most subtle
effects seen in quantum electrodynamics~\cite{Rigden2003}. Using the tools of quantum mechanics, the hydrogen atom is simple enough to explore -- both mathematically and experimentally --
 many of the most important ideas in modern physics.
 It can also teach us what a solution to the problem looks like, in a way that is
 instructive when going on to tackle more complicated problems such as the behavior of heavier atoms.

We argue that this analogy is helpful in thinking about the simple repression
 motif as a foundation for launching into the study of more complicated regulatory
 architectures. One of Aesop's fables
recounts the exploits of a braggart who after a trip to the island of Rhodes
claimed to have made a long jump that could not be equaled by others.  A witness
to the braggart's commentary replied ``{\it hic rhodus, hic salta}'' meaning,
``Here is your Rhodes, jump now''. The simple repression motif is
our Rhodes. Here, we take the leap to see the extent to which we
can construct predictive theoretical models for how this regulatory
circuit behaves.

\begin{figure}
\centering{\includegraphics[width=4.0truein]{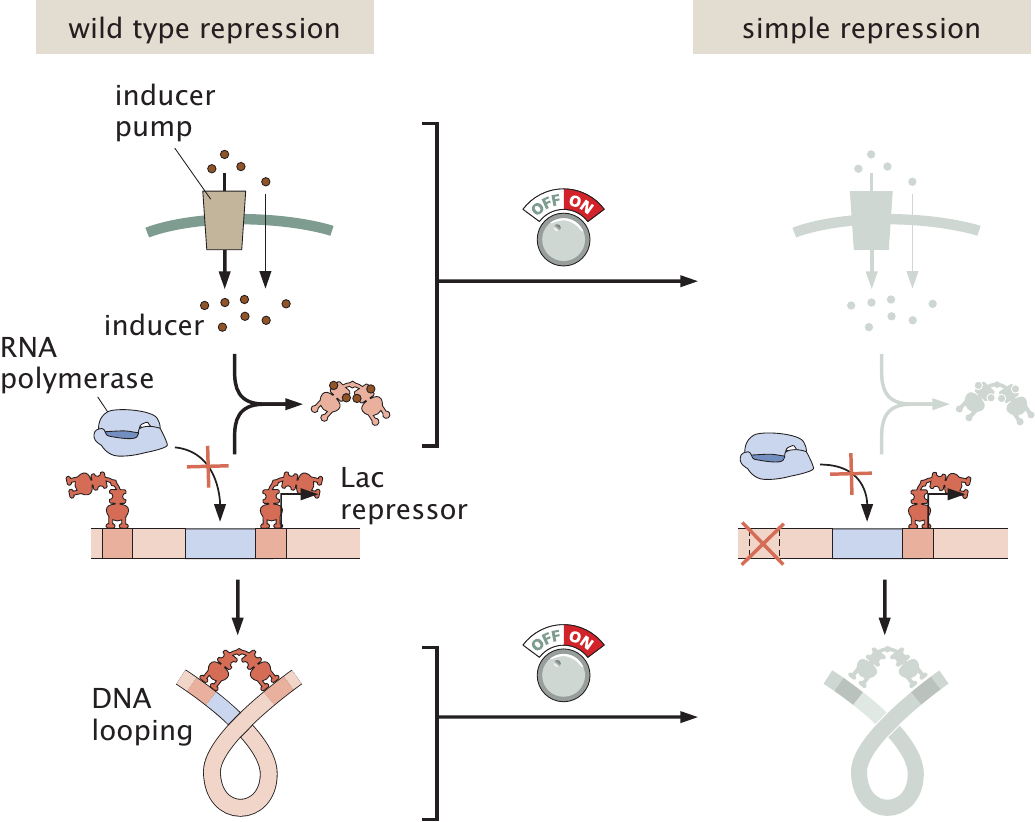}}
\caption{Deconstructing the {\it lac} operon to make the simple repression
hydrogen atom.  Key features of the wild-type {\it lac} operon such as DNA looping are
removed from the architecture to turn it into a model (0,1) architecture.
\label{fig:lacoperonComplexity}}
\end{figure}
\begin{figure}
\centering{\includegraphics[width=3.0truein]{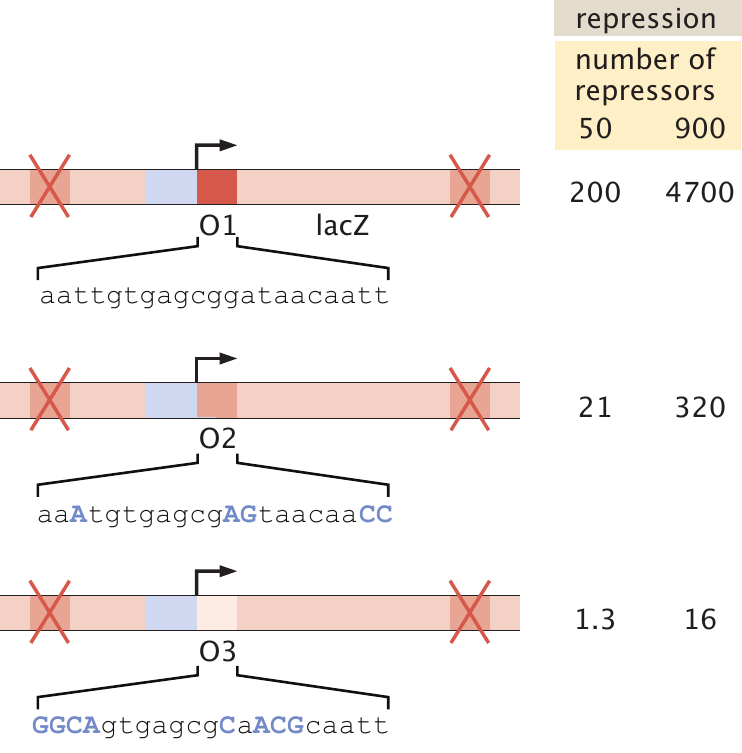}}
\caption{Classic experiments reveal key regulatory knobs of the simple repression motif.  Oehler {\it et al.} deleted the auxiliary binding sites in the {\it lac} operon rendering it into a simple
repression architecture~\cite{Oehler1990, Oehler1994}.  Different operators were used as the repressor binding site
and several different repressor counts were tuned resulting in different values of
the repression, defined as the ratio of gene expression with no repressors present to the level
of expression with repressors present. Changes to operator sequence with respect to the O1 operator are highlighted in blue.
\label{fig:Oehler}}
\end{figure}

The simple repression motif that forms the basis of our work was originally constructed by Oehler and colleagues \cite{Oehler1990, Oehler1994}. In a set of now classic experiments, they pared down the complex {\it lac} operon and rewired it as a powerful
model system, stripped of all but its most essential features. As shown in Figure~\ref{fig:lacoperonComplexity}, Oehler {\it et al.} reduced the number of repressor binding sites (operators) from three to one, creating precisely the repressor-operator
model originally envisaged by Jacob and Monod.
This remaining binding site was placed so as to compete directly
with RNA polymerase for promoter binding.
Oehler {\it et al.} furthermore recognized
the key control parameters for the simple repression motif --
the repressor copy number and the operator binding strength --  and figured out
how to manipulate them over different values in parameter space as shown in
Figure~\ref{fig:Oehler}. Using the DNA sequence of the binding site as a way to manipulate its affinity,
they could then tune the strength of repression, providing a well-conceived model
system for testing the theoretical predictions of various modeling frameworks
aimed at describing transcriptional regulation.  We now consider the kinds of
theoretical predictions needed to carry out the experiment-theory
dialogue advocated here.

\section{Mathematicizing Transcriptional Regulation}

\begin{figure}
\centering{\includegraphics[width=5.0truein]{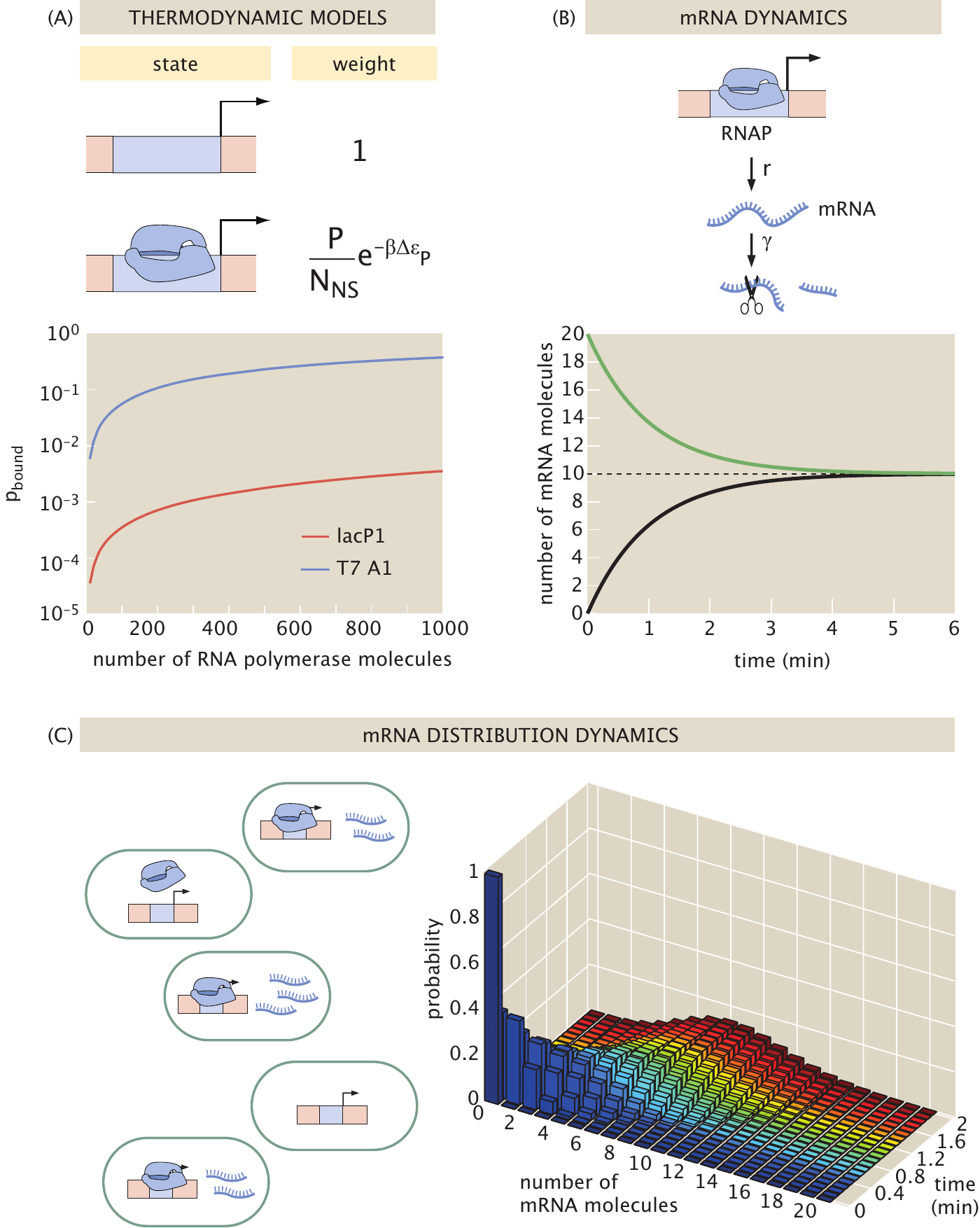}}
\caption{Summary of approaches to computing the level of expression from
the simple unregulated promoter.  These same approaches
can be used for computing the response of more complex regulatory
architectures such as the simple repression motif that is the central preoccupation
of this article. (A) Thermodynamic models compute the probability of promoter occupancy using
the Boltzmann distribution. The graph shows the probability of promoter occupancy for a weak
({\it lacP1}) and a strong (T7 A1) promoter sequence as a function of the number of polymerases.
(B) Dynamics of mean expression using kinetic models.  The graph shows the number of mRNA molecules
as a function of time with the steady-state number shown as a dashed black line. The mRNA dynamics
corresponding to two different initial conditions are shown. (C) Dynamics of mRNA distribution
using the chemical master equation approach. The bar graph shows how the {\it distribution} of
mRNA copy numbers changes over time, ultimately settling on a steady-state Poisson distribution.
(A, adapted from \cite{Bintu2005a}; B and C, $r=10$~mRNA/min and $\gamma = 1~\mbox{min}^{-1}$)
\label{fig:ComputingRepression}}
\end{figure}

While some may say that Figure~\ref{fig:HighSchoolLac} makes
predictions as to when gene expression will be turned ``on" or ``off", we
protest this loose use of the term ``prediction" which in our minds has a very
special meaning. To earn the title of ``the hydrogen atom of X'', one must understand the system
not only qualitatively,  but  with quantitative precision as well.
In this article,
``prediction" is used with care to emphasize the quantitative concreteness of our
thinking. Our aim in the coming sections is to examine the myriad of different physical/mathematical
approaches that have been set forth to think about gene regulation in a
predictive fashion. Figure~\ref{fig:ComputingRepression} shows the different
classes of models that will be entertained in the remainder of the article as
a result of their prevalence in the literature and their impact on the field
itself.  Figure~\ref{fig:ComputingRepression}(A) provides a schematic of how
thermodynamic models are used to compute promoter occupancy, an approach that
will be described in greater detail below.
Figure~\ref{fig:ComputingRepression}(B) focuses instead on mRNA dynamics using
differential equations to account for the mean number of mRNA as a function of
time given the microscopic processes that lead to both an increase and
decrease in the number of mRNAs.  An even more ambitious strategy is presented
in Figure~\ref{fig:ComputingRepression}(C) which focuses on the dynamics of
the full distribution $p(m,t)$, which is defined as the probability of finding
$m$ mRNAs at time $t$. To be concrete, our strategy is to focus on the use of
each of these different methods in the specific case of simple repression with
special focus on what the different classes of models say and how experiments
have been used to test those predictions.

\subsection{The Occupancy Hypothesis and Thermodynamic Models}

The thermodynamic models presented schematically in Figure~\ref{fig:ComputingRepression}(A)
implicitly assume one of the most important and ubiquitous assumptions in all of regulatory
biology, namely, the {\it occupancy hypothesis}.  This hypothesis, which will be described, criticized and contrasted with
experiments in detail in Appendix~\ref{sec:OccupancyHypothesis}, informs approaches
ranging from the bioinformatic search for transcription factors to the use of ChIP-Seq
experiments to the kinds of thermodynamic models that are our focus here. Stated simply,
the central assumption is that the rate of mRNA production is proportional to the probability
of RNA polymerase occupancy at the promoter,
\begin{equation}\label{eq:dmdtpbound}
{dm \over dt} = rp_{bound},
\end{equation}
where we introduce the notation $p_{bound}$ for the probability that RNA polymerase is
bound to the promoter of interest.
More generally, if we have $N$ transcriptionally active states (e.g. polymerase by itself,
polymerase and activator together), then we write
\begin{equation}
{dm \over dt} = \sum_{i=1}^N r_ip_i.
\end{equation}
The idea behind this equation is that the net average rate of transcription is given by the fraction of time the promoter spends in each transcriptionally active state, $p_i$, multiplied by the rate of transcription corresponding to that state, $r_i$.

\begin{figure}
 \centering
\centering{\includegraphics[width=5.0truein]{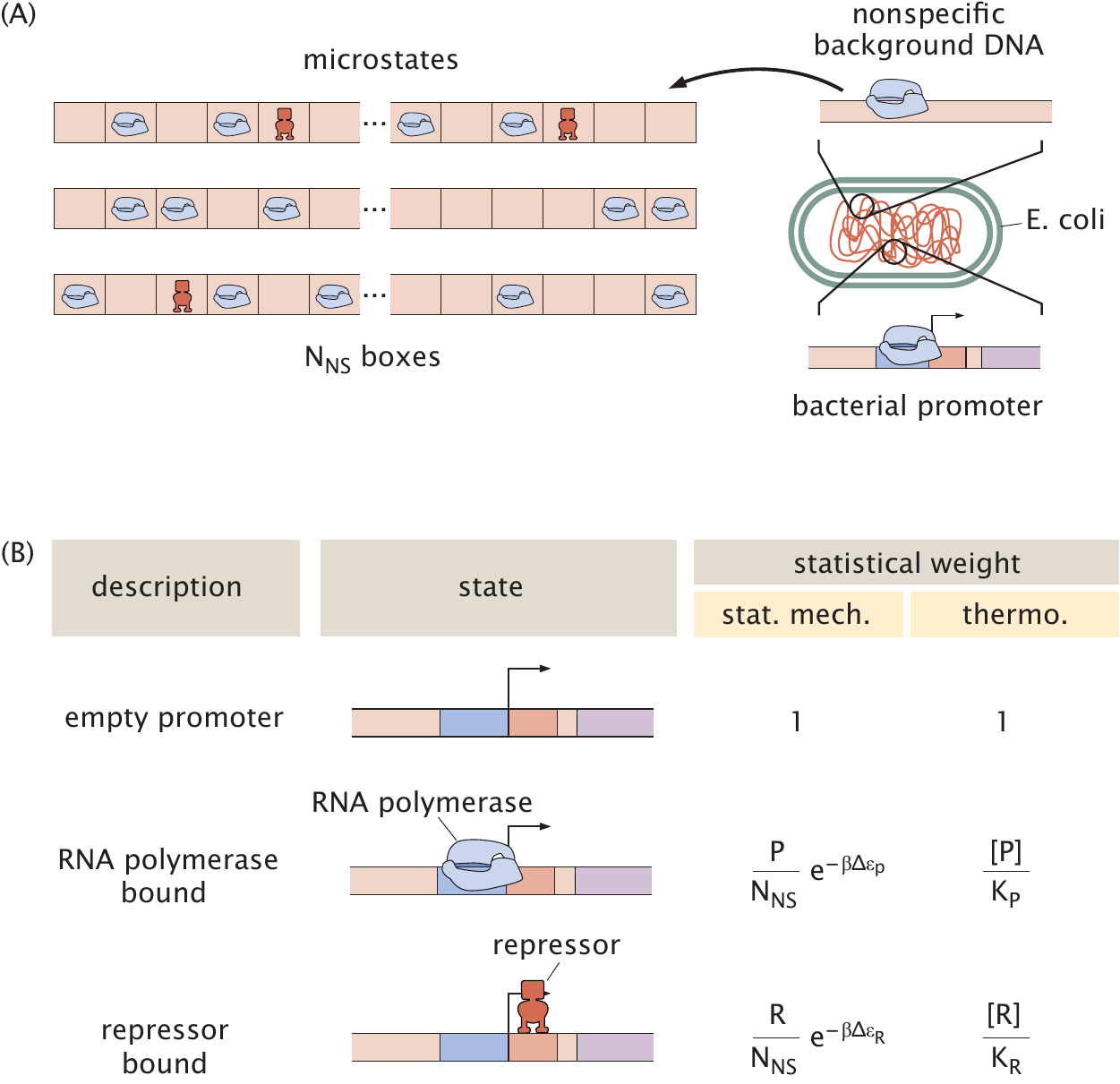}}
 \caption{States and weights for the simple repression motif. (A) Our regulatory
        system is assumed to consist of $P$ RNA polymerase (blue) and $R$ repressors
        (red) per cell that either bind nonspecifically to the genomic background DNA
        (our reference energy state) or compete for binding to our promoter of interest.
        The genomic background is discretized by assuming a number of potential binding
        sites, $N_{NS}$, that is given by the length of the genome ($N_{NS} = 4.6 \times
        10^6$ for \textit{E. coli}). (B) The different regulatory states of our simple
        repression promoter. The statisical weight associated with each state is shown
        using the statistical mechanical and thermodynamic formulations. The binding
        energies of the $R$ repressors and $P$ RNA polymerase to their binding sites on the
        promoter are given by $\Delta\varepsilon_R$ and $\Delta\varepsilon_P$,
        respectively. These energies are given relative to the energy of nonspecific binding to the genomic
        background. In the thermodynamic formulation, $[P]$ and $[R]$ are the cellular
        concentrations of the RNA polymerase and repressor, respectively. Their dissociation
        constants are given by $K_P$ and $K_R$. $N_{NS}$ represents the number of
        nonspecific binding sites for both RNA polymerase and repressor. }
 \label{fig_states}
\end{figure}

But before we can use this result, we need to
know the physical nature of the individual states and how to compute their probabilities. We adopt
notation in which  the probability of the $i^{th}$ transcriptionally
active state can be thought of as
\begin{equation}
p_i=p_i([TF_1],[TF_2], \cdots)
\end{equation}
where the notation indicates that this probability is a function that reflects the occupancy
of the regulatory DNA by the various transcription factors (i.e. regulatory
proteins) that interact
with the regulatory apparatus of the gene of interest.  Hence, each transcriptionally-active state denoted
by the label ``$i$'', corresponds to a different state of the promoter characterized
by a different constellation of bound transcription factors.
These ideas were first put into play in the gene regulatory setting by
Ackers and coworkers and have since been explored more deeply by number of groups~\cite{Ackers1982, Shea1985, Buchler2003a,Vilar2003a,Vilar2003b, Bintu2005a,Bintu2005b, Gertz2009,Sherman2012, Saiz2013}.
For the case of the simple repression motif, the thermodynamic model is illustrated
in Figure~\ref{fig_states}.

As in Figure~\ref{fig:ComputingRepression}(A), the idea
is to identify the relevant microscopic states of the promoter and to assign to each such state its corresponding statistical weight.
The details of how to use statistical mechanics to compute this probability has been described  elsewhere \cite{Bintu2005a, Phillips2015}
so here we resort to simply quoting the central
result of the thermodynamic models for the simple repression motif, namely,
the probability of finding RNA polymerase bound to the promoter given by
\begin{equation}\label{eq:pboundPR}
    p_{bound} = { {P \over N_{NS}} e^{-\beta \Delta \varepsilon_P} \over 1 + {P \over N_{NS}} e^{-\beta \Delta \varepsilon_P} + {R \over N_{NS}} e^{-\beta \Delta \varepsilon_R}},
\end{equation}
where $R$ is the number of repressors, $N_{NS}$ is the size of the genome (i.e. number
of nonspecific sites) and $\Delta \varepsilon_R$ is the binding energy of repressor
to its operator. Similarly, $P$ is the number of RNA polymerase molecules and $\Delta \varepsilon_P$ is its binding energy to the promoter.

In the language of these models, we can now relate the experimentally measurable repression, which is obtained by quantifying the rate of mRNA production, or the steady state levels of mRNA or protein, in the presence and absence of repressor,  to the theoretically calculable quantity $p_{bound}$ such that
\begin{equation}
    \mbox{repression}={dm/dt(R=0) \over dm/dt(R \ne 0)} = {r p_{bound}(R=0) \over r p_{bound}(R \ne 0)} = {p_{bound}(R=0) \over p_{bound}(R \ne 0)},
\end{equation}
Alternatively, we can write the fold-change as
\begin{equation}
    \mbox{fold-change}={p_{bound}(R \ne 0) \over p_{bound}(R = 0)},
\end{equation}
where we have made used of the occupancy hypothesis introduced in Equation~\ref{eq:dmdtpbound}. We now use the expression for $p_{bound}$ from Equation~\ref{eq:pboundPR} and obtain
\begin{equation}\label{eq:FoldChangeFull}
    \mbox{fold-change} = {1 + {P \over N_{NS}} e^{-\beta \Delta \varepsilon_P} \over
    1 + {P \over N_{NS}} e^{-\beta \Delta \varepsilon_P} + {R \over N_{NS}} e^{-\beta \Delta \varepsilon_P}}.
\end{equation}
Finally, we assume that binding of RNA polymerase to the promoter is weak such that $P/N_{NS} e^{-\beta \Delta \varepsilon_P} \ll 1$. In the context of this weak promoter approximation, which is discussed in detail in \cite{Bintu2005a,Garcia2011c},
 the fold-change reduces to
\begin{equation}\label{eq:FoldChangeSimpleRepStatMech}
\mbox{fold-change}={1 \over 1+{R \over N_{NS}} e^{-\beta \Delta \varepsilon_R}}.
\end{equation}

\begin{figure}
\centering{\includegraphics[width=5.5truein]{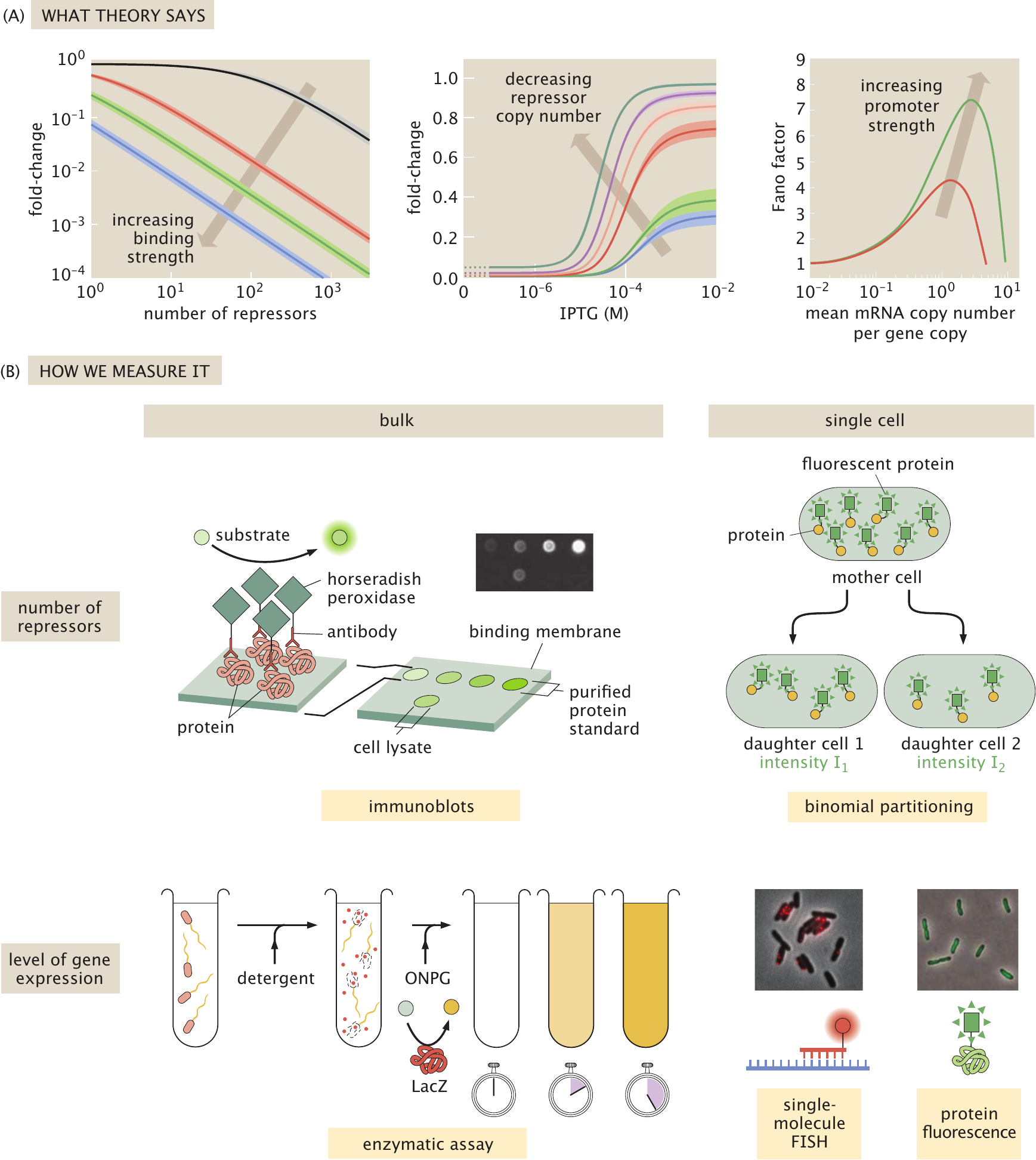}}
\caption{Theory/experiment dialogue in simple repression. (A) Three examples of predictions about the simple repression motif that can be subjected to experimental scrutiny using precision measurements.  The left figure shows the fold-change in gene expression as a function of repressor copy number for different operators, the middle panel shows predictions of induction for different numbers of
repressors and the right panel shows how gene expression noise (Fano factor $=$ variance/mean) varies as a function of the mean gene expression level for different promoter strengths. (B) Bulk and single-cell measurements of both repressor copy number and gene expression.  For copy number, bulk measurements can be done using immunoblotting, while counting statistics can be used
at the single-cell level. To measure gene expression, bulk enzymatic assays have excellent dynamic range. Single-cell measurements can be done either by examining the level of mRNA or protein gene product.
\label{fig:AztecPyramidExperiments}}
\end{figure}

The conceptual backdrop to this result is shown
in Figure~\ref{fig_states}.
As we will describe in great detail later in this article and in Appendix \ref{sec:ThermoVsStatMech}, there is much
confusion about the mapping between statistical mechanics language, which
we believe is more microscopically transparent, and thermodynamic language using
dissociation constants. In that language, our
result for fold-change can be written as
\begin{equation}\label{eq:SimpleRepressionFoldChangeKd}
\mbox{fold-change}={1 \over 1+{[R] \over K_R}},
\end{equation}
where $[R]$ is the concentration of repressor and $K_R$ its dissociation constant to operator DNA. This equation for the fold-change is precisely what is plotted as a theory
prediction in the top left pane of  Figure~\ref{fig:AztecPyramidExperiments}(A).

\subsection{Beyond the Mean: Kinetic Treatments of Transcription}\label{sec:BeyondMean}

Up to this point, we have examined the simple repression architecture
in a manner that describes the steady-state mean level of expression. But this is not to say that mRNA dynamics or the mRNA
distribution are not of interest; quite the opposite. Knowledge of the
higher moments of the distribution provide great insight into the kinetics of the system, and we turn to these now.

\begin{figure}
\centering{\includegraphics[width=5truein]{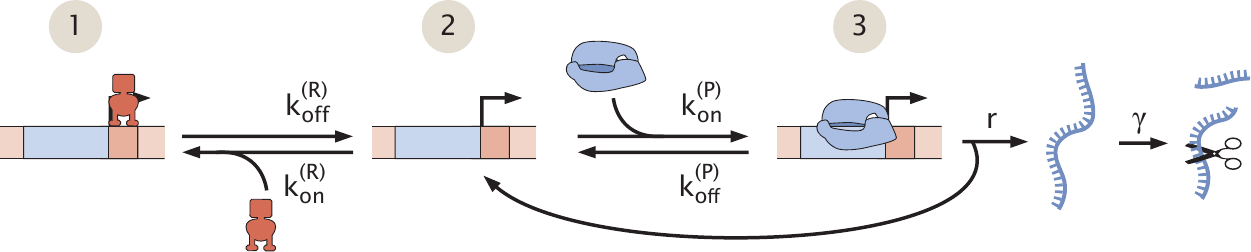}}
\caption{Kinetic model of simple repression. The promoter can be empty, occupied by repressor, or occupied by RNA polymerase. Transitions between the different states are
characterized by rate constants shown in
the caption.  Note that when transcription commences from state $3$, the promoter returns to the empty state (state 2).
\label{fig:EquilibriumAssumptionRepressed}}
\end{figure}

We begin by considering a dynamic description of repression that can be used to calculate the temporal
evolution of the number of mRNA molecules as shown for the case of the constitutive promoter in
Figure~\ref{fig:ComputingRepression}(B). Specifically, we think of simple repression using the
kinetic scheme presented in Figure~\ref{fig:EquilibriumAssumptionRepressed}. For the kinetics of the first state, in which
the promoter is occupied by the repressor molecule, the linear
reaction scheme shows that there is only one way to enter and exit this state
and that is through the ``empty'' state (state 2). This results in the dynamical equation
\begin{equation}
{dp(1) \over dt}= k_{on}^{(R)} p(2) - k_{off}^{(R)}p(1).
\end{equation}
The dynamics of the empty state (state 2) are more complicated because this state
is accessible to both the repressor and the polymerase, meaning
that the dynamics can be written as
\begin{equation}
{dp(2) \over dt}= -k_{on}^{(R)} p(2) +k_{off}^{(R)}p(1)-k_{on}^{(P)} p(2) + k_{off}^{(P)}p(3)+rp(3).
\end{equation}
Note that the final term in this equation reflects the fact that mRNA is produced at rate $r$ from state 3 and once mRNA production begins, polymerase leaves the promoter and
hence the system goes back to state 2.
The state with polymerase occupying the promoter evolves similarly
as can be seen  by writing
\begin{equation}
{dp(3) \over dt}= k_{on}^{(P)} p(2) - k_{off}^{(P)}p(3)-rp(3).
\end{equation}
To close the loop and come full circle to the real question of interest, namely,
the production of mRNA itself, we have
\begin{equation}
{dm \over dt} =-\gamma m + r p(3).
\end{equation}
What this equation tells us is that the promoter is only transcriptionally active
in the third state, namely, that state in which the polymerase binds
the promoter. The above equations can be solved in order to obtain the temporal dynamics of mRNA concentration, as we have illustrated
in Figure~\ref{fig:ComputingRepression}(B) for the unregulated promoter.

An interesting feature of
the kinetic description of simple repression presented here is that enables us to go beyond the
steady-state and equilibrium assumptions that were invoked to calculate the fold-change in gene expression in
Equations~\ref{eq:FoldChangeSimpleRepStatMech}~and~\ref{eq:SimpleRepressionFoldChangeKd}. Instead, we can
 use the kinetic scheme shown in Figure~\ref{fig:EquilibriumAssumptionRepressed} to solve for the fold-change,
but now only invoking steady-state by setting the left side in each equation above to zero.
We begin by solving for the steady-state level of mRNA, $m_{ss}$, and find
\begin{equation}
m_{ss}={r p(3) \over \gamma}.
\end{equation}
But what is $p(3)$? In seeking the unknown steady-state probabilities,
we must respect the constraint that the probabilities sum to one, namely,
\begin{equation}
p(1)+p(2)+p(3)=1.
\end{equation}
We will not go into the details of the  algebra of resolving these three linear
equations as these details are described in \cite{Phillips2015}. Instead, we will simply quote the result as
\begin{equation}\label{eq:p3}
p(3)={1   \over 1+{{(k_{off}^{(P)}+r) \over k_{on}^{(P)} } \left(1+ {k_{on}^{(R)} \over k_{off}^{(R)}} \right)  } },
\end{equation}
which enables us to make contact with the types of experiments discussed earlier, by computing
the fold-change:
\begin{equation}\label{eq:FoldChangeKinetic}
\mbox{fold-change}= {m_{ss}(R \ne 0)  \over m_{ss}(R = 0) }={1 \over
1 + {{(k_{off}^{(P)}+r) \over k_{on}^{(P)} } \over 1+{(k_{off}^{(P)}+r) \over k_{on}^{(P)} } }  {k_{on}^{(R)} \over k_{off}^{(R)}}} .
\end{equation}
Note that we can write $k^{(R)}_{on}=k_{+}^{(R)} R$, where
we have acknowledged that the on-rate for the repressor is proportional to the number of
repressors present in the cell.
Interestingly, we see that this implies that the functional
form of the fold-change is the same even in this steady-state
context as it was in the thermodynamic model framework,
though now at the price of having to introduce an
effective $K_d^{eff}$, resulting in
\begin{equation}
\mbox{fold-change}= {1 \over \left(1+{R \over K_d^{eff}}\right)}.
\label{eqn:SimpleRepressionFoldChangeKinetics}
\end{equation}
By comparing Equations~\ref{eq:SimpleRepressionFoldChangeKd}
and ~\ref{eqn:SimpleRepressionFoldChangeKinetics}, we see
that their scaling with repressor number is identical.  To further
explore the common features between these two expressions
for fold-change, note that we can write
\begin{equation}
K_d^{eff}={k_{off}^{(R)} \over k_{on}^{(P)}} {\left(1 + {(k_{off}^{(P)}+r) \over k_{on}^{(P)}}\right)
\over {(k_{off}^{(P)}+r) \over k_{on}^{(P)}}}.
\end{equation}
 We can simplify this further by noting
that we can write $K_d^{(R)}=k_{off}^{(R)}/k_{+}^{(R)}$ resulting in
\begin{equation}\label{eq:KdEff}
K_d^{eff}=K_d^{(R)} {\left(1 + {(k_{off}^{(P)}+r) \over k_{on}^{(P)}}\right)
\over {(k_{off}^{(P)}+r) \over k_{on}^{(P)}}}.
\end{equation}
This equation reveals that the thermodynamic and kinetic treatments of simple repression have some interesting differences and clearly shows the consequences of imposing the equilibrium assumption in the thermodynamic calculation. The validity of this assumption will be explored in detail in Appendix~\ref{sec:Equilibrium}.

An alternative way of viewing these same problems is by going beyond the
description of the dynamics of the mean mRNA number and appealing to the kinetic
theory of transcription in order to work out the time-evolution of the
probabilities of the different states~\cite{Ko1991,Peccoud1995,Record1996,
Kepler2001, Sanchez2008,Shahrezaei2008,Garcia2010, Michel2010}. Our goal is to
write equations that describe the time evolution of the probability of finding
$m$ mRNA molecules at time $t$. This means that we need to define three coupled
differential equations for the mRNA distribution in each of the three states,
namely,  $p_1(m, t)$, $p_2(m, t)$ and $p_3(m, t)$. Intuitively, if we are thinking
about the possible changes that can alter state $1$, there are only several
transitions that can occur: i) the promoter can switch from state $1$ to state $2$,
ii) the promoter can switch from state $2$ to state $1$ and iii) an mRNA
molecule can decay resulting in a change in $m$. These transitions are expressed using the master equation
formalism and the rate constants defined in Figure~\ref{fig:EquilibriumAssumptionRepressed} as
\begin{equation}\label{ThreeState1}
  {d p_1(m,t) \over dt} =
  - \underbrace{k_{off}^{(R)} p_1(m, t)}_{(1) \rightarrow (2)} 
  + \underbrace{k_{on}^{(R)} p_2(m, t)}_{(2) \rightarrow (1)} 
  + \underbrace{\gamma (m + 1) p_1(m+1 , t)}_{m+1 \rightarrow m} 
  - \underbrace{\gamma m p_1(m , t)}_{m \rightarrow m-1}. 
\end{equation}
The case of state $2$ includes the same transitions between state $1$ and
state $2$, as well as the transitions between state $2$ and $3$ as a result of polymerase unbinding or promoter escape due to transcriptional initiation.  Incorporating
these ideas leads to an equation of the form
\begin{align}\label{ThreeState2}
  {d p_2(m,t) \over dt} = &
    \underbrace{k_{off}^{(R)} p_1(m, t)}_{(1) \rightarrow (2)} 
  - \underbrace{k_{on}^{(R)} p_2(m, t)}_{(2) \rightarrow (1)} 
  + \underbrace{k_{off}^{(P)} p_3(m, t)}_{(3) \rightarrow (2)} 
  - \underbrace{k_{on}^{(P)} p_2(m, t)}_{(2) \rightarrow (3)} 
  + \underbrace{r p_3(m-1,t)}_{\substack{m-1 \to m \\ (3) \to (2)}} \\ 
  & 
  + \underbrace{\gamma (m + 1) p_2(m+1 , t)}_{m+1 \rightarrow m} 
  - \underbrace{\gamma m p_2(m , t)}_{m \rightarrow m-1}. \notag 
\end{align}
Finally for state $3$ we must account for the transitions between state $2$
and state $3$, and the mRNA production at a rate $r$. Bringing all of these
transitions together results in
\begin{align}\label{ThreeState3}
  {d p_3(m,t) \over dt} =&
  - \underbrace{k_{off}^{(P)} p_3(m, t)}_{(3) \rightarrow (2)} 
  + \underbrace{k_{on}^{(P)} p_2(m, t)}_{(2) \rightarrow (3)} 
  - \underbrace{r p_3(m , t)}_{\substack{m \to m+1 \\ (3) \to (2)}} 
  + \underbrace{\gamma (m + 1) p_3(m+1 , t)}_{m+1 \rightarrow m} 
  - \underbrace{\gamma m p_3(m , t)}_{m \rightarrow m-1}. 
\end{align}
This set of coupled equations describes the time evolution of the probability
distribution $p(m,t)$.

As described in the following sections, the equations written above imply a steady-state mRNA
distribution that can be used to compute both the mean and variance in gene expression. In order to render the different
theoretical descriptions self-consistent, the thermodynamic parameters such
as the repressor binding energy $\Delta\varepsilon_R$ must constrain the values
that the repressor rates $k_{off}^{(R)}$ and $k_{on}^{(R)}$ can take.  Now that we have
seen how theory can be used to sharpen our thinking, we turn to how experiments
can be designed to test those theoretical ideas.

%
%


\section{``Spectroscopy'' for the Simple Repression Hydrogen Atom: Precision Measurements
on Gene Expression}

Figure~\ref{fig:AztecPyramidExperiments} provides a picture of how theory
and experiment come together in thinking about the simple repression motif.
As part (B) of that figure shows, there are a variety of approaches that can
be taken to count the repressors and to  measure the level of gene expression.
Expression levels can be quantified using enzymatic or fluorescence assays.
Note that by choosing to measure the ratio of level of gene expression (i.e. the fold-change)
rather than the absolute value of the gene expression itself, the system is
further simplified since various categories of context dependence such
as the position of the gene on the genome are normalized away.
This is not to say that the description of such effects on the
absolute level of expression are uninteresting, but rather the focus on a predictive
understanding of the fold-change in gene expression
reflects the spirit of little steps for little feet that are required to progressively
develop a rigorous view of these problems.

There are many facets to the regulatory response of the simple repression motif
that can be subjected to experimental scrutiny in order to compare them to the results of theoretical predictions, as shown in Figure~\ref{fig:AztecPyramidExperiments}(A).
Indeed, the seeds of this review were planted by many wonderful earlier works
that explored various aspects of the theoretical
and experimental strategies laid out in Figure~\ref{fig:AztecPyramidExperiments}.
Experimentally, as noted above,  Muller-Hill and Oehler  led the way in the
{\it lac} system (see Figure~\ref{fig:Oehler}) as did Schleif in the context of the arabinose operon \cite{Schleif1975, Ogden1980, Dunn1984}.
On the theory side, Ackers and Shea laid the groundwork for thermodynamic
models which allow us to predict the mean level of expression~\cite{Ackers1982, Shea1985}.
 These models were pushed even further by Buchler, Gerland and Hwa \cite{Buchler2003a} and by Vilar,
 Saiz and Leibler \cite{Vilar2003a,Vilar2003b,Saiz2008b}.  Besides the thermodynamic model
 approach~\cite{Bintu2005a, Bintu2005b,Gertz2009,Sherman2012}, others have been interested
in gene expression noise, which demands kinetic models.  These
approaches to transcription have offered numerous insights of their own~\cite{Ko1991,Peccoud1995,Record1996,
Kepler2001, Sanchez2008,Michel2010}. Much of the work presented here draws inspiration from modern quantitative
dissections of the wild-type {\it lac} operon \cite{Setty2003,Kuhlman2007}, as well as from efforts that made it possible to measure
 gene regulatory functions at the single cell level \cite{Golding2005,Rosenfeld2005}, and from research that
embodies the same interplay between theory and experiment featured in this article but in the context of other gene-regulatory
architectures \cite{Bakk2004,Zeng2010,Cui2013,Sepulveda2016}.

\section{Climbing the Simple Repression Pyramid:  A Minimal Parameter Set to Rule Them All}

In the previous sections we outlined how different kinds of theoretical
frameworks enable us to formalizing our ``pathetic thinking'' in order to
to refine our prejudices about how a complex system behaves~\cite{Gunawardena2014}.
One of the key requirements we insist on in using such
theoretical frameworks to describe simple repression is that a single set of parameters applies across
all different situations, as illustrated in Figure~\ref{fig:AztecPyramidExperiments}.
There is a long tradition of developing phenomenological theories that describe
broad classes of behaviors, in which the underlying microscopic processes
that give rise to material response are captured in the form of a small
set of phenomenological, but measurable, parameters.  Consider the steel used to build our bridges and skyscrapers,
or the aluminum used to build our airplane wings: several elastic constants,
a yield stress and a fracture toughness often suffice to fully characterize
the material response under a broad array of geometries and loading conditions~\cite{JonesAshby2012}.
Importantly, each time we go out and use those materials for
something new, we don't have to introduce a new set of parameters.
It is critical to realize that there is no requirement
whatsoever for an underlying ``mechanistic
theory'' of what determines those parameters in order for
a phenomenological theory to be both beautiful and far-reaching in its predictive
value. Though perhaps the ``microscopic mechanism''  of,
for example, how the interactions between the nucleotides on the DNA and residues on
the repressor dictate binding energy is attractive to some investigators, we do not need
 a microscopic understanding of these atomic-level ``mechanisms'' to
 construct a predictive theory of gene regulation.  Indeed, though much progress has
been made in constructing a microscopic basis for these parameters, we generally cannot predict these material parameters from first principles.

\begin{figure}
\centering{\includegraphics[width=6.0truein]{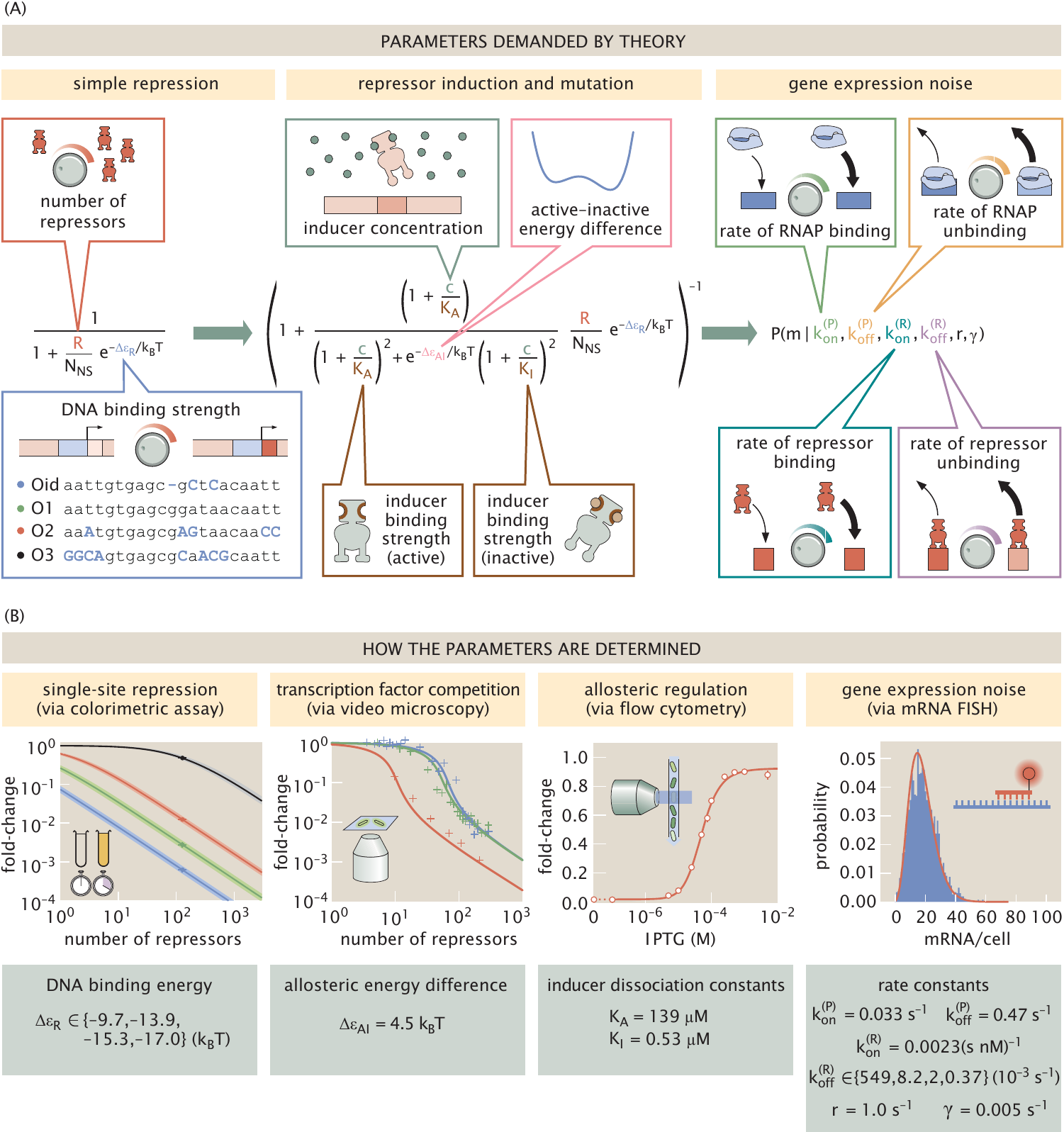}}
\caption{Determination of the minimal parameter set for describing simple repression
across a broad array of experimental approaches and simple repression
regulatory scenarios.  (A) Parameters that are introduced
in the description of simple repression fold-change measurements, induction experiments
and in the context of gene expression noise. (B) Experiments used to determine
the minimal parameter set. (B, left panel adapted from \cite{Garcia2011c}, middle panels adapted from \cite{Razo2018}, right panel adapted from \cite{JonesBrewster2014})
\label{fig:AztecPyramidParameters}}
\end{figure}

Here, we adopt a phenomenological mindset in the context of the gene regulatory response.
Though it is clear that there are a huge variety of complicated processes
taking place within the cell that we do not understand, we address whether it is nonetheless possible to introduce a few key
effective parameters that will allow us to characterize the regulatory response
of the simple repression motif under a broad array of different circumstances.
Figure~\ref{fig:AztecPyramidParameters}(A) shows us how
the theoretical ideas highlighted in the previous sections demand
a small number of parameters before we can use them predictively.
For example, in the simple repression motif, we require a binding
energy $\Delta \varepsilon_R$ (or equivalently a $K_R$) to characterize
the strength of repressor binding to operator.  Similarly, when
describing the induction response of transcription factors to inducer,
we require parameters $K_A$ and $K_I$ that describe the affinity
of inducer to the transcription factor when it is in its active
and inactive states, respectively~\cite{Razo2018}. We also require a free energy
difference $\Delta \varepsilon_{AI}$ that characterizes the relative stability of
the active and inactive states in the absence of inducer.
 Finally, when describing gene expression dynamics,
 we require rate constants for mRNA degradation ($\gamma$), transcript initiation ($r$),
 and the on and off rates of repressor and RNA polymerase to their binding sites ($k_{on}^{(R)}$ and $k_{off}^{(R)}$
 for the repressor, and $k_{on}^{(P)}$ and $k_{off}^{(P)}$ for RNA polymerase).
 The question we ask is: once we have established this minimal set of
 parameters, how well can we then range in our predictions across different classes of
 experiments involving the simple repression motif?

We now show how it is possible to ascend the simple repression pyramid
introduced in Figure~\ref{fig:AztecPyramid}. In Figure~\ref{fig:AztecPyramidParameters}(B), we outline
how we fully determined a single minimal set of parameters needed to characterize a host of regulatory responses.
Note that others have also made complete parameter
determinations, but did so across different experiments~\cite{Saiz2013}. The lefthand panel of Figure~\ref{fig:AztecPyramidParameters}(B) illustrates how
experiments like those of Oehler {\it et al.}, with one particular repressor
copy number and a specific operator sequence, can be used
to determine the parameter $\Delta \varepsilon_R$ (or $K_R$).
The second experiment highlighted in Figure~\ref{fig:AztecPyramidParameters}(B)
shows how the transcription factor titration effect can be used to
determine the parameter $\Delta \varepsilon_{AI}$ (or alternatively $L = e^{-\beta \Delta \varepsilon_{AI}}$)
which characterizes the equilibrium between the inactive and active states
of repressor in the absence of inducer.  The third panel in the figure
shows how a single induction response curve can fix the parameters $K_A$ and
$K_I$ that determine the binding of inducer to the repressor in the active
and inactive states, respectively.
Finally, the righthand panel demonstrates how, by going beyond
the mean and looking at the full mRNA distributions for the
constitutive promoter and the simple repression motif, it is possible to
infer the rates of RNA polymerase and Lac repressor
binding and unbinding as well as the rates of mRNA production and degradation.

This kinetic approach takes advantage of the known closed
form of the full mRNA distribution for a two-state promoter
\cite{Peccoud1995}. Using this expression for the distribution we can perform a
Bayesian parameter inference to obtain values for the polymerase rates
$k^{(P)}_{on}$ and $k^{(P)}_{off}$, as well as for the mRNA production rate $r$ that
fit the single molecule mRNA count data from \cite{JonesBrewster2014}. The
kinetic rates for the repressor are obtained by assuming that  $k^{(R)}_{on}$
is diffusion-limited \cite{JonesBrewster2014}, and demanding that $k^{(R)}_{off}$ be consistent
with the binding energies obtained in the lefthand panel of Figure~\ref{fig:AztecPyramidParameters}(B).
We note, however, that this model differs from the one presented
in Figure~\ref{fig:EquilibriumAssumptionRepressed} in the sense that upon
initiation of transcription at a rate $r$, the system does not transition from
state 3 to state 2. Further
comparison between this model and the model presented in Figure~\ref{fig:EquilibriumAssumptionRepressed}
is still needed and will be explored in future work \cite{Razo-Mejia2018}.

%

\begin{figure}
\centering{\includegraphics[width=4.0truein]{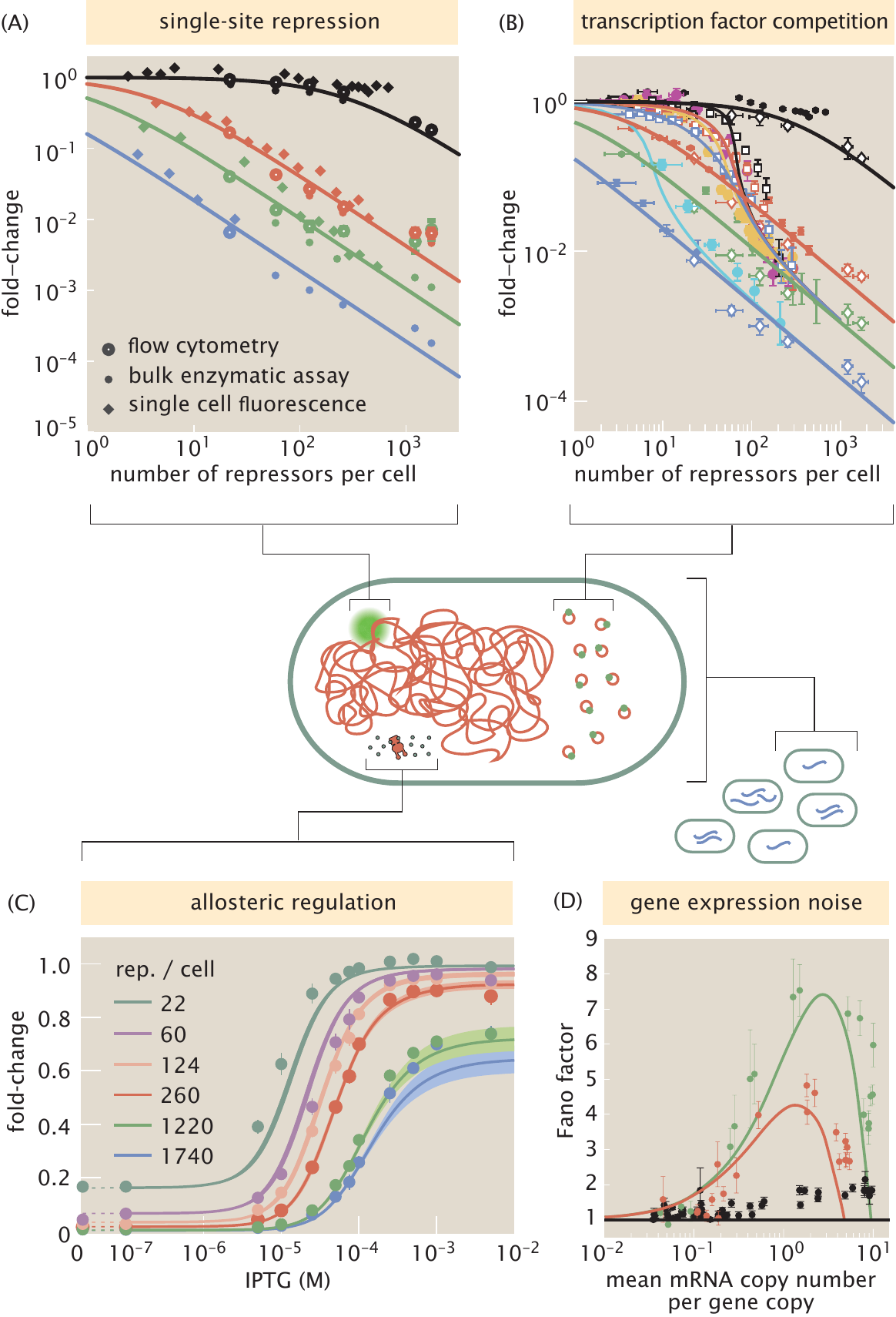}}
\caption{Experiment-theory dialogue in simple repression.  All curves are parameter-free
predictions based upon
the minimal parameter set introduced in Figure~\ref{fig:AztecPyramidParameters}. (A) Fold-change for simple repression
as a function of repressor copy number and operator strength for a single gene copy~\cite{Garcia2011c,
Weinert2014}.
(B) Fold-change for simple repression
as a function of repressor copy number and operator strength with repressor titration effect~\cite{Weinert2014}.
(C) Induction of the simple repression motif for different numbers of copies
of the repressor~\cite{Razo2018}. (D) Measurement of
gene expression noise for simple repression motif as reported by
 the Fano factor (variance/mean)~\cite{JonesBrewster2014}.
\label{fig:ExperimentTheoryDialogue}}
\end{figure}

With our single minimal parameter set in hand, it is now time to take the leap and to see
whether the theoretical framework that has been used to describe
various facets of the simple repression architecture actually works. Figure~\ref{fig:ExperimentTheoryDialogue}
shows the diversity of
predictions and corresponding measurements that partner with the
predictions given at the top of Figure~\ref{fig:AztecPyramidExperiments}.
In fact, the understanding summarized in this figure was developed sequentially
rather than with the ``all at once'' appearance conjured up
by Figure~\ref{fig:AztecPyramidParameters}.  Indeed, that is the principal reason that the discussion is so self-referential since over
the last decade, inspired by the many successes of others \cite{Oehler1994, Muller1996,Vilar2003a,Saiz2005, Kuhlman2007,Saiz2013},
we undertook a systematic effort to design experiments that
allowed us to control the various knobs of transcription already highlighted, to construct
the strains that make this possible and then to do the highest precision measurements
we could in order to test these predictions.

Figure~\ref{fig:ExperimentTheoryDialogue}(A)
shows a modern and predictive incarnation
of the experiments done by Oehler {\it et al.} to determine the response of the simple
repression motif to changes in repressor numbers and operator sequence (we showcased their results earlier
in Figure~\ref{fig:Oehler}). In this set of experiments, our ambition was to
control both the copy number of repressors and operator binding strengths and systematically measure the resultant expression over the entire suite of
different constructs, using only one repressor copy number for
each DNA binding strength to determine the parameter $\Delta \varepsilon_R$ as described
above. The measurements were taken in multiple ways: we used
both enzymatic and fluorescent reporters to read out the level
of gene expression, and we separately counted the number of repressors using quantitative immunoblotting and fluorescence measurements.
One of our central interests is in whether or not
different experimental approaches to ostensibly identical measurements yield
the same outcomes. We were encouraged, at least in this case, to find reasonable
concordance between them.

\begin{figure}
\centering{\includegraphics[width=4.0truein]{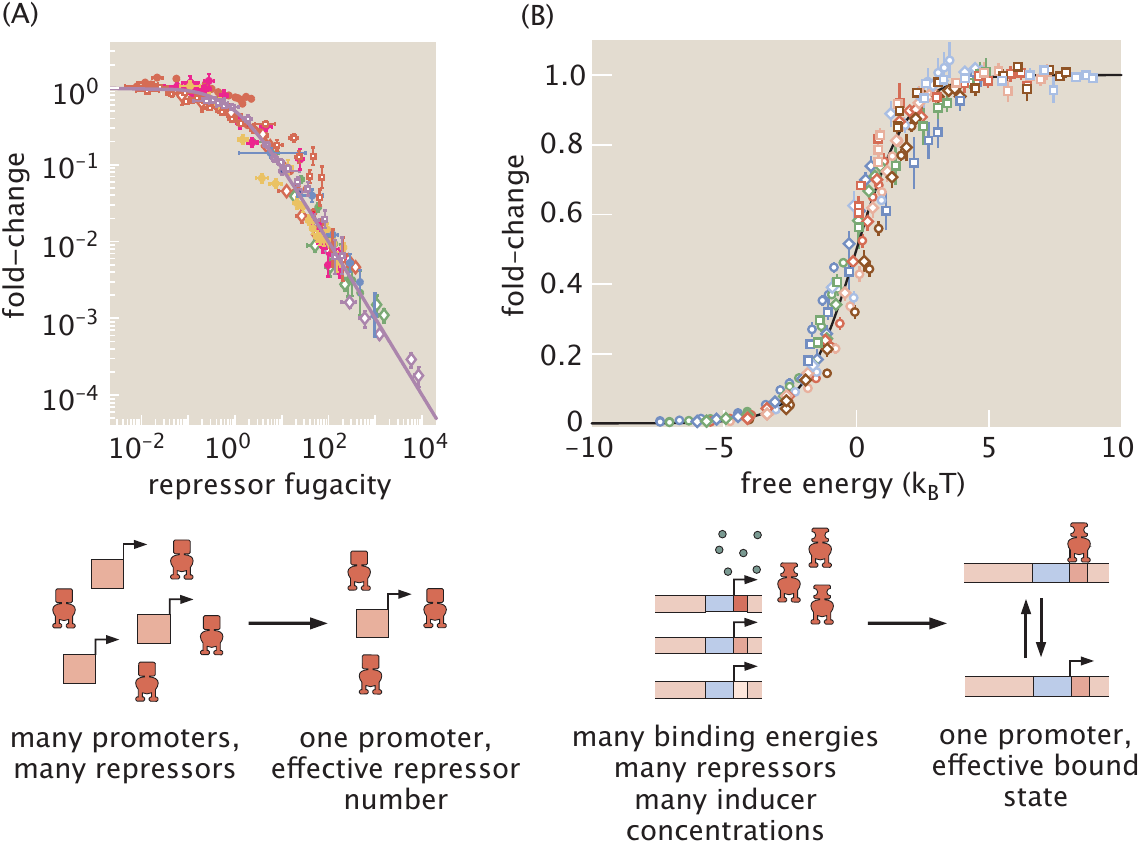}}
\caption{Data collapse of all data from the simple repression architecture. (A) Gene expression in the simple repression motif is dictated by an effective repressor copy number~\cite{Weinert2014-2}. (B)
Level of induction depends upon inducer concentration, repressor copy number
and repressor binding strength all of which fold into the free energy difference
between active and inactive forms of the repressor~\cite{Razo2018}.
\label{fig:DataCollapse}}
\end{figure}

The level of expression from our simple repression promoter can be significantly affected if the represors are
enticed away from it by other binding sites. The results of this much more demanding set of predictions surrounding the transcription factor
titration effect~\cite{Weinert2014} are shown in the next panel of the figure, Figure~\ref{fig:ExperimentTheoryDialogue}(B).
There are a number of ways to titrate away repressors: we can put extra copies of our gene of interest
on the chromosome or on plasmids (shown in the schematic
below the data), or use plasmids to simply introduce
decoy binding sites for the repressor that have no explicit regulatory role other
than pulling it out of circulation, effectively tuning the chemical potential of the repressor.
Note that in this case, the fold-change has a particularly rich behavior and this is on a log-log plot, where functional forms often appear as straight lines.
Figure~\ref{fig:DataCollapse}(A) brings together all of the data from
Figure~\ref{fig:ExperimentTheoryDialogue}(A) and (B) under one simple conceptual roof by determining the natural scaling
variable of the simple repression motif. This data collapse implies that any
combination of repressor concentration, binding site strength, and number and strength of competing
binding sites can be replaced by an equivalent effective promoter consisting of one binding site and an effective repressor number.

The middle panel of Figure~\ref{fig:AztecPyramidExperiments}(A) highlights
the next level in the hierarchy of theoretical predictions that can be made
about the simple repression motif, namely, how this motif responds to
inducer.  In Figure~\ref{fig:ExperimentTheoryDialogue}(C) we show
one example (from a much larger set of predictions~\cite{Razo2018}) of how the induction
response can be predicted for different operator strengths and repressor
copy numbers.  Here we highlight predictions for the $O2$ operator ranging
over the same repressor copy numbers already shown in Figure~\ref{fig:ExperimentTheoryDialogue}(A).
As with our ability to introduce the natural variables of the problem in
Figure~\ref{fig:DataCollapse}(A), induction responses also have a scaling
form that permits us to collapse all data onto a single curve
(Figure~\ref{fig:DataCollapse}(B)). Once again, the emergence of this natural scaling variable tells us that any set of repressor number, binding energies and inducer concentrations can be mapped onto a simple repression architecture with a corresponding effective binding energy.

The final part of our comparison of theory and experiment in the context of
the simple repression motif is shown in Figure~\ref{fig:ExperimentTheoryDialogue}(D).
The predictions about gene expression noise were already highlighted in the right panel of Figure~\ref{fig:AztecPyramidExperiments}(A).  Here what we see is that the Fano factor (i.e.
the variance normalized by the mean) is quite different for constitutive promoters
and promoters subject to repression in the simple repression motif~\cite{JonesBrewster2014}.
Discrepancies between the noise revealed in different regulatory architectures
remain to be resolved~\cite{So2011}.

The hierarchical analysis presented in Figure~\ref{fig:ExperimentTheoryDialogue} illustrates the unity of outlook and parameters afforded by performing
all experiments in the same strains. When experimental consistency is placed front and center, one minimal set of parameters appears to
serve as a predictive foundation for thinking about a broad variety of different constructs
and conditions over a host of different experimental scenarios and methods.


Nearly fifty years ago, Theodosius Dobzhansky wrote a beautiful article in {\it
American Biology Teacher} entitled ``Nothing in biology makes sense except in
the light of evolution" \cite{Dobzhansky1973}. This phrase, now an oft-quoted tenet of modern
biology, has resulted in evolution as the capstone to numerous biological
pyramids. As such, there is a reason we talk about
{\it climbing} the simple repression pyramid rather than saying that {\it we have climbed} it. Though the evolutionary aspects
of transcription are represented by the smallest part of
the pyramid in Figure~\ref{fig:AztecPyramid}, they are perhaps the most
daunting. At the time of writing, many different groups are still working to construct this section of the pyramid for simple repression \cite{razo-mejia2014, poelwijk2011, poelwijk2011a, dawid2010, Dekel2005, tugrul2015}.

To make meaningful predictions about the evolutionary potential of the simple repression motif, it is a requirement that we have a thorough knowledge of the minimal parameter set described in the preceding section. For example, we have shown that the sequence of the operator strongly influences the maximum level of gene expression given an input such as the concentration of inducer. One could extend this conclusion to make predictions of how the various properties of the induction profiles could
change due to mutation. It is reasonable to assume that mutations in the DNA binding pocket would alter only the strength of DNA binding and leave the inducer binding constants the same as the wild type. Conversely, mutations in the inducer binding domain would alter only the inducer binding constants. With quantitative knowledge of the single mutants, the theoretical underpinnings allow us to assume a na{\"i}ve hypothesis in which the two mutations are additive, resulting in a predictable change in the induction profile. Measurements of this flavor have been performed and published \cite{Daber2011}, however without knowledge of the parameters,
the predictive power is extremely limited.\\

\section{A Critical Analysis of Theories of Transcription}


Thus far, we have painted a rosy picture of the dialogue between
theory and experiment in the study of transcription in the simple
repression motif.  It is now time
to critique these approaches and see what such critiques imply
about future efforts to dissect the regulatory genome.  In the
paragraphs that follow, we have amassed a series of worthy
critiques of the program laid out thus far in the paper, and in each
case, we set ourselves the task of sizing up these critiques to see what we can learn
from them.
Our strategy is to discuss the high points of the analysis in the main body of
the text and to relegate the technical details behind that analysis to the appendices.



\begin{figure}
\centering{\includegraphics[width=5.0truein]{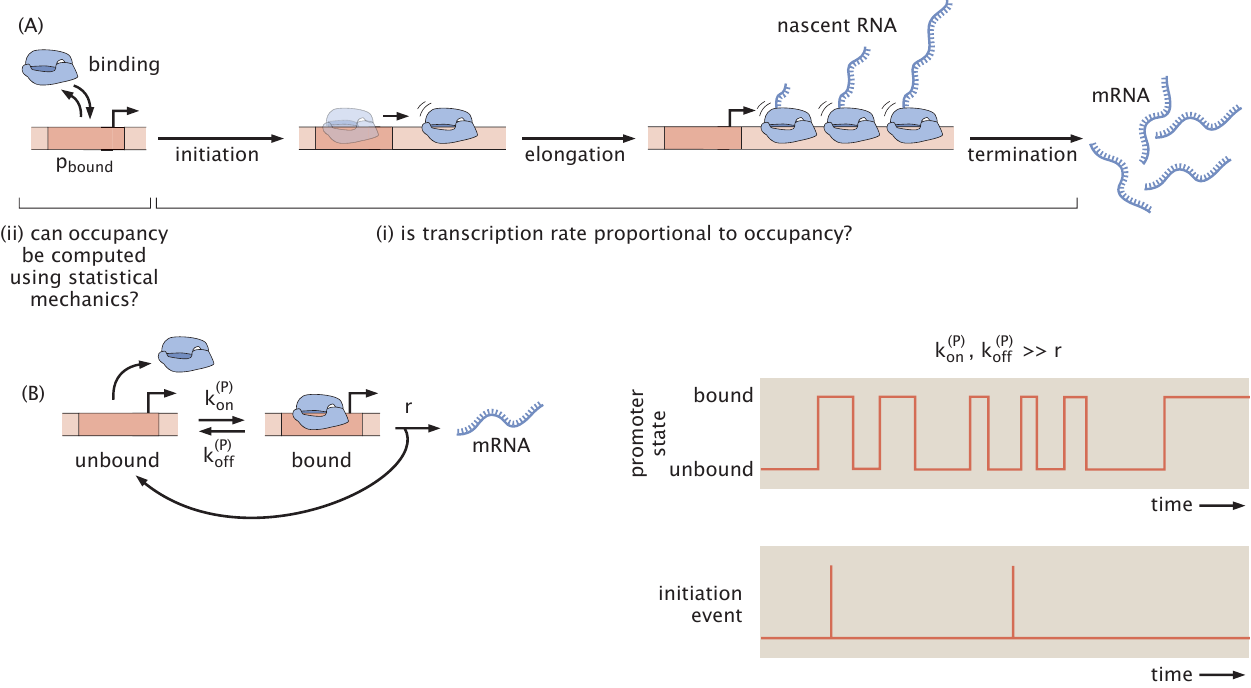}}
\caption{The occupancy hypothesis and the equilibrium assumption. (A) The
multiple steps between RNA polymerase binding and the termination of an mRNA raise
the question of whether the binding probability (occupancy) of RNA polymerase to
the promoter can be used as a proxy for the quantity of mRNA produced, and whether RNA polymerase
binding is in quasi-equilibrium such that the tools of statistical mechanics can be
used to compute this quantity. (B) The equilibrium assumption is fulfilled if
the rates of RNA polymerase binding and unbinding ($k_{on}^{(P)}$ and
$k_{off}^{(P)}$, respectively) are much faster than the rate of transcriptional
initiation $r$. See Appendix~\ref{sec:Equilibrium} for details on this simulation.
\label{fig:EquilibriumValidity}}
\end{figure}

{\it The Equilibrium Assumption in Thermodynamic Models.}
As already seen in Figure~\ref{fig:ComputingRepression}, there are multiple
 approaches to modeling transcription.  One broad class of models
 sometimes goes under the heading of ``thermodynamic models'', but
we would rather refer to them as models founded upon the {\it occupancy hypothesis}.
We can examine two critical questions about such models, shown diagrammatically in
Figure~\ref{fig:EquilibriumValidity}(A): (i) to what extent is it true that the rate of
transcription is proportional to the probability of promoter occupancy and (ii) can promoter
occupancy be fruitfully computed using the quasi-equilibrium assumption?

Recall that the assumption that the rate of transcription is proportional to the probability of RNA polymerase binding
to the promoter is central to the thermodynamic models.
Indeed, this assumption makes it possible to connect a theoretically accessible quantity, $p_{bound}$, to an experimentally measurable quantity, $dm/dt$.  This connection  can be used to test the  predictions stemming from these models. To answer the question of whether
the rate of transcription is proportional to $p_{bound}$, we must remember that, as shown in Figure~\ref{fig:EquilibriumValidity}(A),
there is a plethora of kinetic steps  between the binding of RNA polymerase and transcription factors to the DNA, and the ultimate
production of an mRNA molecule. Further, steps such as ``initiation'' in the figure are an oversimplification, as the process leading to promoter clearance and the initiation of productive transcription is composed of multiple
intermediate steps \cite{Record1996}.
 In Appendix~\ref{sec:OccupancyHypothesis} we explore the conditions under which this occupancy hypothesis is fulfilled. In particular, we consider a situation where the transition rates between intermediate steps correspond to zero-order reactions. To illustrate this, we refer to the first transition in Figure~\ref{fig:EquilibriumValidity}(A), which shows that the fraction of RNA polymerase molecules initiating transcription, denoted by  $I$, is related to $p_{bound}$. In a zero-order reaction scheme, the temporal evolution of $I$ is given by
\begin{equation}
    {dI \over dt} = r_i p_{bound},
\end{equation}
where $r_i$ is the rate of transcriptional initiation. In this scenario, the rate of change in the fraction of molecules initiating transcription is proportional to  the fraction of molecules bound to the promoter. As described in the Appendix, under this assumption, Equation~\ref{eq:dmdtpbound} can be used to relate the probability of finding RNA polymerase bound to the promoter to the rate of mRNA production.

Putting the occupancy hypothesis to a direct and stringent test requires us to have the
ability to simultaneously measure RNA polymerase promoter occupancy and output transcriptional activity.
The development of new approaches to directly measure DNA-binding protein occupancy in the vicinity of a
promoter and relate this binding to output transcriptional activity will make it possible to realize such
a test in the near future~\cite{Elf2007,Hammar2014,Xu2015,Cho2016}. While technology catches up to the
demands of our theoretical models, an indirect strategy for testing the occupancy hypothesis is to simply
ask how well the thermodynamic models do for the various predictions highlighted throughout the paper. Figure~\ref{fig:ExperimentTheoryDialogue} suggests
that, for the {\it lac} operon, the occupancy hypothesis is valid. However, it is
important to note that there are cases where this hypothesis has been explicitly called into question
both in the {\it lac} operon \cite{Garcia2012a,Hammar2014} and other regulatory contexts~\cite{Leung2004,Meijsing2009}.
As a result, the validity of the occupancy hypothesis should be critically examined on a system-by-system basis.

The second key assumption to be considered is the extent to which the system can be
viewed as being in ``equilibrium'', such that the tools of statistical mechanics can
be applied to calculate $p_{bound}$ and the fold-change. This equilibrium assumption permeates
the vast majority of the work presented here. In Appendix~\ref{sec:Equilibrium} we
dissect it in the context of the kinetic rates revealed in Figure~\ref{fig:AztecPyramidParameters}(B).
As we showed in Figure~\ref{fig:EquilibriumValidity}(B), in order for equilibrium to be a valid assumption
when calculating $p_{bound}$ for the constitutive promoter, the rates of RNA polymerase binding and
unbinding ($k_{on}^{(P)}$ and $k_{off}^{(P)}$, respectively) need to be much larger than the rate of
initiation $r$. However, we find that the inferred rates do not justify the use of the equilibrium
assumption: the rate of RNA polymerase unbinding from the promoter
is not much faster than the subsequent rate of initiation, such that the system does not
get to cycle through its various binding states and equilibrate before a transcript is produced.
However, our calculations reveal that, given these same rates, the fold-change in gene expression
can be calculated based on the equilibrium assumption. As discussed in detail in the Appendix, if $k_{on}^{(P)} \ll k_{off}^{(P)} + r$,
then when the system transitions to the polymerase-bound state,
it will quickly revert back to the unbound state either by unbinding or through transcription initiation.
As a result of this separation of times scales, the repressor gets to explore the bound and unbound states such
that its binding is equilibrated even if the RNA polymerase binding is not.

Finally, it is important to note that our conclusions about the applicability of equilibrium rely on committing to the kinetic scheme presented in Figure~\ref{fig:EquilibriumAssumptionRepressed} and on the inferred parameters shown in Figure~\ref{fig:AztecPyramidParameters}(B). Changes to the molecular picture of the processes underlying repression and gene expression could significantly affect our conclusions. Indeed, researchers have cast doubt on the applicability of equilibrium to describe the {\it lac} operon \cite{Hammar2014} as well as other gene-regulatory systems \cite{Estrada2016b,Li2018}.



{\it Reconciling Thermodynamic Models and Statistical Mechanical Models.}
Thermodynamic models of transcription can be formulated either directly
in the language of statistical mechanics, by invoking binding energies and
explicitly acknowledging the various microscopic states available to the system,
or in the language of thermodynamics, in which DNA-protein interactions
are characterized using dissociation constants. The literature
is not always clear about the relation between these two perspectives
and our central argument (fleshed out in detail in Appendix~\ref{sec:ThermoVsStatMech}) is that
they are equivalent. That argument was really already made in Figure~\ref{fig_states}, in which we saw
that the statistical weights of the three states of the simple repression motif
can be written in either of these languages.

We personally favor the statistical mechanical language because we find
that, in going to new regulatory architectures, it is more microscopically transparent
to enumerate the microscopic states and their corresponding energies than to
invoke dissociations constants that combine these microscopic interactions into an
effective parameter as shown below for the case of the nonspecific background.
One related point of possible confusion  concerns the use of parameters such as $N_{NS}$ in the statistical mechanical
approach to occupancy models of transcriptional regulation.
In the Appendix, we demonstrate that the dissociation constant $K_d$ is given by
\begin{equation}
K_d={N_{NS} \over V_{cell}} e^{-\beta \Delta \varepsilon}.
\end{equation}
This equivalence shows that the parameter
$N_{NS}$, which reflects the genome size and hence the size of
the nonspecific background binding landscape, is in fact just a contribution to the standard state
concentration used in conjunction with the dissociation constant $K_d$ in disguise.

%


{\it The Energy of Nonspecific Binding.} One of the key simplifying assumptions
often invoked in the context of thermodynamic models of transcription is the
treatment of the binding of transcription factors to the nonspecific
background as though all such nonspecific sites are equivalent. For transcription
factors such as LacI, there is wide-ranging evidence from diverse types of experiments (e.g. measurements of the
protein content of genome-free mini-cells and imaging using modern microscopy techniques) that these transcription
factors are almost
always bound to the genome rather than free in cytoplasm~\cite{Runzi1976,Kao-Huang1977,GarzadeLeon2017}.
As such, when computing the probability of promoter occupancy by either polymerase
or repressors, we need to account for the distribution of these molecules across
the remainder of the genome.

\begin{figure}
\centering{\includegraphics[width=3.5truein]{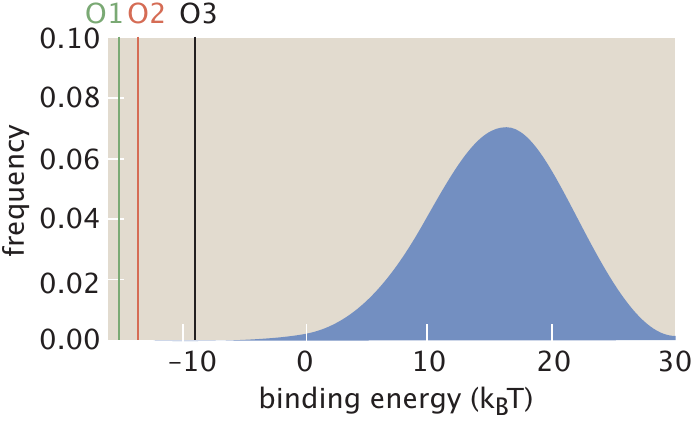}}
\caption{Distribution of nonspecific binding energies. The distribution shows the predicted binding energies for LacI to
all possible 21 bp sequences on the {\it E. coli} genome (strain MG1655, GenBank: U00096.3). Binding energies
were calculated using an energy matrix obtained by Sort-Seq on the LacI simple repression
architecture~\cite{Barnes2018}. \label{fig:NonSpecificReservoir}}
\end{figure}

 With an
approximately $5 \times 10^6$ basepair genome as in {\it E. coli}, it at first blush
seems ridiculous to proceed as though $5 \times 10^6 - 1$ of those sites have the exact same
energy, $\varepsilon_{NS}$.  To explore the distribution of nonspecific energies, one
idea is to slide an energy matrix, much like those
determined through Sort-Seq~\cite{Kinney2010, Brewster2012,Kuhlman2013,Belliveau2018},
across the entire genome, base pair by base pair,  to get the full distribution. Such a distribution is
shown in Figure~\ref{fig:NonSpecificReservoir} where the energy matrix for
the LacI repressor was applied to the entire {\it E. coli} genome.  We see
immediately that an exceedingly
small number of sites have a negative binding energy, meaning more preferable
binding than the vast majority of sites, which are found to be positive. The three native {\it lac}
operators, shown as black, red, and green vertical lines, have highly negative
binding energies compared to the rest of the sites. With knowledge of the distribution, it is tempting to  use this directly in the thermodynamic  calculations to possibly get a better
treatment of the nonspecific background. However, for now, it is a  luxury
to have an accurate energy matrix that reports the binding energy  of a given
transcription factor to a DNA binding
site {\it in vivo}. We certainly don't know the binding energy matrix for all
transcription factors that would permit the determination of the
distribution of nonspecific binding energies.

But more interestingly, as we show in detail in Appendix~\ref{sec:NSGenomicBackground},
there really is no difference between using
the complete distribution of binding energies versus an effective energy of
the entire genome. This concept is explored in detail in the Appendix and
agrees with more sophisticated treatments using concepts from statistical physics~\cite{Gerland2002b, Sengupta2002}. We treat this problem using the three toy models shown in Figure~\ref{fig:ThreeBindingEnergyDistributions}.
 First, we assume that there is a uniform binding energy
distribution in which all binding sites have the same energy. By definition, this is
the simplest approach where this energy can be used directly in the partition function.
The second example is the extreme case in which there are only two nonspecific binding energies, $\varepsilon_1$ and $\varepsilon_2$,
which are evenly distributed about the genome. In this case we can show the nonspecific
background behaves as though it has a single effective binding energy of the form
\begin{equation}
\epsilon_{NS} = {\epsilon_1 - \epsilon_2 \over 2},
\end{equation}
showing that the effective nonspecific binding energy $\epsilon_{NS}$  tells the
exact same story as using the full distribution. Finally, we take the more realistic
case in which we assume a Gaussian distribution of binding energies across the genome with
mean $\bar{\epsilon}$ and standard deviation $\sigma$,
much like what is seen in Figure~\ref{fig:NonSpecificReservoir}. Here, a few more mathematical
steps outlined in the Appendix delivers us to the result,
\begin{equation}
\epsilon_{eff.} = \bar{\epsilon} - {\beta\sigma^2 \over 2}.
\end{equation}
Note that this shows that even if we have a Gaussian distribution of nonspecific binding energies, it can be treated
exactly as a uniform distribution with a single effective energy.

{\it Promoter competition against nonspecific DNA-binding
proteins.} Up until this point, we have considered the effect of LacI
nonspecific binding throughout the genome on its regulatory action in the
context of the simple repression motif. However, just like in the simple
repression motif where the promoter and operator constitute the specific binding sites
for RNA polymerase and repressor, respectively, these same sequences serve as
substrates for the nonspecific binding of other DNA-binding proteins that
decorate the bacterial genome. In Appendix~\ref{sec:NSPromoterBinding}, we show
how the effect of these nonspecific competitors can be absorbed into an
effective number of nonspecific binding sites $N_{NS}$ such that the theoretical
models describing the simple repression motif retain their predictive power.
Interestingly, the calculations presented in the Appendix also suggest that, as
the concentrations of these DNA-binding proteins is modulated due to changes in
growth rate, the effect of these competitors on the rescaled $N_{NS}$ remains
unaltered. This indifference to growth rate stems from the fact that, as growth
rate increases, both the overall protein concentration and the cell's DNA
content increase. This simultaneous increase in protein and DNA concentration
leads to a relatively constant number of proteins per DNA target in the cell
irrespectively of growth conditions.

\begin{figure}
\centering{\includegraphics[width=6.0truein]{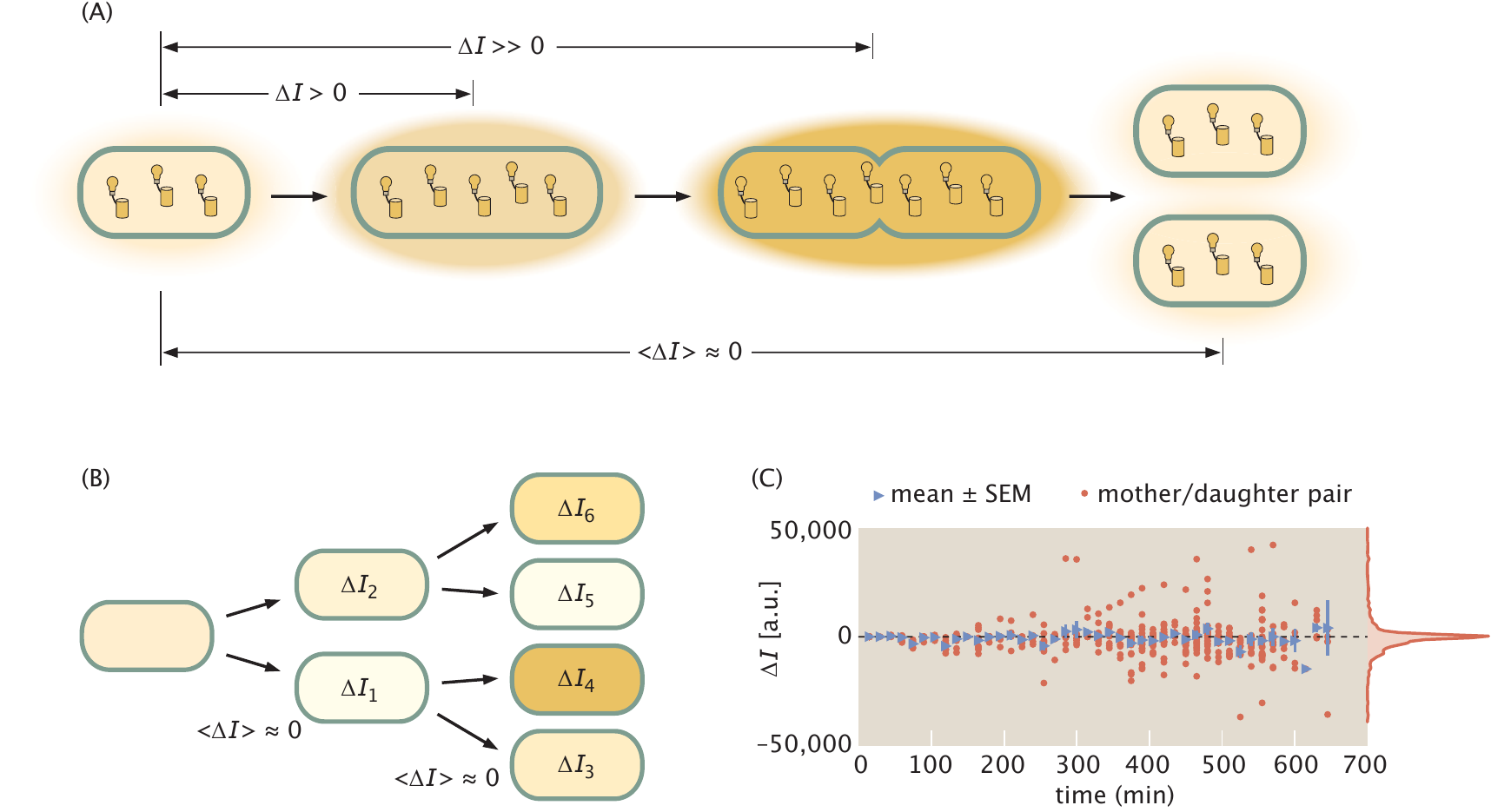}}
\caption{Test of the idea of steady-state gene expression for cells in exponential phase. (A)
Diagrammatic view of protein dilution through cell division. As cells grow,
the expression of fluorescent proteins marches on. As the cell approaches
division, the total detected fluorescence is much larger than detected at the
cells' birth. On average, the proteins are split evenly among the daughter
cells, resulting in a comparable fluorescence level as the original mother
cell. (B) Schematic of experimental measurement. To test the steady-state hypothesis, we monitored the growth of several bacterial microcolonies originating
from single cells and tracked the difference in intensity with respect to their mother cell as a function of
time for each daughter cell through the family tree. (C) Fluorescence intensity difference between mother/daughter pairs as
a function of time. Red points indicated individual daughter/mother pairings in a given lineage.
Blue triangles represent the average difference at that time point. Error
bars on blue points are the standard error of the mean. A kernel density
estimation of the $\Delta I$ distribution is shown on the right-hand side of
the plot. The black dashed line is at zero. \label{fig:SteadyState}}
\end{figure}

{\it Is Gene Expression in Steady State?} 
A critical assumption in our experimental
measurements of gene expression is that gene expression is in steady-state.
The definition of steady-state has different meanings depending on the method
of measurement. For mRNA FISH, for example, we assume that the mRNAs are
produced at rate $r$ that matches the rate of degradation $\gamma m_{ss}$,
where $m_{ss}$ is the steady-state level of mRNA. When measuring protein
expression, we assume that the protein accrued over the cell cycle is negated
by the dilution of these proteins into the daughter cells upon division, as is
shown in Figure~\ref{fig:SteadyState}(A). Through this assumption, we are able
to state that, on average, a single measurement represents the level of
expression for that particular time point rather than integrating over the
entire life history of the cell. A typical rule-of-thumb is that steady-state
expression is reached when the cells enter the exponential phase of growth.

We put this hypothesis to the test by directly measuring the expression
level of exponential phase \textit{E. coli} over time. Using video microscopy, we monitored the growth of cells constitutively
expressing YFP in exponential phase (OD$_{\mbox{600nm}}$ $\sim 0.3 - 0.4$) in
minimal medium with a doubling time of approximately an hour (Figure~\ref{fig:SteadyState}(A)) following the experimental approach undertaken by Brewster {\it et al.} \cite{Weinert2014}.
Starting from a
single cell, we tracked the lineages as the microcolony developed and compared
the fluorescence in arbitrary units of each cell to the founding mother cell.
If steady-state gene expression has been achieved, this approach,
schematized in Figure~\ref{fig:SteadyState}(B), will result in
an average difference in fluroescence $\Delta I$ of zero. The results of this experiment are shown in
Figure~\ref{fig:SteadyState}(C). In the figure, we see that individual measurements
(red points) are scattered about zero but that, once the mean
difference in intensity is considered (blue triangles), the data becomes very tightly
distributed about zero (black dashed line). These results show that, when
cells are growing in exponential phase, gene expression levels are in steady-state and
the reporter is not accrued over the life history of the cell lineage.


{\it Allosteric Models vs Hill Functions.}  Though
many thermodynamic models of gene regulation attempt
to enumerate the entirety of microscopic states and assign
them their appropriate statistical weights, it is also
extremely popular to adopt a strictly phenomenological
model of binding described by Hill functions.   It is undeniable that the Hill
function features prominently in the analysis of many biological processes
(for interesting examples see, \cite{Rogers2015, Rohlhill2017, Setty2003, Dekel2005}). However,
treating allosteric systems with Hill functions often abstracts away the
important physical meaning of the parameters and replaces them with
combinations of polynomials often referred to as ``lumped parameters". For
example, one could treat the induction profiles of LacI discussed earlier
in this work using a Hill equation of the form,
\begin{equation}
\text{fold-change} = \text{leakiness} + \text{dynamic range} {\left({c \over K_D}\right)^n \over 1 + \left(c \over K_D\right)^n},
\label{eq:hill_function}
\end{equation}
where the leakiness is set as the zero point of expression.  With increasing concentration
$c$ of ligand, the leakiness is modified
 by an expression describing the activity of the repressor using
 a Hill function. In this
expression, $c$ corresponds to the concentration of inducer. {$n$ is the
Hill coefficient which describes the cooperativity of repression, and
$K_D$ is an effective dissociation constant \cite{Kuhlman2007}.

Note that nowhere in this expression is any treatment of the allosteric
nature of the protein! While structural biology has demonstrated that this repressor can exist
in  active and inactive states, each of which has its own dissociation
constant for the inducer, all of these details have been lumped
into the $K_D$ parameter.
Figure~\ref{fig:hill_v_mwc}(A) shows  Equation~\ref{eq:hill_function} applied to
an induction profile of the {\it lac} simple repression motif with an O2
operator and 260 repressors per cell. Unsurprisingly, this equation can fit
the data very nicely when all of the coefficients are properly determined.  In
fact, this fit is nearly indistinguishable from that obtained through a Monod-Wyman-Changeux
(MWC) model inspired approach \cite{Razo2018}, as is shown in Figure~\ref{fig:hill_v_mwc}(B). However, fitting a Hill function results in a single curve.
In the Hill framework, for each induction
profile we must fit Equation~\ref{eq:hill_function} once again for  all
parameters. As the parameters for an allosteric model have a direct
connection to the biological properties of the repressor molecule, we  can use
the parameter values determined from one experimental circumstance to predict  a wide swath
of other induction profiles. Examples of such curves are  shown as gray
profiles in Figure~\ref{fig:hill_v_mwc}(B).

What distinguishes allosteric models such as MWC and Koshland-Nemethy-Filmer (KNF, \cite{Koshland1966})
from Hill functions is that they make a
tangible connection with what structural biology has taught us about the conformational states
of proteins.  The existence of inactive and active states implies that activity curves will be
very special ratio of polynomials. While an individual fit may be comparable
in quality to that obtained by a Hill function, the loss of this physical
context results in a fit that has no predictive ability. The MWC and KNF
models, however, open the door to a huge suite of predictions not only about
experiments like those described in this review but also for biochemical
experiments at the level of single molecules. For example, the allosteric
treatment of induction hints at how mutating the repressor directly would
change the behavior of the system. It's easy to hypothesize that mutations in
the DNA binding domain would alter the binding energy of the repressor to the
DNA $\Delta\varepsilon_R$ whereas mutations in the inducer domain would
alter the $K_A$ and $K_I$ (Figure~\ref{fig:AztecPyramidParameters}(A)). If we were to redo the analysis by fitting
phenomenological Hill equations, we would be left in the dark as to how
to predict the effect of either of these perturbations.


\begin{figure}
\centering{\includegraphics[width=5.0truein]{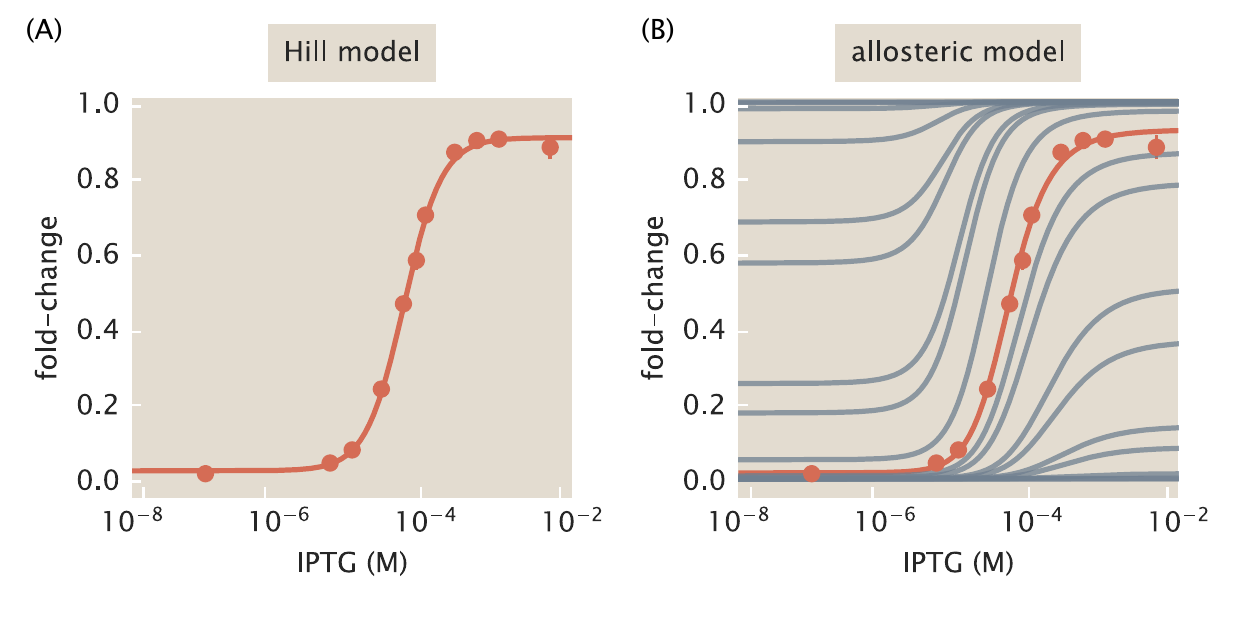}}
\caption{ Predictive versatility of the Hill function versus allosteric models.
(A) Measurements of the fold-change of a  simple repression architecture as a
function of IPTG concentration. Points and error bars represent the mean and standard error
of ten biological replicates of repression of the O2 operator with 260
repressors per cell. (A) Solid line is the best-fit of the standard Hill
function given in Equation~\ref{eq:hill_function}. (B) Best-fit line for the data
using the MWC model of allostery coupled with the thermodynamic model is in
red. Gray lines represent predicted induction profiles of other  combinations
of repressor copy numbers and DNA binding energies. These predictions are made
using only the parameters fit from a single strain.  Tests of these predictions
were shown in Figure~\ref{fig:ExperimentTheoryDialogue}(C).
\label{fig:hill_v_mwc}}
\end{figure}

{\it Two-State vs Three-State Dynamics.} Most of the theoretical work on mRNA distribution dynamics has focused on the
two-state model for a regulated promoter in which the promoter is treated as though it has two available states, inactive and active~\cite{So2011,Peccoud1995,Sanchez2008}.
Indeed,  the predictions from \cite{JonesBrewster2014} shown in Figure~\ref{fig:ExperimentTheoryDialogue}(D) were calculated using this model. However, another  critical question to be examined in the context of theoretical models
of transcriptional noise is the relative merits of the two-state and three-state models (for the three-state model see Figure~\ref{fig:EquilibriumAssumptionRepressed}).
Note that within this framework, the
unregulated promoter itself becomes an effective two-state model
since we now acknowledge both the empty promoter and the promoter
occupied by RNA polymerase.  In this case, the mRNA distribution can be fitted with the
parameters $k_{on}^{(P)}$, $k_{off}^{(P)}$, $r$ and $\gamma$ while still
accounting for the variability in promoter copy number across the cell cycle
and this is the strategy used in the parameter determination described
in Figure~\ref{fig:AztecPyramidParameters}(B).

We have found that  it is possible to fit the full mRNA distribution
using either the two-state or three-state models as already described in
 ref.~\cite{JonesBrewster2014}.  However, to get a fully self-consistent parameter
 set  in which the mean-fold change as described in both the thermodynamic
 and kinetic pictures are identical, it is necessary to resort to the three-state model
  that explicitly accounts for repressor and polymerase binding.
 Specifically,  we demand  that the repressor kinetic parameters $k_{on}^{(R)}$
and $k_{off}^{(R)}$ are consistent with the  repressor copy number $R$ and
the repressor-DNA binding energy $\Delta\varepsilon_R$.
The parameters reported
in Figure~\ref{fig:AztecPyramidParameters}(B) were determined using these constraints,
giving identical results for the mean fold-change under both languages and not surprisingly,
requiring the full three-state model for this self-consistent picture to emerge.

\section{Simple Repression in Other Contexts}

Thus far, we have focused on one realization of the simple repression architecture.
But in fact,
the way that cells use the simple regulatory architecture is
much more diverse, as illustrated in Figure~\ref{fig:OtherSimpleRepression}.
 Variants of this architecture
provide alternative means for the cell to perform signal transduction. Like
LacI, many repressors are inducible, whereby binding of a small-molecule
signaling ligand reduces their ability to bind DNA. The identities of these
ligands are generally related to the physiological role provided by the operon
under control. For example, while LacI binds allolactose and is involved in
lactose utilization, GalR binds galactose and this in turn provides control over
galactose usage~\cite{Semsey2002, Oehler2006}. Among those having a simple repression architecture, MprA has
been reported to bind antimicrobial agents such as 4-dinitrophenol and carbonyl
cyanide m-chlorophenylhydrazone (CCCP), and negatively regulate the expression of
multidrug resistance pumps~\cite{Brooun1999}. A related but opposite logic is also
commonly observed, referred to as co-repression, where binding of a
small-molecule ligand instead will enhance the binding of the repressor to DNA. For
example, TrpR binds tryptophan and provides repression of the tryptophan
biosynthesis pathway, as well as repressing its own expression~\cite{Yang1996}.

\begin{figure}
\centering{\includegraphics[width=6.0truein]{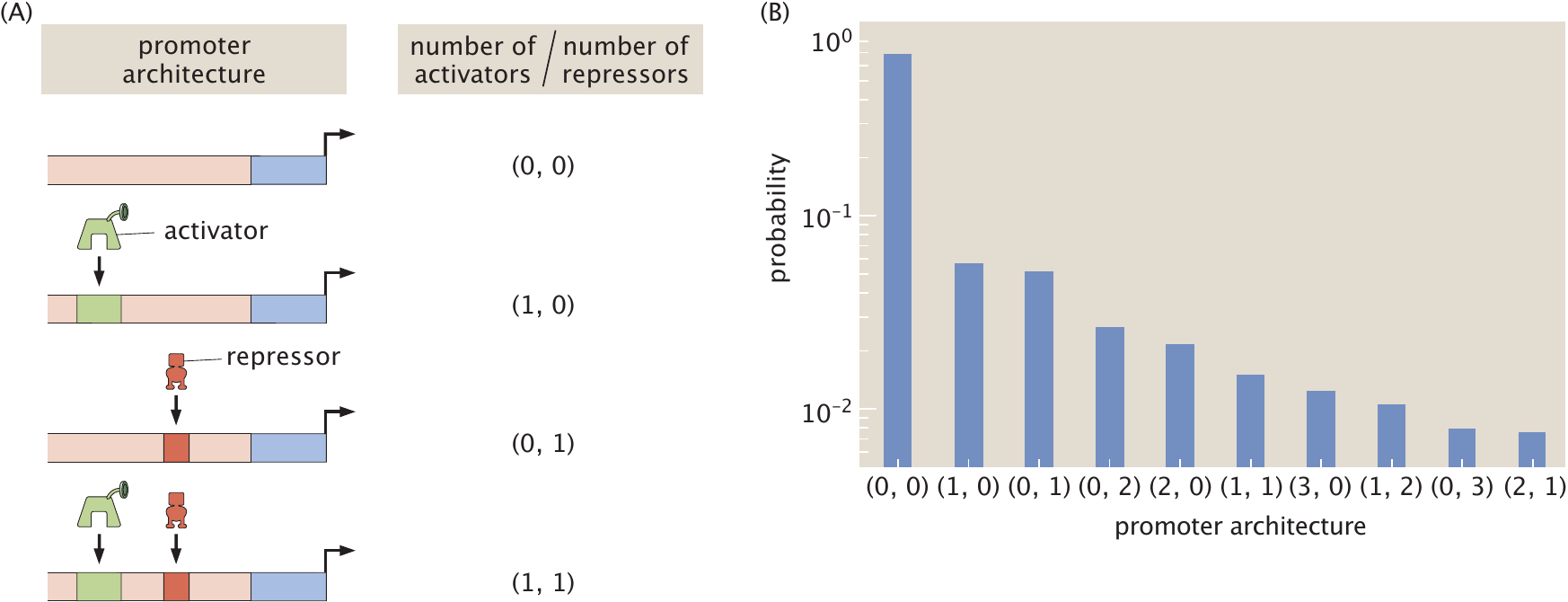}}
\caption{ Simple repression in other contexts. Here we summarize several different modes of regulation that are observed at (0,1) architectures. Like LacI, many transcription factors are inducible and binding by a specific ligand leads to a loss of repression. Conversely, a number of transcription factors undergo co-repression and bind the DNA more strongly upon binding of a ligand to the repressor. For the examples identified, the transcription factor is shown in red text, while the ligand is shown in black. Several transcription factors appear as part of two component signal transduction systems, whose  phosphorylation-dependent DNA-binding strength is changed by the activity of membrane bound sensor kinases. Lastly, repression can be modulated by changing the copy number of the repressor in response to stimuli. This can be achieved through self-cleavage (e.g., LexA), or by cellular proteases (e.g., RelB by Lon).
\label{fig:OtherSimpleRepression}}
\end{figure}

In both induction and co-repression, signaling is achieved by a ligand internal
to the cell. Another approach is to instead monitor the external environment directly,
which is the role provided by two-component signal transduction systems~\cite{Laub2007}. Here, the
signal detection is typically carried out by a transmembrane protein, a sensor
histidine kinase, which then activates a transcription regulator by
phosphorylation. Such sensors that activate repressors involved in simple repression
architectures include PhoR, ArcB, and CpxA, which regulate the DNA binding
activity of PhoB, ArcA, and CpxR, respectively. The repressor PhoB is involved
in regulating phosphorus uptake and metabolism, while ArcA
primarily acts as a repressor under anaerobic conditions~\cite{Wanner1993, Malpica2006}. CpxR appears to act on
at least 100 genes, in response to cell envelope stress, but also plays roles
associated with motility, biofilm development, and multidrug
resistance~\cite{Ruiz2005}.

Cells have also devised ways to rapidly respond to stimuli by actively degrading
regulatory proteins under specific stimuli. The DNA damage, or SOS, response
provides one such example, which is mediated by the repressor LexA~\cite{Little1982}.  Under
conditions of DNA damage, LexA undergoes a self-cleavage reaction that is further catalyzed by the protein RecA, and this provides de-repression of about 40 genes~\cite{Giese2008}.
Toxin-antitoxin systems such as RelB-RelE serve as another example of this.
While the toxin RelE is metabolically stable, with a cellular concentration
dependent on the cell division time, the antitoxin RelB is actively degraded by
the protease Lon and this can lead to a much shorter half-life~\cite{Cataudella2013, Overgaard2009}.

The examples provided here serve as a testbed for signal transduction strategies
that demand further quantitative analysis and can be considered under the
experimental-theoretical framework we have presented in this review.
Table~\ref{tb_summary_O1} gives us another way to get a sense of the diversity
of simple repression motifs in {\it E. coli} by showing us the copy numbers of
the key transcription factors involved in simple repression.

\begin{table}
\footnotesize
\begin{center}
\begin{tabular}{lrrr}
\toprule
protein &  copy number in         &   standard deviation          & coefficient of variation \\
        &  glucose, minimal media &   across 22 growth conditions &   \\
\midrule
     HU &                   87425 &                                           28629 &                      0.37 \\
   H-NS &                   22541 &                                            7181 &                      0.24 \\
   IscR &                    7687 &                                            2603 &                      0.49 \\
    Fur &                    6492 &                                            1707 &                      0.29 \\
    Lrp &                    6092 &                                            1339 &                      0.20 \\
    IHF &                    5018 &                                            1223 &                      0.25 \\
   ArcA &                    3367 &                                            1030 &                      0.24 \\
    CRP &                    2048 &                                             646 &                      0.24 \\
   AlaS &                    1948 &                                             605 &                      0.33 \\
   MprA &                    1085 &                                             516 &                      0.61 \\
   PepA &                    1076 &                                             259 &                      0.23 \\
   MetJ &                     990 &                                             231 &                      0.31 \\
   CpxR &                     933 &                                             158 &                      0.17 \\
   NsrR &                     872 &                                             189 &                      1.78 \\
   PurR &                     826 &                                             165 &                      0.24 \\
    FNR &                     609 &                                             236 &                      0.49 \\
   LexA &                     560 &                                             177 &                      0.32 \\
   CysB &                     523 &                                             124 &                      0.33 \\
   AllR &                     206 &                                              68 &                      0.43 \\
   FadR &                     186 &                                              75 &                      0.34 \\
   RelB &                     178 &                                              61 &                      0.53 \\
   TrpR &                     167 &                                              35 &                      0.22 \\
    Cra &                     148 &                                              87 &                      0.37 \\
   UidR &                     139 &                                             137 &                      1.06 \\
   NagC &                     124 &                                              36 &                      0.26 \\
   LacI &                      23 &                                               8 &                      0.65 \\
   AcrR &                      21 &                                              10 &                      1.08 \\
   DicA &                      20 &                                               6 &                      0.40 \\
   BirA &                      19 &                                               7 &                      0.50 \\
   AscG &                      17 &                                              12 &                      0.62 \\
   NadR &                      16 &                                               4 &                      0.26 \\
   PaaX &                      11 &                                              19 &                      0.64 \\
   PhoB &                       7 &                                               5 &                      0.45 \\
\bottomrule
\end{tabular}
\caption{Summary of transcription factors identified in (0,1) regulatory
architectures. Protein copy numbers are per cell and were determined by mass spectrometry \cite{Schmidt2016}. The
values for HU and IHF were taken as the average of their individual subunits (HupA and HupB for HU, and
IhfA and IhfB for IHF). \label{tb_summary_O1}}
\end{center}
\end{table}


\section{Beyond Simple Repression: Building New Pyramids}

Of course, as we already showed in Figure~\ref{fig:ArchitecturesSummarized},
there is far more to transcriptional regulation than simple repression.
Since the original development of the repressor-operator model by Jacob and Monod,
the regulatory mechanisms of the {\it lac} operon have been resolved in exquisite detail, as shown diagrammatically
on the left-hand side of Figure~\ref{fig:lacoperonComplexity} \cite{Muller-Hill1996}. The picture that has
emerged is a rather complex one, in which Lac repressor monomers assemble into a dimer of dimers. These
repressors can bind to two of the three operators found in the {\it lac} operon simultaneously, resulting
in DNA looping and the stabilization of repressor action. Furthermore, the binding affinity of repressor
to the DNA is modulated by inducer, which can be actively pumped into the cell by the Lac permease, which
is one of the subjects of regulation by the repressor.
This panoply  of regulatory features  calls for a  complex theoretical description of the {\it lac}
operon which can be nevertheless built on the parameters already obtained by building
the simple repression pyramid.

One of the most interesting features of regulation in prokaryotes and eukaryotes alike comes in the form of DNA looping.
Such biological action at a distance is seen in the
wild-type {\it lac} operon itself, allowing us to dissect this ubiquitous regulatory mechanism quantitatively.
Just as it was
possible to engineer pared-down versions of the simple repression motif,
similar exercises have been undertaken in the context of
DNA looping, as shown in Figure~\ref{fig:Looping}(A). Looping has been explored in a wonderful series of
experiments from the M\"uller-Hill lab \cite{Oehler1994,Muller1996}, and has also been elegantly treated using thermodynamic models~\cite{Vilar2005}. These threads of research show
how a pyramid of regulatory understanding for wild-type operons can be constructed, featuring
multiple binding sites and DNA looping.


Using the same minimal parameter set already identified in Figure~\ref{fig:AztecPyramidParameters}(B),
it is possible to make predictions about how the regulatory response
will work in the context of DNA looping.  For example, thermodynamic models of
DNA looping identify one new key parameter with respect to those presented in
Figure~\ref{fig:AztecPyramidParameters}(B): the DNA looping free energy \cite{Vilar2005,Bintu2005a}.
By fitting this model to the repression corresponding to the looping architecture shown in
Figure~\ref{fig:Looping}(A) for a particular number of repressors per cell, the model
predicts the repression value as repressor copy number is systematically varied.
Similarly, it is also possible to do an operator swap experiment in which
the DNA loop itself, and hence the DNA looping free energy,
is unchanged, but instead the binding sites that the repressor
uses to form the loop are varied.  Figure~\ref{fig:Looping}(B) shows the outcome of
such experiments.  In Figure~\ref{fig:Looping}(C), we also show
that the inferred looping free energy is indifferent to the choice of
operators used to induce the loop.  The collection of results shown in
Figure~\ref{fig:Looping} provide further exciting evidence of the transferability of
the minimal parameter set determined in the simple repression architecture.

\begin{figure}
\centering{\includegraphics[width=5.0truein]{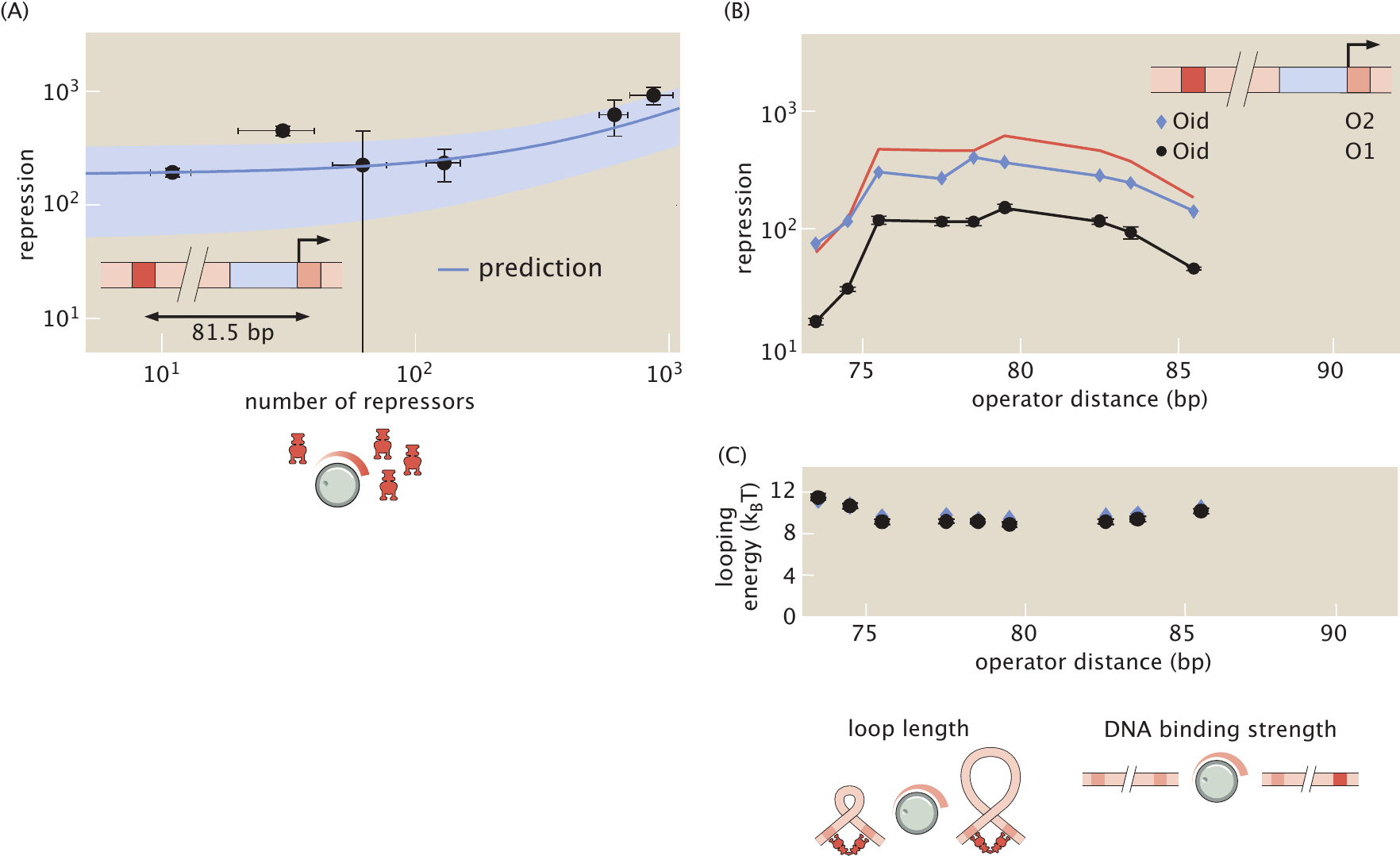}}
\caption{Regulatory action at a distance.  The same minimal parameter
set remains valid in the context of DNA looping, with the only requirement being
to introduce a new parameter that captures the free energy of DNA looping. (A) Repression for
the case of DNA looping as a function of the number of repressors per cell. (B,C) Operator swap
experiment.  In this case, for each DNA loop length, the operators that flank the loop
were changed.  (B) Using the Oid-O1 data to obtain the DNA looping free energy as a function
of operator distance, the thermodynamic model makes a parameter-free prediction of how repression
will work out in this case shown in the red curve.  (C) Inferred looping free energy is the same regardless of
which operators flank the loop.   Adapted from~\cite{Boedicker2013}.
\label{fig:Looping}}
\end{figure}


In our opinion, one of the most surprising aspects about the state of the art in regulatory
biology is our ignorance of regulation across genomes writ large. Even in the best understood of organisms, namely {\it E. coli},
we have no idea how more than half of the annotated genes are rgulated, as we illustrated earlier in Figure~\ref{fig:RegulatoryIgnorance} \cite{Keseler2010, GamaCastro2016, Fang2017}.
There we represented the circular {\it E. coli} genome with those operons for which there
is some regulatory annotation shown in blue and those thus far featuring
no such regulatory knowledge shown in red. Faced with the kind of ignorance revealed in
that figure, there is no prospect of building up
a regulatory dissection like that we have reviewed in the context of
simple repression.  To rectify this, we need to establish methods that will allow us, first of all, to simply draw the cartoons of how a given
gene's regulatory apparatus is wired.  Recent work has begun to develop tools that make
it possible to go from regulatory sequence to the kind of regulatory architecture
cartoons shown in Figure~\ref{fig:ArchitecturesSummarized} \cite{Kinney2010, Belliveau2018, Kosuri2013}.
Figure~\ref{fig:BeyondLac} exemplifies how a combination of
mutagenesis, deep sequencing, mass spectrometry and information theory has made it
possible to take the uncharacterized genes reported in Figure~\ref{fig:RegulatoryIgnorance}
and figure out their regulatory architecture~\cite{Kinney2010, Belliveau2018}. Each time we
identify how a given regulatory architecture is configured we are then poised to construct a new
pyramid based upon minimal parameter sets like the one we describe here.

\begin{figure}
\centering{\includegraphics[width=2.0truein]{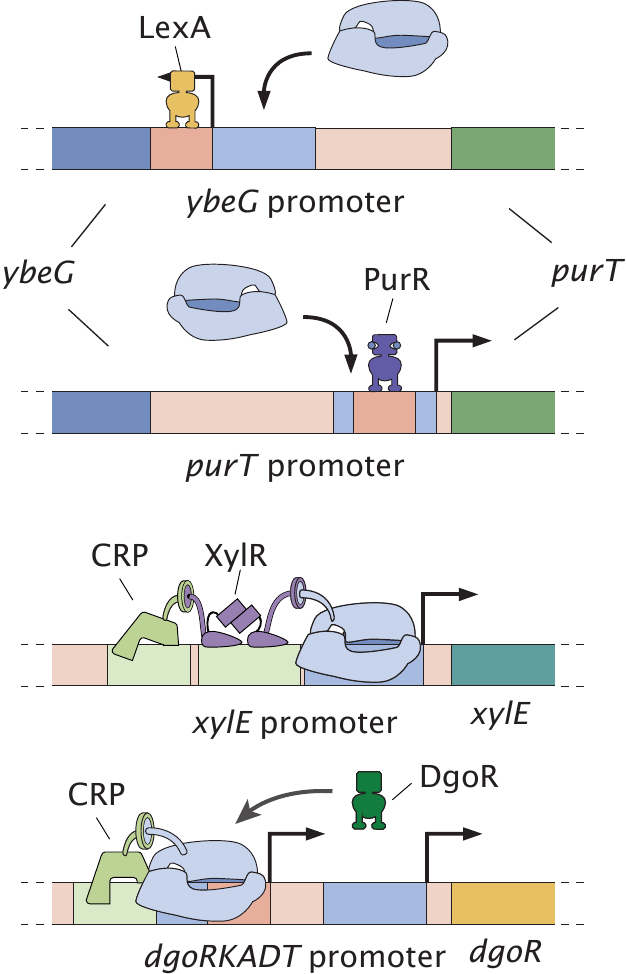}}
\caption{Beyond the {\it lac} operon in regulatory dissection.  Using the Sort-Seq method
it is now possible to identify regulatory architectures and the transcription factors
that mediate them, making it possible to do regulatory dissections like that described
here~\cite{Belliveau2018}.
\label{fig:BeyondLac}}
\end{figure}

\section{Gene regulation and statistical physics: Tactical Success But Strategic Failure?}


An interesting reflection offered on the work presented here is that
it should be viewed as a ``tactical success but a strategic failure''. There are two aspects to this critique, and each is worth addressing.
The first is that the architectures explored here are ``synthetic'' and thus anything
we learn does not apply to the ``real biology''.  In response, we note that
we set out more than a decade ago to understand gene regulation in bacteria in
a quantitative and predictive manner with a view to exporting it to the entire
regulatory genome not only of {\it E. coli}, but other more
complex organisms as well.  However, what we found was that even for the most well
studied regulatory system, we had dispiritingly little quantitative understanding
of how it would behave as the various ``knobs'' that control transcription
were tuned.  This demonstrated that we could not tackle
the complexity of real endogenous promoters with potentially quite
complex regulatory architectures without first proving to ourselves that
we could understand the most basic unit already introduced
in Jacob and Monod's repressor-operator model and denoted here
as the simple repression motif.   Though we  backpedaled from our
original goals  to do the most
simple case, we think the work showcased here demonstrates that we have
laid the groundwork for a full regulatory dissection of
the {\it E. coli} genome.  With the existence of methods like those
highlighted in Figure~\ref{fig:BeyondLac}  we are now poised to extend
these kinds of regulatory dissections to the entire genome and believe that such
work will unearth many generalizable principles~\cite{Belliveau2018}.

The second thrust of the ``tactical success but strategic failure'' critique points out that, although we were able to find a single
self-consistent minimal parameter set to describe regulation of the simple repression motif, it applies only to the particular
conditions in which these specific strains were grown; if the growth conditions are
shifted then we will need to determine the relevant parameters all over again. This might
be true, but to consider its weight
we turn to an analogous example from the long history of the physics of
materials. For a cubic material such as aluminum, we can measure the elastic constants
($C_{11}$, $C_{12}$ and $C_{44}$) of single crystals. Now if we want to use those elastic constants
to compute what will happen to a structure such as an airplane wing, we can confidently
do so.  However, if we alter the temperature of the metal away from that under which the constants were measured, the values of those elastic
constants will change. Figuring out how elastic constants are modified by temperature entailed a great deal of subsequent work~\cite{Phillips2001}.
But acknowledging that a material response is subtle does not
at all invalidate the original theory of linear elasticity, and for the gene regulatory situations considered here, we think it possible
that a similar scenario might reveal itself. The first step is to make predictions and test them to determine whether different conditions do indeed
require different parameters.  The only way to actually {\it know} what happens in complex regulatory circuits is first to master a predictive understanding
of the simplest case and subsequently build out from there.


Despite these worthy critiques, the point of this article was to show
that with sufficient care, it is indeed possible to use a single minimal parameter
set to describe a broad array of different regulatory situations.
In our view,
the results are sufficiently encouraging that it is now time to move to new systems
such as systematic studies of the regulatory landscape of newly sequenced genomes of microbes
from the ocean floor, for example.  Having made the jump
on the simple repression Rhodes, we are excited to see what comes of
efforts of the kind described here in novel microorganisms
and in the more challenging setting of multicellular organisms as well.

\section*{Acknowledgments}

We are grateful to a long list of generous colleagues who have helped
us learn about this topic.  We want to thank Stephanie Barnes,
Lacra Bintu, James Boedicker, Rob Brewster,  Robijn Bruinsma, Nick Buchler, Steve Busby, Jean-Pierre Changeux,  Barak Cohen,  Tal Einav, Uli Gerland,
Ido Golding, Terry Hwa, Bill Ireland, Justin Kinney, Jane Kondev, Tom Kuhlman,  Mitch
Lewis, Sarah Marzen, Leonid Mirny,  Alvaro Sanchez, Eran Segal,  Marc Sherman, Kim Sneppen, Franz Weinert, Ned Wingreen and Jon Widom.  We are grateful to Steve Busby, Ido Golding, Justin Kinney,
Tom Kuhlman, Steve Quake  and Alvaro Sanchez for reading the paper and providing
important feedback. We are especially grateful to Nigel Orme who has worked with us
for years to create illustrations that tell a conceptual and quantitative story
about physical biology.  It has also been a privilege to be entrusted by
the National Science Foundation, the National Institutes of Health, The California
Institute of Technology and La Fondation Pierre Gilles de Gennes with the funds that make this kind of research possible.
Specifically we are grateful to the NIH for support through award numbers DP1
OD000217 (Director's Pioneer Award) and R01 GM085286.
HGG was supported by the Burroughs Wellcome Fund Career Award at the Scientific Interface, the Sloan Research Foundation, the Human Frontiers Science Program, the Searle Scholars Program, the Shurl \& Kay Curci Foundation, the Hellman Foundation, the NIH Director's New Innovator Award (DP2 OD024541-01), and an NSF CAREER Award (1652236).\\

%
%
%
%
%
%
%
%

\begin{flushleft}
{\Large
\textbf{Figure 1 Theory Meets Figure 2 Experiments in the Study of Gene Expression\\
Supplementary Information}
}\\
\bf{Rob Phillips}$^{1,2,\ast}$,\bf{Nathan M. Belliveau}$^2$, \bf{ Griffin Chure}$^2$,  \bf{ Hernan G. Garcia}$^3$, \bf{Manuel Razo-Mejia}$^2$, \bf{Clarissa Scholes}$^4$
\\
\bf{1} Dept. of  Physics, California Institute of Technology, Pasadena, California, U.S.A
\\
\bf{2} Division of Biology and Biological Engineering, California Institute of Technology, Pasadena, California, U.S.A
\\
\bf{3} Department of Molecular \& Cell Biology, Department of Physics, Biophysics Graduate Group and Institute for Quantitative Biosciences-QB3, University of California, Berkeley, California, U.S.A
\\
\bf{4} Department of Systems Biology, Harvard Medical School, Boston, Massachusetts, U.S.A
\\
$\ast$ E-mail: phillips@pboc.caltech.edu
\end{flushleft}

\noindent This Appendix aims to spell out in full detail some of the key technical issues
that arise in the attempt to make quantitative theoretical models of
transcriptional regulation.

\section{The occupancy hypothesis}\label{sec:OccupancyHypothesis}

The theoretical models presented in this work rely on the fundamental
assumption that mRNA copy number can act as a proxy for the occupancy
of the promoter by RNA polymerase. Only through this assumption are we
able to relate experimentally accessible quantities, such as mRNA copy
number or number of fluorescent proteins, to the promoter
states that are considered theoretically. In this section we explore
the validity and reach of this so-called occupancy hypothesis by considering the
mathematical relationship between mRNA copy number, $m$, and the probability of
finding RNA polymerase bound to the promoter, $p_{bound}$.

To make this analysis possible, we consider the simple model of transcription shown
in Figure~\ref{fig:OccupancyHypothesisModel}.  As seen in the figure, we model each
step between polymerase binding and mRNA production as a zero-order transition.
In this context, the fraction of promoters in the process of initiating transcription,
$I$, is given by
\begin{equation}\label{eq:dIdt}
    {dI \over dt} = r_i p_{bound} - r_e I,
\end{equation}
where $r_i$ is the rate of initiation, and $r_e$ is the rate of elongation.   As elongation ensues, we will keep track of which base pair the
polymerase is located on using the fraction of polymerase molecules occupying
base pair $j$, which we denote by $E_j$. The fraction of molecules at the first base pair
can be obtained by solving
\begin{equation}
    {dE_1 \over dt} = r_e I - r_e E_1.
\end{equation}
Similarly, for base pair $j<N$, where $N$ is the length of the gene being
transcribed, we have
\begin{equation}
    {dE_j \over dt} = r_e E_{j-1} - r_e E_j.
\end{equation}
Finally, the fraction of polymerase molecules at the last base pair is given by
\begin{equation}\label{eq:dENdt}
    {dE_N \over dt} = r_e E_{N-1} - r_t E_N,
\end{equation}
where $r_t$ is the rate of termination. Once an mRNA is terminated we assume
that it is subject to degradation at a rate $\gamma$ such that is concentration
$m$ is given by
\begin{equation}\label{eq:dmdtGamma}
    {dm \over dt} = r_t E_N - \gamma m.
\end{equation}
By solving the system of equations shown above, we can then relate the magnitude
predicted by our models, $p_{bound}$, to the measurable number of mRNA molecules
$m$.

\begin{figure}
\centering{\includegraphics[width=5truein]{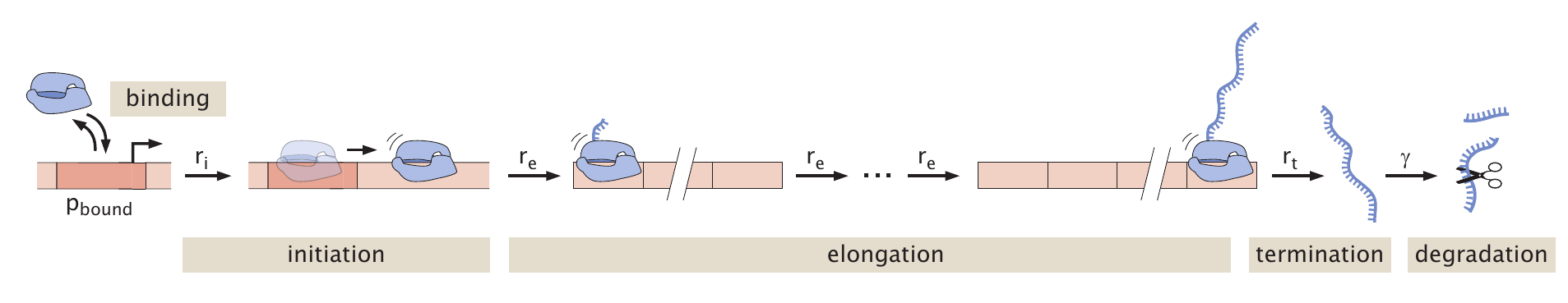}}
\caption{Simple model of mRNA production to probe the occupancy hypothesis. We assume that all steps from RNA polymerase binding to the termination and subsequent degradation of mRNA are described by zero-order kinetics.
\label{fig:OccupancyHypothesisModel}}
\end{figure}

In order to solve for $m$ using the above equations, we will assume steady-state such that all derivatives are zero. Further, due to the fact that every step in the process shown in Figure~\ref{fig:OccupancyHypothesisModel} is linear in the concentrations of the different molecular species, we can make use of a very convenient property of the system of equations. Specifically, we add up all equations together resulting in
\begin{equation}
    0 = r_i p_{bound} - \gamma m
\end{equation}
such that
\begin{equation}
    m = {r_i \over \gamma} p_{bound}.
\end{equation}
This provides us with the first important result. Specifically, under
conditions of steady-state and assuming a transcriptional cascade
composed of zero-order reactions, we find a simple linear relationship between the mRNA copy number and the
occupancy state of the promoter, as determined through $p_{bound}$.

Under slightly different assumptions, the occupancy hypothesis can also be used to relate
$p_{bound}$ to the rate of mRNA production $dm/dt$ as shown in Equation~\ref{eq:dmdtpbound}.
First, we relax the assumption made above that all the processes described by Equations~\ref{eq:dIdt}
through \ref{eq:dmdtGamma} are in steady-state. Instead we posit that only the processes up until
Equation~\ref{eq:dmdtGamma} reached this steady-state. To put this in other words, we
will set only the derivatives in Equations~\ref{eq:dIdt} through~\ref{eq:dENdt} to
zero. If we, once again, add up the system of equations, we arrive at
\begin{equation}\label{eq:dmdtrgamma}
    {dm \over dt} = r_i p_{bound} - \gamma m.
\end{equation}
Finally, we consider that mRNA degradation is negligible. This assumption true
as long as the rate of initiation is faster than the degradation term such that
$r_i \ll \gamma m$. Under this condition, we can neglect the last term on the
right-hand side of Equation~\ref{eq:dmdtrgamma} leading to
\begin{equation}
    {dm \over dt} \approx r_i p_{bound}
\end{equation}
which is Equation~\ref{eq:dmdtpbound} if we identify the
rate of transcriptional initiation $r_i$ with the effective rate of mRNA
production $r$ used throughout the main text.

\section{Equivalence of thermodynamic and statistical mechanical models of
promoter occupancy}\label{sec:ThermoVsStatMech}

We next consider how the the statistical mechanical formulation of expression
(Bintu et al. \cite{Bintu2005a}) compares with alternative thermodynamic
formulations that use the language of dissociation constants (e.g. Buchler,
Gerland, and Hwa \cite{Gerland2002b, Buchler2003a, Kuhlman2007}, and introduced by
Shea and Ackers \cite{Ackers1982, Shea1985}). We begin with the statistical
mechanical formulation of the simple repression architecture and calculate the
probability of RNA polymerase bound to its target promoter, $p_{\text{bound}}$.
We then consider how this formulation relates to thermodynamic formulations
using dissociation constants. In doing so, we are able to show how these
dissociation constants implicitly include a factor $N_{NS}$ that was explicitly present in
the statistical mechanical formulation and accounts for the reservoir of
nonspecific binding sites on the genomic background.


Regardless of how we arrive at our model of transcriptional regulation, these models
are all  founded upon an assumption that the observed expression is proportional to
the binding probability of RNA polymerase and that an assumption of
steady-state is sufficiently valid. Here we begin by outlining the statistical
mechanical formulation of the simple repression architecture \cite{Garcia2011c}.
We effectively treat the genome as a reservoir containing $N_{NS}$ nonspecific
binding sites bound by RNA polymerase and a number of different transcription factors
(Figure \ref{fig_states}(A)). Due to the high concentration of DNA in the cell it
is generally reasonable to assume that most, if not all of the transcription
factors in the cell are bound to the genomic DNA~\cite{Dehaseth1977, Kao-Huang1977}.

Here we would like to estimate the probability that RNA polymerase is bound to our simple
repression promoter, $p_{\text{bound}}$, that is present on the genome. As shown
in Figure~\ref{fig_states}(B), the promoter can either be empty, occupied by RNA polymerase,
or occupied by a repressor (in this case, LacI).  This probability depends on
the difference in free energy associated with each particular state of the
system. We will take as a reference state that where all RNA polymerase and LacI proteins
are bound nonspecifically to the genomic background. Following this approach,
the probability of bound RNA polymerase, $p_{\text{bound}}$ can be found to be given by,

\begin{equation}\label{eq_p_bound_statmech}
p_\text{bound}=\frac{\frac{P}{N_{NS}}e^{-\beta \Delta\varepsilon_P}}{1+
               \frac{R}{N_{NS}}e^{-\beta \Delta\varepsilon_R}+
               \frac{P}{N_{NS}}e^{-\beta\Delta\varepsilon_P}},
\end{equation}
with $\beta = \frac{1}{k_BT}$, where $k_B$ is the Boltzmann constant and $T$ is
the temperature of the system. Here, $\Delta \varepsilon_P$ denotes the difference in binding energy when
repressor binds the promoter, relative to nonspecific binding on the genome.
$\Delta \varepsilon_P$  similarly denotes the difference in binding energy when
RNA polymerase binds the DNA. $R$ and $P$ represent the copy number per cell of repressor
and RNA polymerase, respectively. Note that in our formulation, we have assumed that both
the repressor and RNA polymerase are unable to bind simultaneously.


Now we can consider the thermodynamic approach that was taken by Buchler,
Gerland, and Hwa \cite{Buchler2003a}. In their work, the authors adopted and
generalized the approach in the classic work of Shea and Ackers
\cite{Ackers1982, Shea1985} and so we shall begin there. In that classic work,
Shea and Ackers developed a statistical mechanical model to describe the
bacteriophage lambda switch, enumerating each possible configuration of the
regulatory architecture. Following their approach, we will denote $\Delta\acute G_P$
as the free energy for binding of RNA polymerase to the promoter, and $\Delta\acute G_R$
for binding of LacI to the promoter. In their framework, the probability
that RNA polymerase is bound to the promoter, $p_\text{bound}$, is then given by
\begin{equation}\label{eq_p_bound_definition__}
p_\text{bound}=\frac{[P] e^{-\beta \Delta \acute G_P}}{1+
               [P] e^{-\beta \Delta \acute G_P} +
               [R] e^{-\beta \Delta \acute G_R}},
\end{equation}
where $[P]$ and $[R]$ are the concentrations of unbound RNA polymerase and unbound LacI,
respectively.
The free energies can be related to
corresponding dissociation constants through the standard relationship,
\begin{equation}\label{eq_gibbs_Kd_RNAP}
\Delta \acute G_P = k_BT \ln {K_P \over c_0},
\end{equation}
and
\begin{equation}\label{eq_gibbs_Kd_R}
\Delta \acute G_R = k_BT \ln {K_R \over c_0},
\end{equation}
although note that in each case the argument of the logarithm is normalized
by a standard state concentration $c_0$, normally taken to be $1$~M.
Here $K_P$ is the dissociation constant for binding by RNA polymerase to the promoter, and
$K_R$ is the dissociation constant for binding of LacI to the promoter. These
dissociation constants represent the concentration when each binding site is
half-maximally occupied. Using these
relationships between energy and dissociation constants in
Equation~\ref{eq_gibbs_Kd_RNAP} and Equation~\ref{eq_gibbs_Kd_R}, we can
re-write  $p_\text{bound}$ as,
\begin{equation}\label{eq_p_bound_thermo}
p_\text{bound}=\frac{\frac{[P]}{K_P}}{1+
               \frac{[P]}{K_P} +
               \frac{[R]}{K_R}}.
\end{equation}
This is the thermodynamic representation that would be obtained following the approach
of Buchler, Gerland, and Hwa \cite{Buchler2003a}. Here we see that the probability is still
determined by considering the set of states available to the promoter, but with
the corresponding Boltzmann weight for binding by RNA polymerase defined by $[P]/K_P$, and
that of LacI by $[R]/K_R$.

Comparing the statistical mechanical equation of $p_\text{bound}$ in Equation
\ref{eq_p_bound_statmech} with the thermodynamic representation in Equation
\ref{eq_p_bound_thermo} above, we find that
\begin{equation}\label{eq_Kp_statmech}
 K_P = \frac{N_{NS}}{V_{\text{cell}}} e^{-\beta \Delta \varepsilon_P},
\end{equation}
and
\begin{equation}\label{eq_Kp_statmech_2}
 K_R = \frac{N_{NS}}{V_{\text{cell}}} e^{-\beta \Delta \varepsilon_R}.
\end{equation}
Here $V_{\text{cell}}$ refers to the volume of the cell and is used to translate
between protein copy numbers and concentrations. In the \textit{in vivo} context
considered here, the dissociation constants reflect binding by proteins that are
otherwise assumed to be bound to the nonspecific genomic background, and will
generally differ from what might be obtained from \textit{in vitro} measurements
\cite{Kuhlman2007}.  Hence, we argue that both the statistical mechanical and
thermodynamic formulations represent equivalent descriptions. The main
distinction is that the statistical mechanical formulation is explicit in
describing the nonspecific genomic background through the term $N_{NS}$ and
assuming one copy of the promoter.

\section{The equilibrium assumption}\label{sec:Equilibrium}

Having established the conditions under which we can connect the probability of
finding RNA polymerase bound to the promoter, $p_{bound}$,  with the rate of
mRNA production, we now ask whether it is reasonable to use the tools of
statistical mechanics to calculate $p_{bound}$. While we are encouraged by the
apparent validity of the theory based on the agreement with experimental data
shown throughout the main text, here we will carefully consider the equilibrium
assumption that underlies calculating $p_{bound}$ in the context of our minimal
parameter set (defined in Figure~\ref{fig:AztecPyramidParameters}(B)).
While it will be shown below that the rates of RNA polymerase binding and
unbinding are incompatible with an equilibrium assumption for binding by RNA
polymerase, we will find that under the weak-promoter approximation,
there exists a regime where it is indeed reasonable to apply a statistical
mechanical treatment to calculate $p_{bound}$.

First, we focus on the model of an unregulated promoter shown in
Figure~\ref{fig:EquilibriumAssumptionConstitutive}(A). Here, the promoter can be
unoccupied or occupied by RNA polymerase. The fraction of promoters in each
state is denoted by $p_{unbound}$ and $p_{bound}$, respectively.  When RNA polymerase
is bound it can also initiate transcription at a rate $r$.
Upon RNA polymerase escape from the promoter, the system is taken back to an unoccupied
state. The rate of change in the fraction of occupied promoters is given by
\begin{equation}\label{eq:dpdt2State}
    {d p_{bound} \over dt} = k_{on}^{(P)} p_{unbound} - k_{off}^{(P)} p_{bound} - r \, p_{bound}
\end{equation}
while the rate of mRNA production can be written as
\begin{equation}
    {d m \over dt} = r p_{bound}
\end{equation}
which corresponds to the rate of mRNA production as posited by the occupancy hypothesis.

\begin{figure}
\centering{\includegraphics[width=5truein]{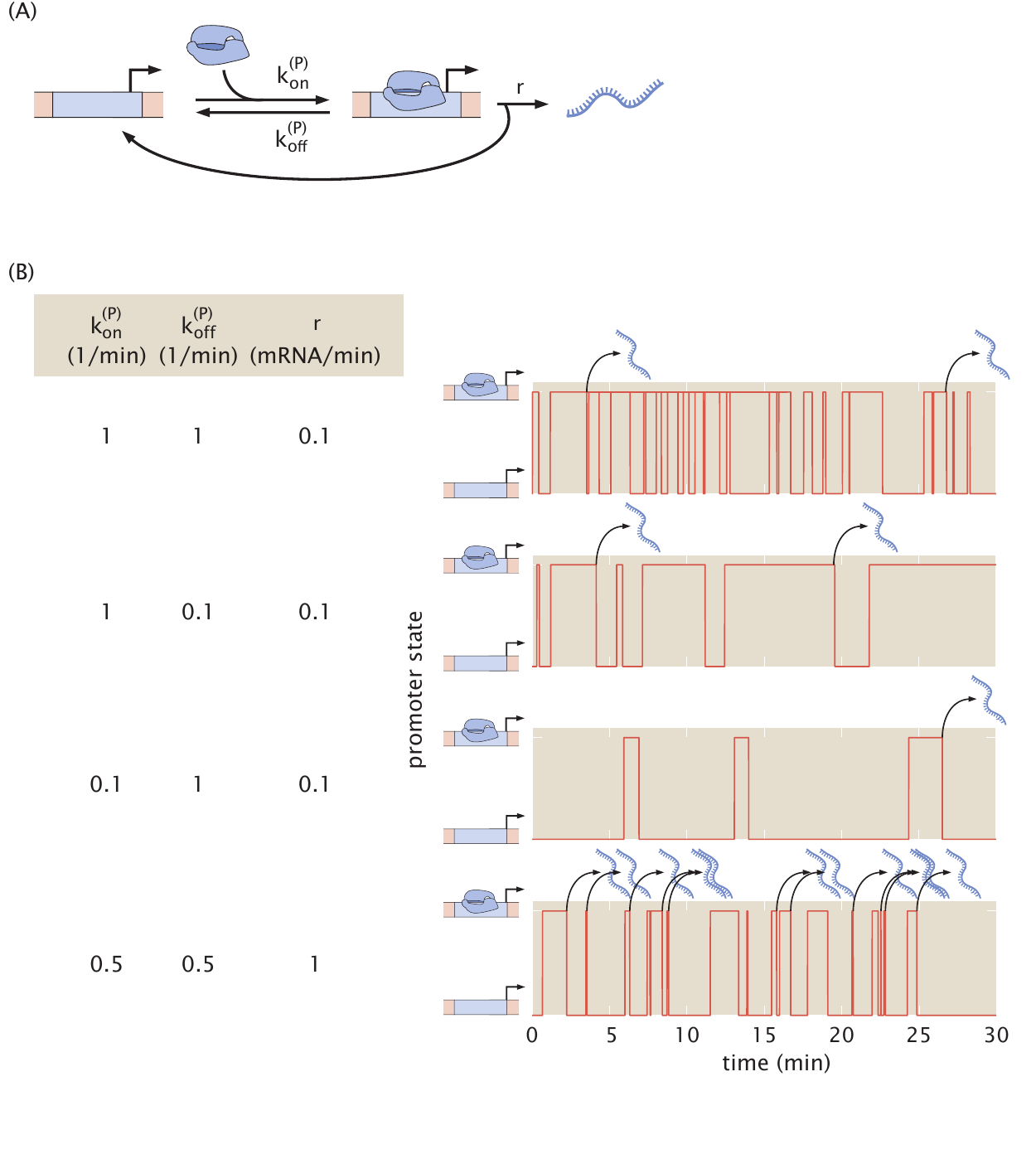}}
\caption{
Exploring the equilibrium assumption for the constitutive promoter. (A) Kinetic scheme for a constitutive promoter. (B) Stochastic simulations of promoter state and initiation events for different parameters of the constitutive promoter.
\label{fig:EquilibriumAssumptionConstitutive}}
\end{figure}

We next seek to establish under what conditions we can calculate $p_{bound}$
using statistical mechanics. In the equilibrium limit, $p_{bound}$ for this
unregulated promoter  can be calculated using the states and weights defined
in Figure~\ref{fig:ComputingRepression}(A) such that
\begin{equation}
    p_{bound}^{equil} = {{P \over N_{NS}} e^{-\beta \Delta \varepsilon_p} \over
        1 + {P \over N_{NS}} e^{-\beta \Delta \varepsilon_p}}.
\end{equation}
In Appendix~\ref{sec:ThermoVsStatMech}, we saw that this same expression can
be written in the thermodynamic language as
\begin{equation}
    p_{bound}^{equil} = {{[P] \over K_P} \over 1 + {[P] \over K_P}},
\end{equation}
where $K_P$ is the dissociation constant between RNA polymerase and the promoter. This
expression for $p_{bound}^{equil}$ can be related to the scheme shown in
Figure~\ref{fig:EquilibriumAssumptionConstitutive}(A) by using $[P]/K_P = k_{on}^{(P)}/k_{off}^{(P)}$
such that
\begin{equation}\label{eq:pboundEquil}
    p_{bound}^{equil} = {k_{on}^{(P)}  \over k_{on}^{(P)} + k_{off}^{(P)}}.
\end{equation}
In order to  calculate $p_{bound}$ without enforcing equilibrium,
we invoke steady-state in the fraction of
occupied and unoccupied promoters such that Equation~\ref{eq:dpdt2State} can be
written as
\begin{equation}\label{eq:dpdt2State_ss}
    0 = k_{on}^{(P)} p_{unbound} - k_{off}^{(P)} p_{bound} - r \, p_{bound}.
\end{equation}
We now make use of the fact that the probabilities are normalized, $p_{bound} + p_{unbound} = 1$ in order to obtain
\begin{equation}\label{eq:pbound}
    p_{bound} = {k_{on}^{(P)} \over k_{on}^{(P)} + k_{off}^{(P)} +r}.
\end{equation}
Clearly, $p_{bound}$ in Equation~\ref{eq:pbound} is not equal to
$p_{bound}^{equil}$ in Equation~\ref{eq:pboundEquil}. The only way to recover
$p_{bound}^{equil}$ is for the rate of initiation $r$ to be much slower that
one of the other rates in the system. Namely, we need $r \ll k_{on}^{(P)}$ or $r
\ll k_{off}^{(P)}$ such that $ k_{on}^{(P)} + k_{off}^{(P)} +r \approx
k_{on}^{(P)} + k_{off}^{(P)}$. These different limits are explored in
Figure~\ref{fig:EquilibriumAssumptionConstitutive}(B) through stochastic
simulations that calculate the promoter state and initiation events as a
function of time. In the first three simulations within
Figure~\ref{fig:EquilibriumAssumptionConstitutive}(B), we show how, when the
conditions described above are met, the promoter cycles multiple times between
its bound and unbound state before an initiation event ensues. This
back-and-forth between the bound and unbound states leads to {\it
quasiequilibrium}. That is, the fact that the transitions between the bound and
unbound states are faster than the rate of initiation allows us to invoke
separation of time scales such that, at each time point, we can use
statistical mechanics to describe the equilibrium between these two states.
However, if $r$ is larger than these transition rates, most instances of the
promoter being bound lead to an initiation event as shown in the last simulation
in the Figure~\ref{fig:EquilibriumAssumptionConstitutive}(B) and there is no longer
a separation of time scales.

Interestingly, the inferred transition rates from
Figure~\ref{fig:AztecPyramidParameters}(B) do not fulfill this condition as
$k_{on}^{(P)}, k_{off}^{(P)} < r$. Thus, at least {\it a priori}, equilibrium
cannot be invoked to describe the transcription of an  {\it unregulated} {\it
lac} promoter. However, the successes of the theory at predicting experiments
suggest that, under certain conditions, we are still allowed to invoke the
quasi-equilibrium assumption for the {\it regulated} {\it lac} promoter.

We next consider the kinetic scheme for the regulated promoter,
shown in Figure~\ref{fig:EquilibriumSimpleRepression}(A).
The reader is reminded that this scheme does not make any assumption about the
relative strength of each transition rate or about equilibrium. In this context, we are first
interested in asking whether the probability of finding RNA polymerase bound to
the promoter $p(3) = p_{bound}$, which we solved for in Equation~\ref{eq:p3}, is
equivalent to the same probability that can be calculated in the equilibrium
case, $p_{bound}^{equil}$, shown in Equation~\ref{eq:pboundPR}.

To make progress, we rewrite $p_{bound}^{equil}$ in Equation~\ref{eq:pboundPR}
in the language of dissociations constants
\begin{equation}\label{eq:pBoundPRKd}
    p_{bound}^{equil} = { {[P] \over K_P} \over 1 + {[P] \over K_P} + {[R] \over K_R}}.
\end{equation}
Invoking the identities introduced in Section~\ref{sec:BeyondMean}
such that $k_{on}^{(R)} = k_{+}^{(R)} [R]$ and $k_{on}^{(P)} = k_{+}^{(P)} [P]$,
and the definition of the dissociations constant for repressor and RNA
polymerase given by $k_{off}^{(R)} / k_+^{(R)} = K_R$ and $k_{off}^{(P)} /
k_+^{(P)} = K_P$, respectively, we obtain
\begin{equation}\label{eq:pBoundPRRates}
    p_{bound}^{equil} = { {k_{on}^{(P)} \over k_{off}^{(P)}} \over 1 + {k_{on}^{(P)} \over k_{off}^{(P)}} + {k_{on}^{(R)} \over k_{off}^{(R)}}}.
\end{equation}
In contrast, $p_{bound}$ from Equation~\ref{eq:p3}, which is absent of any assumption of equilibrium, is given by
\begin{equation}
    p_{bound} = { {k_{on}^{(P)} \over k_{off}^{(P)} + r} \over 1 + {k_{on}^{(P)} \over
        k_{off}^{(P)} + r} + {k_{on}^{(R)} \over k_{off}^{(R)}}}.
\end{equation}
Again, as with the unregulated promoter, we find that the expression for
$p_{bound}$ is not equal to $p_{bound}^{equil}$. One way to alleviate this
discrepancy is through the quasiequilibrium assumption noted above, requiring
that the rate of RNA polymerase unbinding is much faster than the rate of
initiation, $k_{off}^{(P)} \ll r$. However, Figure~\ref{fig:AztecPyramidParameters}(B)
reveals that  $k_{off}^{(P)} \approx
r$ and not $k_{off}^{(P)} \ll r$ as demanded above for the quasiequilibrium
approximation to apply. Interestingly, at least for the case of simple repression considered
here, we will see below that the equilibrium
assumption can still be invoked under certain conditions for the calculation of the
fold-change in gene expression.

In Equation~\ref{eq:FoldChangeKinetic} in the main text, we calculated the
fold-change in gene expression corresponding to the kinetic scheme presented in
Figure~\ref{fig:EquilibriumAssumptionRepressed} and reproduced in
Figure~\ref{fig:EquilibriumSimpleRepression}(A).  This
calculation made no assumption regarding equilibrium and resulted in
\begin{equation}\label{eq:FoldChangeKineticV2}
    \mbox{fold-change} = { 1 + {k_{on}^{(P)} \over k_{off}^{(P)} + r} \over 1 + {k_{on}^{(P)} \over k_{off}^{(P)} + r} + {k_{on}^{(R)} \over k_{off}^{(R)}}}.
\end{equation}
Our objective is then to determine under what limits we can reduce this
fold-change to its equilibrium counterpart obtained in
Equation~\ref{eq:FoldChangeFull} or in the context of the weak-promoter
approximation shown in Equation~\ref{eq:FoldChangeSimpleRepStatMech}.

As expected from our calculations on the applicability of equilibrium to derive
$p_{bound}$, if we assume that $k_{off}^{(P)} \ll r$,
Equation~\ref{eq:FoldChangeKineticV2} reduces to the fold-change in equilibrium
shown in Equation~\ref{eq:FoldChangeFull}. We already saw that this limit is not
consistent with the inferred rates.
However, instead, consider the limit where $k_{on}^{(P)} \ll k_{off}^{(P)} + r$.
In this case, we can neglect the term ${k_{on}^{(P)} \over k_{off}^{(P)} + r}$ in
Equation~\ref{eq:FoldChangeKineticV2} such that the fold-change reduces to
\begin{equation}\label{eq:FoldChangeKineticV3}
    \mbox{fold-change} \approx { 1 \over 1 + {k_{on}^{(R)} \over k_{off}^{(R)}}} = {1 \over 1 + {[R] \over K_R}},
\end{equation}
which corresponds to the fold-change in equilibrium under the weak-promoter
approximation shown in
Equations~\ref{eq:FoldChangeSimpleRepStatMech}~and~\ref{eq:SimpleRepressionFoldChangeKd}.
In Figure~\ref{fig:EquilibriumSimpleRepression}(B) we explore this regime using
stochastic simulations. The simulation reveals that, in this limit, the promoter mostly transitions
between its repressor-occupied state and its empty state. Only rarely will the system
transition to the RNA polymerase-bound state and, on these rare occasions, this event almost
always leads to the initiation of transcription and the return of the promoter to its empty state.
As a result, there is a clear separation of time scales between the process of repressor binding
and unbinding and the subsequent steps in the transcriptional cascade. This separation of time
scales justifies the applicability of the quasiequilibrium assumptions to calculate the fold-change in gene
expression in terms of the probability of repressor binding.

As seen in Figure~\ref{fig:AztecPyramidParameters}(B), our estimates for
$k_{on}^{(R)}$, $k_{off}^{(R)}$ and $r$ suggest that we are in this regime
where the fold-change in gene expression can be calculated using the tools of
statistical mechanics despite the fact that the probability of RNA polymerase
binding to the promoter cannot be obtained using such equilibrium
considerations. Thus, by considering fold-change instead of $p_{bound}$ directly,
we are able to ignore the potentially non-equilibrium behavior of RNA
polymerase.

\begin{figure}
\centering{\includegraphics[width=5truein]{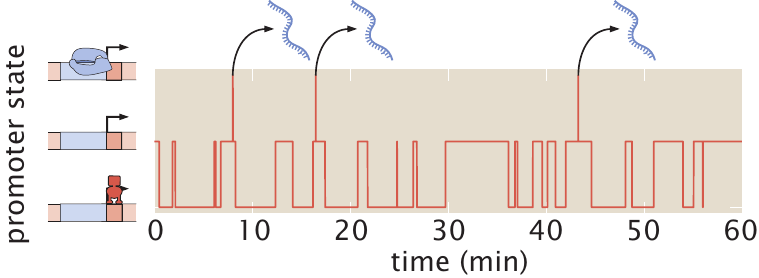}}
\caption{
Exploring the equilibrium assumption for simple repression.
Stochastic simulations of
promoter state and initiation events
for the kinetic scheme introduced in Figure~\ref{fig:EquilibriumAssumptionRepressed}
for different parameters of the regulated
promoter, for the case where $k_{on}^{(P)} \ll k_{off}^{(P)} + r$. Here we
observe many more binding and unbinding events by the repressor than by RNA
polymerase, characteristic of our statistical mechanical description. The parameters
used are $k_{on}^{(P)} = 0.1~\mbox{min}^{-1}$, $k_{off}^{(P)} = 1~\mbox{min}^{-1}$,
$k_{on}^{(R)} = 0.5~\mbox{min}^{-1}$, $k_{off}^{(R)} = 0.5~\mbox{min}^{-1}$, and
$r = 60~\mbox{min}^{-1}$.
\label{fig:EquilibriumSimpleRepression}}
\end{figure}

%

\section{The nonspecific genomic background}\label{sec:NSGenomicBackground}

A simplifying assumption often made in thermodynamic models of transcription is
the idea that the binding of transcription factors to nonspecific sites is characterized
by a single binding energy as shown in Figure~\ref{fig:ThreeBindingEnergyDistributions}(A).
In this case, the partition function for putting $P$ polymerases on the nonspecific background
is
\begin{equation}
Z_{NS}(P,N_{NS})={N_{NS}^P \over P!} e^{-\beta P \varepsilon_{NS}}.
\label{eqn:ZNonspecific}
\end{equation}
Of course, this is a convenient simplifying assumption that is pedagogically helpful,
but raises the question of whether it masks some important effect.  In fact, as we
show in the remainder of this section, even when the nonspecific background is characterized
by a distribution of energies, ultimately, it can be represented by an equation of
the form Equation~\ref{eqn:ZNonspecific}, but with the energy $\varepsilon_{NS}$ replaced
by an {\it effective energy}.

\begin{figure}
\centering{\includegraphics[width=5.5truein]{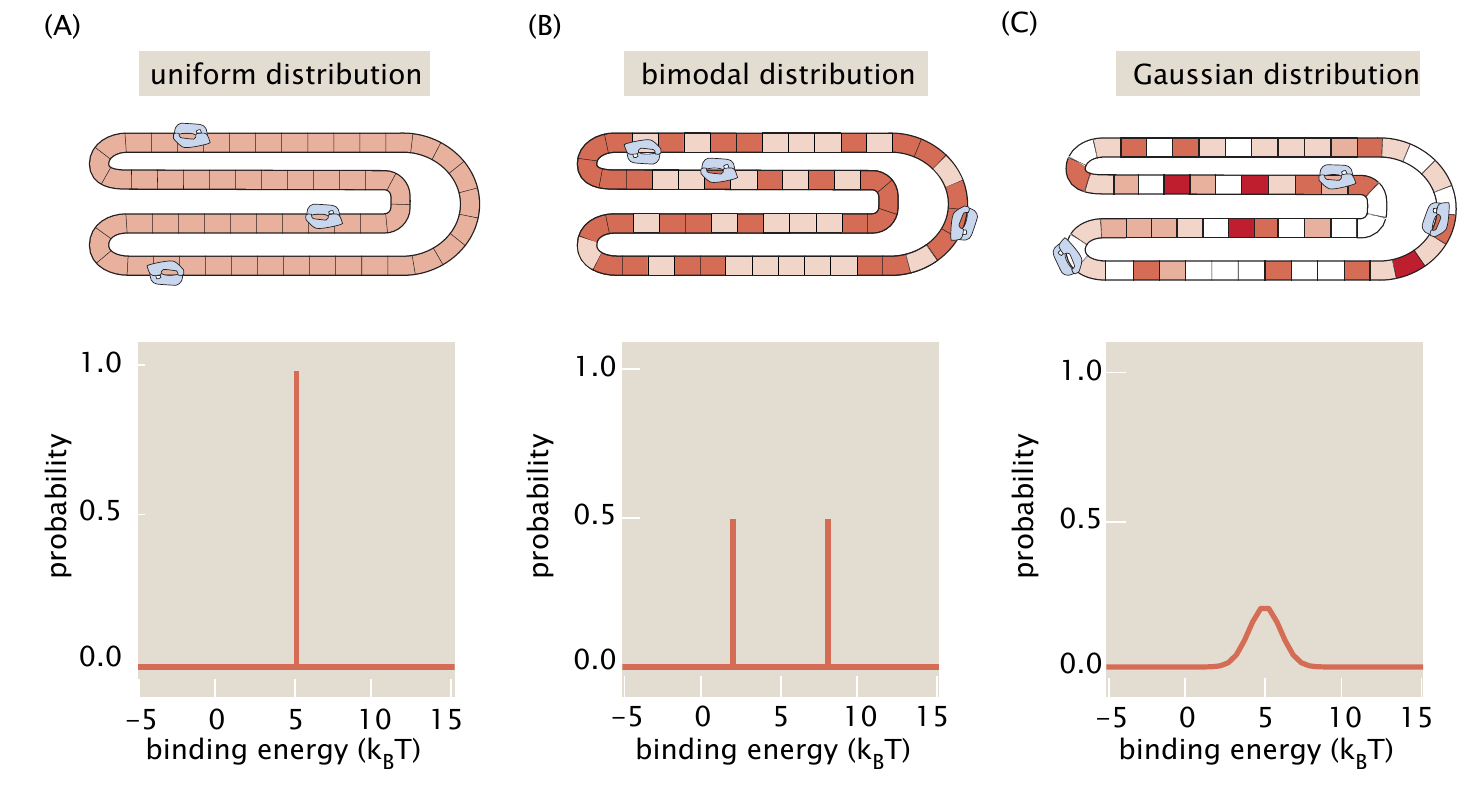}}
\caption{Increasingly sophisticated models of the nonspecific background.
(A) Uniform background. (B) Two-state model of the nonspecific background. (C)
Nonspecific binding energies characterized by a Gaussian distribution.
\label{fig:ThreeBindingEnergyDistributions}}
\end{figure}

To get a feeling for how the effective energy arises, we begin with a toy model of
the nonspecific background as shown in Figure~\ref{fig:ThreeBindingEnergyDistributions}(B).
In this case, the $P$ polymerases are distributed between the $N_{NS}/2$ sites available
with binding energy $\varepsilon_1=\bar{\varepsilon}+\Delta$ and the
$N_{NS}/2$ sites available
with binding energy $\varepsilon_2=\bar{\varepsilon}-\Delta$ such that $\bar{\varepsilon}$ is the mean non-specific binding energy.  To compute the
partition function, we need to sum over {\it all} the ways of distributing  the $P$ polymerases
over the two nonspecific reservoirs.  We imagine that
the number bound on reservoir 1 is $i$ and the number bound on reservoir 2 is $P-i$,
and then sum over all $i$ ranging from $i=0$ all the way to $i=P$, resulting in
\begin{equation}
Z_{NS}=\sum_{i=0}^P g_1(i)g_2(P-i) e^{-\beta[i \varepsilon_1+(P-i) \varepsilon_2]},
\end{equation}
where $g_1(i)$ is the number of ways of distributing $i$ polymerases over the $N_{NS}/2$ sites
of reservoir 1 and $g_2(P-i)$ is the number of ways of distributing $P-i$ polymerases over
the $N_{NS}/2$ sites
of reservoir 2.
Because $i << N_{NS}/2$, we can write $g_1(i)$ as
\begin{equation}
g_1(i) \approx {({N_{NS} \over 2})^i \over i!}
\end{equation}
and similarly write $g_2(P-i)$ as
\begin{equation}
g_2(P-i) \approx {({N_{NS} \over 2})^{P-i} \over (P-i)!}.
\end{equation}
In light of these results, we can now rewrite the partition function for nonspecific binding as
\begin{equation}
Z_{NS}=\sum_{i=0}^P {({N_{NS} \over 2})^{P} \over i!(P-i)!} e^{-\beta[i \varepsilon_1+(P-i) \varepsilon_2]}
\end{equation}
which can be rewritten as
\begin{equation}
Z_{NS}={({N_{NS} \over 2})^{P} \over P!} e^{-\beta P \varepsilon_2} \sum_{i=0}^P {P! \over i!(P-i)!} e^{-\beta i (\varepsilon_1- \varepsilon_2)},
\end{equation}
where we have multiplied the previous expression by $P!/P!=1$ in anticipation of beating our
formula into the form of a binomial.
Indeed, our sum is now of the form of a binomial allowing us to use
\begin{equation}
\sum_{i=0}^P {P! \over i!(P-i)!} x^P =(1+x)^P.
\end{equation}
As a result, we can write our partition function in the form
\begin{equation}
Z_{NS}={N_{NS} ^{P} \over P!} {1 \over 2^P} (e^{-\beta \varepsilon_2} (1+ e^{-\beta  (\varepsilon_1- \varepsilon_2)}))^P.
\label{eqn:Z2reservoirs}
\end{equation}
This should be compared with
\begin{equation}
Z_{NS} ={N_{NS}^P \over P!} e^{-\beta P \varepsilon_{NS}}
\label{eqn:Z1reservoir}
\end{equation}
which is the result for the partition function for the most simple model in which
the nonspecific background is assumed to be uniform.

We now want to see whether our expression given in eqn.~\ref{eqn:Z2reservoirs} is equivalent
to the single reservoir model.
By equating eqn.~\ref{eqn:Z2reservoirs} and eqn.~\ref{eqn:Z1reservoir} and taking the log of both sides we have
\begin{equation}
\varepsilon_{NS}= k_BT ~ \mbox{ln} 2 + \varepsilon_2 -k_BT~\mbox{ln}~(1+e^{-\beta(\varepsilon_1-\varepsilon_2)})
\end{equation}
We can simplify this by noting that the term involving the logarithm can be simplified as
\begin{equation}
\mbox{ln} (1+e^{-\beta(\varepsilon_1-\varepsilon_2)})
= \mbox{ln} (1+e^{-2\beta \Delta}) \approx \mbox{ln} (1+1-2\beta \Delta)
\approx \mbox{ln}~ 2 +\mbox{ln}~(1-\beta \Delta),
\end{equation}
where we have used the fact that $\varepsilon_1-\varepsilon_2=2 \Delta$.
Given that $\beta \Delta << 1$ (i.e. the energy difference between the two states is
small), we can use the Taylor series $\mbox{ln}~(1-x) \approx -x$
with the result that
\begin{equation}
\varepsilon_{NS} = \bar{\varepsilon}
\end{equation}
This result shows us that in the toy model of the nonspecific background of
Figure~\ref{fig:ThreeBindingEnergyDistributions}(B), the two nonspecific backgrounds
are equivalent to a single reservoir with an energy given by the mean of the energies
of the two reservoirs, establishing that in this pedagogically motivated model
we can use a single energy to describe the nonspecific background.
Now let's move to the case of realistic distribution of nonspecific energies.

Figure~\ref{fig:NonSpecificReservoir} shows the  distribution of nonspecific
binding energies obtained by taking the energy matrix describing the binding of
LacI and applying it to all sites across the {\it E. coli} genome
(also see Figure~\ref{fig:ThreeBindingEnergyDistributions}(C) for a comparison with the other
models considered thus far).
Other
examples of the distribution of nonspecific binding energies have
been considered as well with similar outcome~\cite{Gerland2002b, Sengupta2002}.
As a result, we can write the number of binding sites with energy between $E$ and $E+dE$ as
\begin{equation}
n(E)= {N_{NS} \over \sqrt{2 \pi \sigma^2}} e^{-(E-\bar{\varepsilon})^2 /2 \sigma^2},
\end{equation}
where $\bar{\varepsilon}$ is the mean of the distribution of nonspecific binding energies
and $\sigma$ provides a measure of the width of that distribution.

To compute the partition function for the binding of a polymerase, for example,
to this nonuniform genomic background, we need to sum over all the microscopic states
available to the polymerase.  Symbolically, the quantity we need to evaluate is
\begin{equation}
Z_{NS}=\sum_{E} n(E) e^{-\beta E}.
\end{equation}
In fact, since we are assuming a continuous distribution of energies, this really
is an integral of the form
\begin{equation}
Z_{NS}= \int_{\infty}^{\infty} e^{-\beta E} {N_{NS} \over \sqrt{2 \pi \sigma^2}} e^{-(E-\bar{\varepsilon})^2 /2 \sigma^2} dE.
\end{equation}
This result
can be rewritten as
\begin{equation}
Z_{NS}={N_{NS} \over \sqrt{2 \pi \sigma^2}} e^{-\bar{\varepsilon}^2 /2 \sigma^2}
\int_{\infty}^{\infty}  e^{-{1 \over 2 \sigma^2}[E^2  - 2E \bar{\varepsilon}+ 2 \sigma^2 \beta E]} dE.
\end{equation}
By completing the square, this integral results in
\begin{equation}
Z_{NS}=N_{NS} e^{-\beta \bar{\varepsilon}} e^{\beta^2 \sigma^2 /2}
\end{equation}
which should be compared with the result we would get if we assumed a
homogeneous nonspecific background with only a single binding energy $\varepsilon_{NS}$
resulting in the form
\begin{equation}
Z_{NS}=N_{NS} e^{-\beta \varepsilon_{NS}}
\end{equation}
By equating these two expressions, we find that we can treat the nonuniform background
as though it were a homogenous genomic background with effective binding energy
\begin{equation}
\varepsilon_{eff}=\bar{\varepsilon}-{\beta \sigma^2 \over 2}.
\end{equation}

The result above considered a single polymerase or repressor molecule bound to
the nonuniform nonspecific background.  What happens in the case where we have
$P$ polymerases bound nonspecifically?  Because each of those
polymerases binds independently of the others (because the number of polymerases
is of order $10^3 - 10^4$ and the genome size is greater than $10^6$ we don't need
to worry about polymerases interfering with each other), the total partition function
for all of these polymerases bound to the nonspecific background is given by
\begin{equation}
Z_{NS}(P,N_{NS})= {(\int_{\infty}^{\infty} e^{-\beta E} {N_{NS} \over \sqrt{2 \pi \sigma^2}} e^{-(E-\bar{\varepsilon})^2 /2 \sigma^2} dE)^P  \over P! }=  {N_{NS}^P e^{-\beta P \varepsilon_{eff}}  \over P!},
\end{equation}
where once again $\varepsilon_{eff}=\bar{\varepsilon}-{\beta \sigma^2 \over 2}$ and this result
shows that if the distribution of binding energies is Gaussian, then we can
treat the nonspecific background as being equivalent to a uniform nonspecific
background with energy $\varepsilon_{eff}$.  The point of all of this analysis was simply
to examine the validity of the convenient simplifying assumption of some thermodynamic
models of treating the nonspecific background as uniform.   As shown elsewhere~\cite{Gerland2002b,Sengupta2002}, this approximation is quite reasonable.\\

\section{Accounting for the effect of nonspecific promoter occupancy}\label{sec:NSPromoterBinding}

So far our statistical mechanical treatment of the simple repression
architecture has treated the RNA polymerase and LacI proteins as isolated from the pool of
other transcription factors that are also littered across the genomic DNA. In
Figure~\ref{fig:schmidt_lac} we plot the abundance of DNA-binding
proteins per cell across a number of growth conditions using the proteomic study
from Schmidt et al. \cite{Schmidt2016}. These values include nucleoid-associated
proteins that also bind the genomic DNA. For growth in M9 minimal media with
0.5\% glucose, we find that there are about $3 \times 10^5$ DNA-binding proteins
per cell and we can use this to make a simple estimate of genomic occupancy by
these proteins. Let us assume that each transcription factor binds the DNA as a
dimer (this will vary with the transcription factor species) and occupies a DNA
length of 15 bp (this varies from 7 bp to 38 bp in \textit{E. coli} for
transcription factors listed on RegulonDB; \cite{GamaCastro2016}). For growth
in 0.5\% glucose, we find that about 2.3 Mbp or about half the genome is
occupied (15 bp$\times$3$\times$10$^5$DNA-binding proteins$\times$1/2 dimers per
protein).

Given the high occupancy of DNA-binding proteins on the genomic DNA estimated
above, there might be some expectation that, in contrast to our current model of
simple repression, the occupancy of the genome by these other DNA-binding
proteins cannot be ignored. Here we consider the effect of their occupancy by
adding an explicit set of states to represent the case where these additional
DNA-binding proteins can occupy the roughly 60 bp promoter region of our simple repression
architecture. For simplicity we assume that these proteins only bind nonspecifically,
ignoring any potential sequence-specific effects.
In Figure~\ref{fig_statesandweights_nonspecificTF}(A) we show the states and weights
of the simple repression promoter, where we have included this additional set of states. We could have
extended this further, either by treating each additional DNA-binding protein
species separately, or by being more careful about our specification of these additional  states. However,
the point of this exercise is to see what effect the pool of nonspecifically
bound DNA-proteins might have on our model. We can calculate $p_\text{bound}$ which,
if we invoke the weak promoter approximation $\left({P \over N_{NS}} e^{-\beta
\Delta\varepsilon_{P}} \ll 1 \right)$, is given by
\begin{equation}
  p_\text{bound} = {{P \over N_{NS}} e^{-\beta \Delta\varepsilon_{P}} \over 1 + L \cdot {C_{NS} \over N_{NS}}
                + {R \over N_{NS}} e^{-\beta \Delta\varepsilon_{R}}}.
\end{equation}
$L$ represents the number of ways other DNA-binding proteins may bind the
promoter nonspecifically, and for simplicity is taken as the length of the
promoter region ($L$ $\approx$ $60$ bp). $C_{NS}$ represents the copy number of
all other DNA-binding proteins bound to the genome that we noted earlier.
Fold-change, which is the ratio of $p_\text{bound}(R \geq
0)$ to $p_\text{bound}(R = 0)$, will then be given by
\begin{equation}\label{eq_fold_change_new}
  \text{fold-change} = \frac{1 + L \cdot {C_{NS} \over N_{NS}}}{{P \over N_{NS}} e^{-\beta \Delta\varepsilon_{P}}} \cdot
  \frac{{P \over N_{NS}} e^{-\beta \Delta\varepsilon_{P}}}{1 + L \cdot {C_{NS} \over N_{NS}} + {R \over N_{NS}} e^{-\beta \Delta\varepsilon_{R}}}.
\end{equation}
Here, the RNA polymerase components ${P \over N_{NS}} e^{-\beta \Delta\varepsilon_{P}}$ cancel out and upon some
rearrangement, we find that
\begin{equation}\label{eq_fold_change_new2}
  \text{fold-change} = \frac{1}{1 + {R \over N_{NS}} e^{-\beta \Delta\varepsilon_{R}}(1 + L \cdot {C_{NS} \over N_{NS}})^{-1}}.
\end{equation}
Using $C_{NS} \approx 1.5 \times 10^5$, which is based on our estimate of the
total DNA-binding protein  copy number found above for growth in glucose (bound as dimers),  we
calculate a value of $L \cdot {C_{NS} \over N_{NS}} \approx 2$. Importantly, we
find that this additional term in our fold-change equation does not depend on the key parameters of our
simple repression architecture, namely the repressor copy number or repressor
binding energy, and we can arrive back to our original form of fold-change
by a defining $N_{NS}' =
N_{NS} \times (1+ L \cdot {C_{ns} \over N_{NS}})$.

\begin{figure}
\centering
\includegraphics[scale=1]{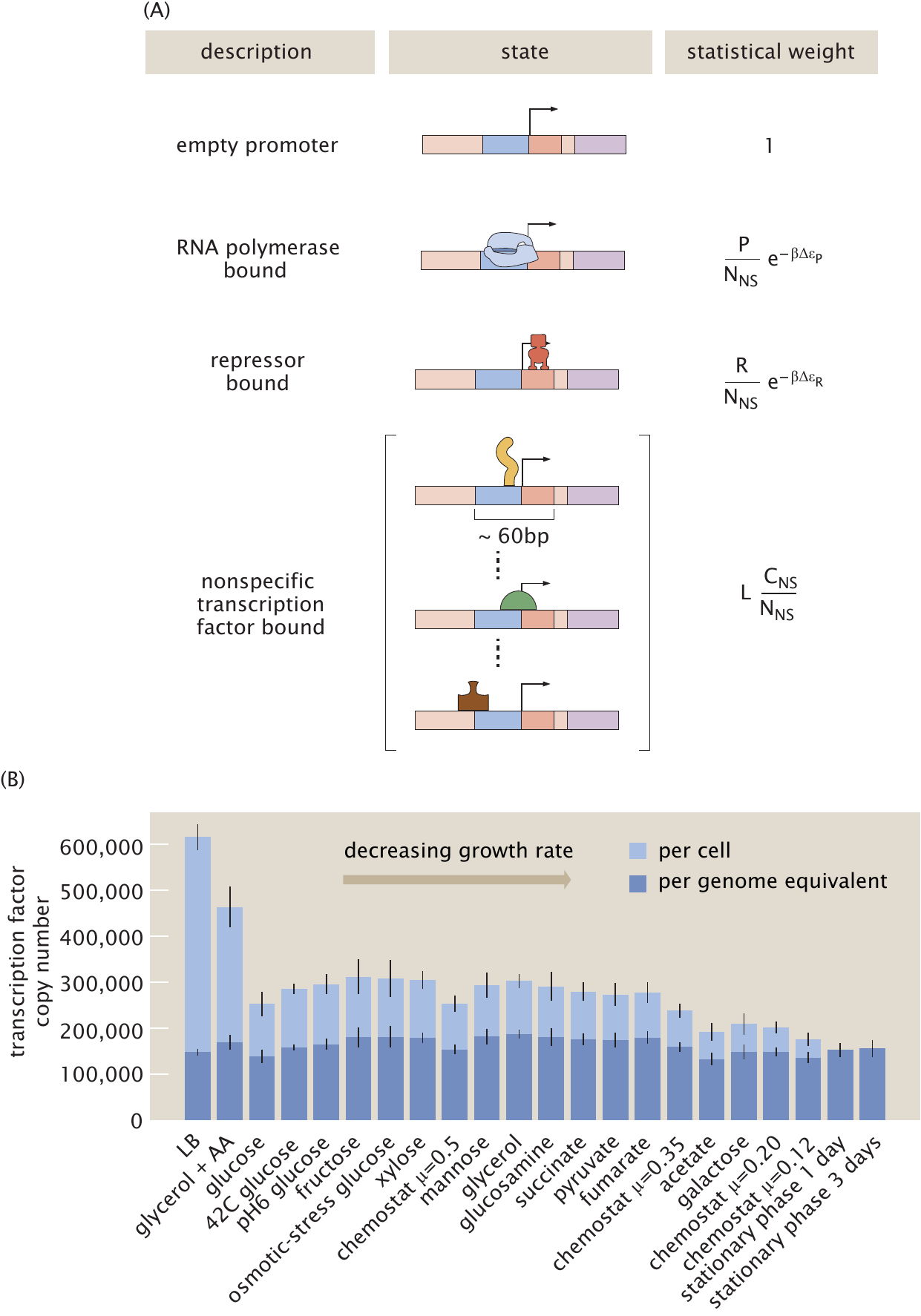}
\caption{A crowded chromosome.  (A) States and Weights for simple repression
with a pool of nonspecific  DNA binding proteins.   RNA polymerase (light blue),
a repressor, and other nonspecific DNA binding proteins compete for binding to a
promoter.  The $R$ repressors and $P$ RNA polymerase bind with energies
$\Delta\varepsilon_R$ and $\Delta\varepsilon_P$, respectively. In addition,
there are $C_{NS}$ DNA binding proteins per cell that can bind the promoter of
length $L \approx 60$ bp. These proteins  bind nonspecifically and therefore
only contribute an entropic term. $N_{NS}$ represents the number of nonspecific
binding sites on the genome. (B) Measured protein copy numbers are shown for DNA
binding proteins in {\it E. coli} across 22 growth conditions. Protein copy
numbers per cell were determined by Schmidt et al. \cite{Schmidt2016} with
proteins identified based on their annotation in EcoCyc. Error bars are
propagated from the reported standard deviations. Protein copy numbers per
genome equivalent were calculated by estimating the total genomic content as a
function of growth rate using Cooper and Helmstetter's model of {\it E. coli}
chromosomal replication \cite{Cooper1968,Dennis2008,Kuhlman2012}.
}
\label{fig_statesandweights_nonspecificTF}
\index{figures}
\end{figure}

The estimates so far were based on assuming that cells grow in 0.5\% glucose at
a particular doubling rate. In different media, the growth rate will change
leading also to a modulation in the total number of transcription factors:
faster growing cells have a larger protein complement than their slower-growing
counterparts. However, faster growing cells also have more copies of the genome
as a means to keep up with the fast replication pace.
Figure~\ref{fig_statesandweights_nonspecificTF}(B) shows that these two effects
cancel each other out. Specifically, variations in the number of transcription
factors as a result of changes in growth rate are counteracted by the
corresponding change in the average genome copy number per cell such that the
number of nonspecific binding proteins per base pair remains approximately
constant throughout a wide range of growth conditions. As a result, the small
effect of considering all nonspecifically bound transcription factors remains
unaltered regardless of growth rate.\\


%
%
%
%
%
%
%
%

\noindent{\bf References}\\

Only a limited number of references from this vast field could be cited
due to space considerations. \\

\bibliographystyle{model1-num-names}
\bibliography{PaperLibraryAnnRevFinal}


\end{document}